\documentclass[11pt]{article}

\newcommand{\prj}[2]{{\mr p}^{#1}_{#2}}
\newcommand{\inj}[2]{{\mr i}_{#1}^{#2}}
\newcommand{\fnlsum}[1]{\Sigma(#1)}

\newcommand{\join}[3]{[#1 \cdot #2,#3]}
\newcommand{\ld}[3]{[#1 \ast #2,#3]}
\newcommand{\dcq}[3]{\hat\chi(#1)_{#2}^{#3}}
\newcommand{\dcqc}[3]{\hat\chi^\CC(#1)_{#2}^{#3}}
\newcommand{\cq}[3]{\chi(#1)_{#2}^{#3}}

\newcommand{\rc}[4]{R(#1)^{#2 #3}_{#4}}
\newcommand{\rcneu}[6]{R(#1)^{#2 #3 , #6}_{#4 , #5}}
\newcommand{\rct}[4]{R(#1)_{#2 #3}^{#4}}
\newcommand{\rctneu}[6]{R(#1)_{#2 #3 , #6}^{#4 , #5}}
\newcommand{\weight}{{\tt w}}
\newcommand{\llangle}{\text{$\langle$}\!\text{$\langle$}}
\newcommand{\rrangle}{\text{$\rangle$}\!\text{$\rangle$}}

\newcommand{\bigket}[1]{\big|#1\big\rangle}

\newcommand{\bigbra}[1]{\big\langle#1\big|}

\newcommand{\bigbraket}[2]{\big\langle#1\big|#2\big\rangle}

\newcommand{\Bigket}[1]{\Big|#1\Big\rangle}
\newcommand{\Bigbra}[1]{\Big\langle#1\Big|}

\newcommand{\BF}[4]{\big(\chi_{_{\scriptstyle #1,#2}}\big){}^{#4}_{#3}}
\newcommand{\BFC}[4]{\big(\chi_{_{\scriptstyle #1,#2}}^{\CC}\big){}^{#4}_{#3}}
\newcommand{\dBF}[4]{\big(\hat\chi_{_{\scriptstyle #1,#2}}\big){}^{#4}_{#3}}
\newcommand{\dBFC}[4]{\big(\hat\chi_{_{\scriptstyle #1,#2}}^{\CC}\big){}^{#4}_{#3}}

\newcommand{\ccg}[4]{C(#1)^{#2}_{#3,#4}}
\newcommand{\ocg}[7]{C^{#1,#2,#3,#4}_{#5,#6,#7}}
\newcommand{\ocgs}[6]{C^{#1,#2,#3}_{#4,#5,#6}}

\usepackage[all]{xy}

\usepackage{geometry} 
\usepackage{graphicx}
\usepackage{tabularx}
\usepackage{typearea}
\usepackage{wrapfig}
\usepackage{fancyhdr}
\usepackage{amsmath}
\usepackage{amssymb}
\usepackage{amsthm}
\usepackage{hyperref}
\usepackage{accents}
\usepackage[ansinew]{inputenc}
\usepackage{enumerate}
\usepackage{bbm}
\usepackage{mathtools}
\usepackage{cite}
\usepackage{tikz}
\usepackage{etex}
\usepackage{xcolor}
\usepackage{tikz-qtree}
\usetikzlibrary{trees}
\usetikzlibrary{calc}
\usepackage{binarytree}

\numberwithin{equation}{section}

\theoremstyle{theorem}
\newtheorem{Theorem}{Theorem}[section]
\newtheorem{Proposition}[Theorem]{Proposition}
\newtheorem{Lemma}[Theorem]{Lemma}
\newtheorem{Corollary}[Theorem]{Corollary}
\newtheorem{Definition}[Theorem]{Definition}

\theoremstyle{remark}
\newtheorem{Remark}[Theorem]{Remark}
\newtheorem{Example}[Theorem]{Example}
\newcommand{\bco}{\begin{Corollary}}
\newcommand{\eco}{\end{Corollary}}

\newcommand{\bpr}{\begin{Proposition}}
\newcommand{\epr}{\end{Proposition}}

\newcommand{\btm}{\begin{Theorem}}
\newcommand{\etm}{\end{Theorem}}

\newcommand{\ben}{\begin{enumerate}}
\newcommand{\een}{\end{enumerate}}

\newcommand{\bit}{\begin{itemize}}
\newcommand{\eit}{\end{itemize}}

\newcommand{\bca}{\begin{cases}}
\newcommand{\eca}{\end{cases}}

\newcommand{\bre}{\begin{Remark}\rm}
\newcommand{\ere}{\end{Remark}}

\newcommand*{\bbm}{\begin{Remark}}
\newcommand*{\ebm}{\end{Remark}}

\newcommand{\ble}{\begin{Lemma}}
\newcommand{\ele}{\end{Lemma}}

\newcommand*{\bsz}{\begin{Proposition}}
\newcommand*{\esz}{\end{Proposition}}

\newcommand{\beq}{\begin{equation}}
\newcommand{\eeq}{\end{equation}}

\newcommand{\bbma}{\begin{bmatrix}}
\newcommand{\ebma}{\end{bmatrix}}

\newcommand{\bBma}{\begin{Bmatrix}}
\newcommand{\eBma}{\end{Bmatrix}}

\newcommand{\bpma}{\begin{pmatrix}}
\newcommand{\epma}{\end{pmatrix}}

\newcommand*{\bbs}{\begin{Example}}
\newcommand*{\ebs}{\end{Example}}

\newcommand*{\bfg}{\begin{Corollary}}
\newcommand*{\efg}{\end{Corollary}}

\newcommand*{\bdf}{\begin{Definition}}
\newcommand*{\edf}{\end{Definition}}

\newcommand*{\bbw}{\begin{proof}}
\newcommand*{\ebw}{\end{proof}}

\newcommand*{\bpf}{\begin{proof}}
\newcommand*{\epf}{\end{proof}}

\newtagform{simple}{}{}
\newtagform{standard}{(}{)}

\newcommand{\eqqeb}{\usetagform{simple}\tag{$\lozenge$}}

\allowdisplaybreaks

\newcommand{\II}{\mathbbm{1}}

\newcommand{\CC}{{\mathbb{C}}}

\newcommand{\NN}{{\mathbb{N}}}

\newcommand{\rt}{{\tt r}}

\newcommand{\unit}{{\rm 1\hspace*{-0.4ex}%
\rule{0.1ex}{1.52ex}\hspace*{0.2ex}}}
\def\bcdot{\mbox{{$\cdot$}}} 

\def\bcdot{\mbox{\scalebox{1.4}{$\cdot$}}}

\def\one{\hbox{{1}\kern-.25em\hbox{l}}}

\newcommand*{\res}{\upharpoonright}

\newcommand{\sitem}{\rm\item\it}

\newcommand{\Sl}{{\mr{SL}}}
\newcommand{\SL}{{\mr{SL}}}
\renewcommand{\sl}{\mf{sl}}

\newcommand{\SU}{{\mr{SU}}}
\newcommand{\su}{\mf{su}}

\newcommand{\al}[1]{\begin{align} #1 \end{align}}
\newcommand{\ala}[1]{\begin{align*} #1 \end{align*}}

\DeclareMathOperator{\id}{id}

\DeclareMathOperator{\tr}{tr}

\newcommand{\mc}[1]{\mathcal{#1}}
\newcommand{\mf}[1]{\mathfrak{#1}}
\newcommand{\mr}[1]{\mathrm{#1}}

\newcommand{\comment}[1]{}
\newcommand{\verweis}[1]{}
\newcommand{\todo}[1]{}

\newcommand{\group}{G}

\newcommand{\pha}{{\mc P}}

\newcommand{\ket}[1]{|#1\rangle}
\newcommand{\bra}[1]{\langle#1|}
\newcommand{\braket}[2]{\langle#1|#2\rangle}

\newcommand{\ctg}{\mr T^\ast}

\newcommand{\rref}[1]{{\rm \ref{#1}}}

\newcommand{\ol}[1]{\overline{#1}}
\newcommand{\ul}[1]{\underline{#1}}

\newcommand{\abs}{\hspace*{2.5mm}}
\newcommand*{\qeb}{\nopagebreak\hspace*{0.1em}\hspace*{\fill}{\mbox{$\lozenge$}}}

\makeatletter
\newcommand{\douwidehat}[2]{%
  \sbox0{$\m@th#1\widehat{\hphantom{#2}}$}%
  \sbox2{$\m@th#1x$}
  \sbox4{$\m@th#1#2$}
  \dimen0=\ht0
  \advance\dimen0 -.8\ht2
  \dimen2=\dp4
  \rlap{%
    \raisebox{\dimexpr\dimen0-\dimen2}{%
      \scalebox{1}[-1]{\box0}%
    }%
  }%
  {#2}%
}
\makeatother  

\begin{document}

\setcounter{secnumdepth}{3}

\title{Quasicharacters, recoupling calculus and costratifications of lattice quantum gauge theory}

 \author{
P. \ D.\ Jarvis$^\dagger$, G.\ Rudolph$^\ast$, M.\ Schmidt$^\ast$$^\circ$
\\[5pt]
$^\dagger$ School of Natural Sciences (Mathematics and Physics), University of Tasmania,
\\
Private Bag 37, GPO, Hobart Tas 7001, Australia
\\[5pt]
$^\ast$ Institute for Theoretical Physics, University of Leipzig, 
\\
P.O. Box 100 920, D-4109 Leipzig, Germany
\\[5pt]
$^\circ$ Corresponding author
}

\date{\today}

\maketitle

\begin{abstract}

We study the algebra $\cal R$ of $G$-invariant representative functions over the $N$-fold Cartesian product of copies of a compact Lie group $G$ modulo the action of conjugation by the diagonal subgroup. Using the representation theory of $G$ on the Hilbert space $\mc H = L^2(G^N)^G$, we construct a subset of $G$-invariant representative functions which, by standard theorems, span  $\mc H$ and thus generate $\cal R$. The elements of this basis will be referred to as quasicharacters. For $N=1$, they coincide with the ordinary irreducible group characters of $G$. The form of the quasicharacters depends on the choice of a certain unitary $G$-representation isomorphism, or reduction scheme, for every  isomorphism class of irreps of $G$. We determine the multiplication law of $\cal R$ in terms of the quasicharacters with structure constants. Next, we use the one-to-one correspondence between complete bracketing schemes for the reduction of multiple tensor products of $G$-representations and rooted binary trees. This provides a link to the recoupling theory for $G$-representations. Using these tools, we prove that the structure constants of the algebra $\cal R$ are given by a certain type of recoupling coefficients of $G$-representations. For these recouplings we derive a reduction law in terms of a product over primitive elements of $9j$ symbol type. The latter may be further expressed in terms of sums over products of Clebsch-Gordan coefficients of $G$. For $G = \SU(2)$, everything boils down to combinatorics of angular momentum theory. In the final part, we show that the above calculus enables us to calculate the matrix elements of bi-invariant operators occuring in quantum lattice gauge theory. In particular, both the quantum Hamiltonian and the orbit type relations may be dealt with in this way, thus, reducing both the construction of the costratification and the study of the spectral problem to numerical problems in linear algebra. We spell out the spectral problem for $G = \SU(2)$ and we present sample calculations of matrix elements for the gauge groups $\SU(2)$ and $\SU(3)$. 
The methods developed in this paper may be useful in the study of virtually any quantum model with polynomial constraints related to some symmetry.

\end{abstract}

\vspace{2cm}

\tableofcontents
\vfill
\newpage


\section{Introduction}
\label{AA-ResProg}


This paper builds on previous work \cite{FuRS,FJRS}, where we have developed 
tools for the implementation and the study of the classical gauge orbit stratification in lattice quantum gauge theory with gauge group $G = \SU(2)$. Our work is part of a program which aims at constructing a non-perturbative quantum theory of gauge fields in the Hamiltonian framework, with special emphasis on the role of non-generic gauge orbit types. The starting point is a classical finite-dimensional Hamiltonian system with symmetries, obtained by lattice approximation of the gauge model under consideration. The quantum theory is obtained via canonical quantization. It is best described in the language of $C^*$-algebras with a field algebra which (for a pure gauge theory) may be identified with the algebra of compact operators on the Hilbert space of square-integrable functions over the product $G^N$ of a number of copies of the gauge group manifold $G$. Correspondingly, the observable algebra is obtained via gauge symmetry reduction. We refer to \cite{qcd1,qcd2,qcd3,RS} for details. For the construction of the thermodynamical limit, see \cite{GR,GR2}. In the present paper, we extend our tools to the case of an arbitrary compact Lie group $G$.  

Let us recall that for a nonabelian gauge group the action of the group of local gauge transformations has more than one orbit type. Accordingly, the reduced phase space obtained by symplectic reduction is a stratified symplectic space 
\cite{SjamaarLerman,OrtegaRatiu,RS}. The stratification is given by the orbit type strata. It consists of an open and dense principal stratum, and several secondary strata. Each of these strata is invariant under the dynamics with respect to any invariant Hamiltonian. For case studies we refer to  \cite{cfg, cfgtop,FRS}. 
To study the influence of the classical gauge orbit type stratification on quantum level, we combine the concept of costratification of the quantum Hilbert space as developed by Huebsch\-mann \cite{Hue:Quantization} with a localization concept taken from the theory of coherent states. Loosely speaking, a costratification is given by a family of closed subspaces ${\cal H_\tau}$, one for each stratum.  The closed subspace associated with a certain classical stratum consists of the wave functions which are optimally localized at that stratum, in the sense that they are orthogonal to all states vanishing at that stratum. The vanishing condition can be given sense in the framework of holomorphic quantization, where wave functions are true functions and not just classes of functions. In \cite{HRS} we have constructed this costratification for a toy model with gauge group $\SU(2)$ on a single lattice plaquette. As physical effects, we have found a nontrivial overlap between distant strata and, for a certain range of the coupling, a very large overlap between the ground state of the lattice Hamiltonian and one of the two secondary strata.

Every classical gauge orbit stratum may be characterized by a set of polynomial relations. Within the above mentioned holomorphic picture, the latter may be implemented on quantum level as follows. In this picture, each element $\tau$ of the stratification corresponds to the zero locus of a finite subset $\{r_1, \ldots, r_\ell\}$ of the algebra $\mc R$ of $G$-invariant representative functions on $G^N_\CC$, where $G_\CC$ denotes the complexification of $G$. For the construction of the subspaces $\cal H_\tau$, we have to study the matrix elements 
of the quantum counterparts of the invariants $r_i$. The latter are presented as multiplication operators $\hat r_i$ on the physical Hilbert space $\mc H= L^2(G^N)^G$. 
For that study,  one needs deeper insight into the structure of the algebra 
$\cal R$. To accomplish that we proceed as follows. Using the representation theory of $G$ on the Hilbert space $\mc H$, we construct a subset of $G$-invariant representative functions which, by standard theorems, span  $\mc H$ and thus generate $\cal R$. The elements of this basis will be referred to as quasicharacters. For $N=1$, they coincide with the ordinary irreducible group characters of $G$. The form of the quasicharacters depends on the choice of a certain unitary $G$-representation isomorphism, or reduction scheme, for every $\ul \lambda \in \widehat G^N$, where $\widehat G$ denotes the set of isomorphism classes of irreps of $G$.  We determine the multiplication law of $\cal R$ in terms of the quasicharacters. Next, we use the one-to-one correspondence between complete bracketing schemes for  the reduction of multiple tensor products of $G$-representations and rooted binary trees. This provides a link to the recoupling theory for $G$-representations. In particular, via this link, the choice of the reduction scheme acquires an interpretation in terms of binary trees. Using these tools, we prove that the structure constants of the algebra $\cal R$ are given by a certain type of recoupling coefficients of $G$-representations. For these recouplings we derive a reduction law in terms of a product over primitive elements of $9j$ symbol type. The latter may be further expressed in terms of sums over products of Clebsch-Gordan coefficients of $G$. For $G = \SU(2)$, everything boils down to combinatorics of angular momentum theory.

Finally, in Section \ref{QuantumOp}, we show that the above calculus implies a calculus for bi-invariant operators ocurring in 
quantum lattice gauge theory. These bi-invariant operators come in three classes: 
\begin{enumerate}
\item 
The elements of the algebra $\mc R$ of $G$-invariant representative functions acting as multiplication operators on $\cal H$.
\item
The  bi-invariant linear differential operators on $\cal H$. 
\item
Any linear combination of operators of the above type. 
\end{enumerate}
We show that our calculus enables us to calculate the matrix elements of such operators explicitly. In particular, both the 
quantum Hamiltonian and the orbit type relations may be dealt with in this way, thus, reducing both the construction of the 
costratification and the study of the spectral problem to numerical problems in linear algebra. We spell out the spectral problem for 
$G = \SU(2)$ and we present sample calculations of matrix elements for the gauge groups $\SU(2)$ and $\SU(3)$. 
This opens the door for the  study of models with these gauge groups. A systematic discussion of these issues  will be presented in separate papers. 

Finally, we note that the methods developed in this paper may be useful in the study of virtually all quantum models with polynomial constraints related to some symmetry. For this reason we present the methods first, without reference to lattice gauge theory.


\section{Quasicharacters}
\label{A-reprF}\label{A-QC}


In this section, we recall the construction of an orthonormal basis in $L^2(G^N)^G$ and use it to analyze the multiplication law in the commutative algebra of $G$-invariant representative functions on $G^N$, see \cite{FJRS}.

Let $\mf R(G^N)$ denote the commutative algebra of representative functions on $G^N$ and let $\mc R := \mf R(G^N)^G$ be the subalgebra of $G$-invariant elements. Since $G_\CC^N$ is the complexification of the compact Lie group $G^N$, the proposition and Theorem 3  in Section 8.7.2 of \cite{Procesi:LG} imply that $\mf R(G^N)$ coincides with the coordinate ring of $G_\CC^N$, viewed as a complex affine variety, and that $\mf R(G^N)$ coincides with the algebra of representative functions on $G^N_\CC$. As a consequence, $\mc R$ coincides with the algebra of $G$-invariant representative functions on $G^N_\CC$, where the relation is given by restriction and analytic continuation, respectively.  

Let $\widehat G$ denote the set of isomorphism classes of finite-dimensional irreps of $G$. Given a finite-dimensional unitary representation $(H, D)$ of $G$, let $C(G)_D \subset \mf R(G)$ denote the subspace of representative functions\footnote{The subspace spanned by all matrix coefficients $\langle \zeta , D(\cdot) v \rangle $ with $v \in H$ and $\zeta \in H^\ast$.} of $D$ and let $\chi_D \in C(G)_D$ be the character of $D$, defined by $\chi_D(a) := \tr\big(D(a)\big)$. The same notation will be used for the Lie group $G^N$. Below, all representations are assumed to be continuous and unitary without further notice. The elements of $\widehat G$ will be labelled by the corresponding highest weights $\lambda$ relative to some chosen Cartan subalgebra and some chosen dominant Weyl chamber. Assume that for every $\lambda \in \widehat G$ a concrete unitary irrep $(H_\lambda,D^\lambda)$ of highest weight $\lambda$ in the Hilbert space $H_\lambda$ has been chosen. Given $\ul\lambda = (\lambda^1,\dots,\lambda^N) \in \widehat G^N$, we define a representation $(H_{\ul\lambda},D^{\ul\lambda})$ of $G^N$ by 
\beq
\label{G-irrepsGN}
H_{\ul\lambda} = \bigotimes_{i = 1}^N H_{\lambda^i}
\,,\quad 
D^{\ul\lambda}(\ul a) = \bigotimes_{i=1}^N D^{\lambda^i}(a_i)\,,
\eeq
where $\ul a = (a_1 , a_2 , \dots , a_N)$. This representation is irreducible and we have 
$$
C(G^N)_{D^{\ul\lambda}} \cong \bigotimes_{i = 1}^N C(G)_{D^{\lambda^i}}
\,,
$$ 
isometrically with respect to the $L^2$-norms. Using this, together with the Peter-Weyl theorem for $G$, we obtain that $\bigoplus_{\ul\lambda \in \widehat G^N} C(G^N)_{D^{\ul\lambda}}$ is dense in $L^2 (G^N, {\rm d}^N a)$, where $\mr d a$ denotes the normalized Haar  measure on $G$. Since 
$$
\bigoplus_{\ul\lambda \in \widehat G^N} 
C(G^N)_{D^{\ul\lambda}} \subset \bigoplus_{D \in \widehat{G^N}} C(G^N)_D
\,,
$$
this implies

\ble\label{L-ProdReps}

Every irreducible representation of $G^N$ is equivalent to a product representation $(H_{\ul\lambda},D^{\ul\lambda})$ with $\ul\lambda \in \widehat G^N$. If $(H_{\ul\lambda},D^{\ul\lambda})$ and $(H_{\ul\lambda'},D^{\ul\lambda'})$ are isomorphic, then $\ul\lambda = \ul\lambda'$.
\qed

\ele

Given $\ul\lambda \in \widehat G^N$, let $D_d^{\ul\lambda}$ denote the representation of $G$ on $H_{\ul\lambda}$ defined by  
\beq\label{pi-d}
D_d^{\ul\lambda}(a) := D^{\ul\lambda} (a, \ldots, a)
\,.
\eeq
This representation will be referred to as the diagonal representation induced by $D^{\ul\lambda}$. It is reducible and has the isotypical decomposition
$$
H_{\ul\lambda}
 =
\bigoplus_{\lambda \in \widehat G} H_{\ul\lambda,\lambda}
$$
into uniquely determined subspaces $H_{\ul\lambda,\lambda}$. Recall that these subspaces may be obtained as the images of the orthogonal projectors 
\beq
\label{Proj-irrep}
{\mathbb P}_\lambda
 := 
\dim(H_\lambda) \int_G \ol{\chi_{D^\lambda}}(a) \, D^{\ul\lambda}_d(a) \, \mr d a 
\eeq
on $H_{\ul\lambda}$. These projectors commute with one another and with $D_d^{\ul\lambda}$. If an isotypical subspace $H_{\ul\lambda,\lambda}$ is reducible, we can further decompose it in a non-unique way into irreducible subspaces of isomorphism type $\lambda$. Let $m_{\ul\lambda}(\lambda)$ denote the number of these irreducible subspaces (the multiplicity of $D^\lambda$ in $D_d^{\ul\lambda}$) and let $\widehat G_{\ul\lambda}$ denote the subset of $\widehat G$ consisting of the highest weights $\lambda$ such that $m_{\ul\lambda}(\lambda) > 0$. In this way, we obtain a unitary $G$-representation isomorphism 
\beq\label{G-D-vp}
H_{\ul\lambda}
 ~\cong~
\bigoplus_{\lambda \in \widehat G_{\ul\lambda}} 
 \,
\bigoplus_{l=1}^{m_{\ul\lambda}(\lambda)} 
H_\lambda
\,.
\eeq
Let us assume that we have chosen such an isomorphism for every $\ul\lambda \in \widehat G^n$ and for every $n=2,3,\dots$. Composing this isomorphism with the natural projections and injections of the direct sum, we obtain projections and injections
\beq\label{G-D-p-i}
\prj{\ul\lambda,\lambda,l}{}
:
H_{\ul\lambda}
\to 
H_\lambda
\,,\qquad
\inj{\ul\lambda,\lambda,l}{}
:
H_\lambda
\to 
H_{\ul\lambda}
\,,\qquad
\lambda \in \widehat G_{\ul\lambda}
\,,\qquad
l = 1,\cdots, m_{\ul \lambda}(\lambda)
\,.
\eeq
We have 
$$
\sum_{\lambda,l} 
\inj{\ul\lambda,\lambda,l}{} \circ \prj{\ul\lambda,\lambda,l}{}
=
\one_{H_{\ul\lambda}} 
\,,\qquad
\prj{\ul\lambda,\lambda,l}{} \circ \inj{\ul\lambda,\lambda',l'}{}
=
\delta_l^{l'} \, \delta_\lambda^{\lambda'} \, \one_{H_\lambda}
\,.
$$
For every $\lambda \in \widehat G_{\ul\lambda}$ and every $l,l' = 1, \dots , m_{\ul\lambda}(\lambda)$, we define an operator on $H_{\ul\lambda}$,
\beq
\label{A-T}
(A_{\ul\lambda,\lambda})^{l'}_l
 := 
\frac{1}{\sqrt{\dim(H_\lambda)}}
 ~
\inj{\ul\lambda,\lambda,l}{} \circ \prj{\ul\lambda,\lambda,l'}{}
\,,
\eeq
which is a $G$-representation endomorphism of $D_d^{\ul\lambda}$, and a $G$-invariant function $\BF{\ul\lambda}{\lambda}{l}{l'}$ on $G^N$ by 
\beq
\label{ReprF-Hom}
\BF{\ul\lambda}{\lambda}{l}{l'}(\ul a)
 := 
\sqrt{\dim(H_{\ul\lambda})}
\,
\tr\left(D^{\ul\lambda}(\ul a) (A_{\ul\lambda,\lambda})^{l'}_l \right)
\,.
\eeq
In the sequel, the functions $\BF{\ul\lambda}{\lambda}{l}{l'}$ will be referred 
to as quasicharacters. 

\bsz\label{S-ReprF}

The family of functions 
$$
\left\{
\BF{\ul\lambda}{\lambda}{l}{l'}
~:~
\ul\lambda \in \widehat G^N
,~
\lambda \in \widehat G_{\ul\lambda}
\,,~
l,l' = 1 , \dots , m_{\ul\lambda}(\lambda)
\right\}
$$
constitutes an orthonormal basis in  $L^2(G^N)^G$.

\esz

\bbw

See Proposition 3.7 in \cite{FJRS}.
\ebw

By analytic continuation, the irreps $D^\lambda$ of $G$ induce irreps $D_\CC^\lambda$ of $G_\CC$, the irreps $D^{\ul\lambda}$ of $G^N$ induce irreps $D_\CC^{\ul\lambda}$ of $G^N_\CC$, and the functions \smash{$\BF{\ul\lambda}{\lambda}{l}{l'}$} on $G^N$ induce holomorphic functions \smash{$\BFC{\ul\lambda}{\lambda}{l}{l'}$} on \smash{$G^N_\CC$}. Then, \eqref{G-irrepsGN}, \eqref{pi-d} and \eqref{ReprF-Hom} hold with \smash{$D^{\ul\lambda}$, $D^\lambda$} and \smash{$\BF{\ul\lambda}{\lambda}{l}{l'}$} replaced by, respectively, \smash{$D^{\ul\lambda}_\CC$}, \smash{$D^\lambda_\CC$} and \smash{$\BFC{\ul\lambda}{\lambda}{l}{l'}$}.

\bfg
\label{F-ReprF}

The family of functions
$$
\left\{
\BFC{\ul\lambda}{\lambda}{l}{l'}
~:~
\ul\lambda \in \widehat G^N
,~
\lambda \in \widehat G_{\ul\lambda}
\,,~
l,l' = 1 , \dots , m_{\ul\lambda}(\lambda)
\right\}
$$
constitutes an orthogonal basis in $H$. The norms are
\beq
\label{G-norm}
\|\BFC{\ul\lambda}{\lambda}{l}{l'}\|^2
 =
\prod_{r=1}^N C_{\lambda^r}
 \,,\qquad
C_{\lambda^r} = (\hbar\pi)^{\dim(\group)/2}\mr e^{\hbar|\lambda^r+\rho|^2},
\eeq
where $\rho$ denotes half the sum of the positive roots. The expansion coefficients of $f \in H$ with respect to this basis are given by the scalar products $\bigbraket{\BF{\ul\lambda}{\lambda}{l}{l'}}{f_{\res G^N}}$ in $L^2(G^N)^G$.

\efg

\bbw

See Corollary 3.8 in \cite{FJRS}. 
\ebw

It follows that the quasicharacters span the algebra $\mf R(G^N)$. Hence, to study the multiplicative structure of this algebra, it suffices to find the multiplication law for quasicharacters. To get rid of dimension factors, we will pass to the modified quasicharacters
$$
\dBF{\ul\lambda}{\lambda}{l}{l'}
:=
\sqrt{\frac{\dim(H_\lambda)}{\dim(H_{\ul\lambda})}}
\BF{\ul\lambda}{\lambda}{l}{l'}
\,.
$$
We assume that a unitary $G$-representation isomorphism \eqref{G-D-vp} has been chosen for every $\ul\lambda \in \widehat G^N$ and every $N$. Writing 
\al{\nonumber
&
\dBF{\ul\lambda_1}{\lambda_1}{l_1}{l'_1}(\ul a)
\,
\dBF{\ul\lambda_2}{\lambda_2}{l_2}{l'_2}(\ul a)
\\ \label{G-MF-1}
& \hspace{0.25cm}
=
\sqrt{\dim(H_{\lambda_1}) \dim(H_{\lambda_2})}
 \,
\tr
 \left(
 \left(
\big(A_{\ul\lambda_1,\lambda_1}\big)^{l'_1}_{l_1} 
\otimes 
\big(A_{\ul\lambda_2,\lambda_2}\big)^{l'_2}_{l_2} 
 \right)
 \circ
\Big(D^{\ul\lambda_1}(\ul a) \otimes D^{\ul\lambda_2}(\ul a)\Big)
 \right)
,
 }
we see that in order to expand the product $\dBF{\ul\lambda_1}{\lambda_1}{l_1}{l'_1} \cdot \dBF{\ul\lambda_2}{\lambda_2}{l_2}{l'_2}$ in terms of the basis functions $\dBF{\ul\lambda}{\lambda}{l}{l'}$, a reasonable strategy is to decompose the $G^N$-representation $D^{\ul\lambda_1} \otimes D^{\ul\lambda_2}$ into $G^N$-irreps $\ul\lambda$ and then relate these $G^N$-irreps to the basis functions using the chosen $G$-representation isomorphisms \eqref{G-D-vp}. To implement this, we define two different decompositions of the diagonal representation $D^{\ul\lambda_1}_d \otimes D^{\ul\lambda_2}_d$ into irreps. The first one is adapted to the tensor product on the right hand side of \eqref{G-MF-1}. It is defined by the projections
\beq
\label{eq: decomp12}
\xymatrix{
H_{\ul\lambda_1} \otimes H_{\ul\lambda_2}
\ar[rrr]^{
\prj{\ul\lambda_1,\lambda_1,l_1}{} 
\otimes 
\prj{\ul\lambda_2,\lambda_2,l_2}{}
} 
& & &
H_{\lambda_1}\otimes H_{\lambda_2}
\ar[rrr]^{~~~~\prj{(\lambda_1,\lambda_2),\lambda,k}{}} 
& & &
H_\lambda
\,,
}
\eeq
where $\lambda_i \in \widehat G^N_{\ul\lambda_i}$, 
$l_i = 1 , \dots , m_{\ul\lambda_i}(\lambda_i)$ 
and 
$\lambda \in \widehat G^2_{(\lambda_1,\lambda_2)}$, 
$k = 1 , \dots , m_{(\lambda_1,\lambda_2)}(\lambda)$. 
The second decomposition is adapted to the definition of the basis functions $\dBF{\ul\lambda}{\lambda}{k}{l}$. It is defined by the projections 
\beq
\label{eq: decomp21}
\xymatrix{
H_{\ul\lambda_1} \otimes H_{\ul\lambda_2}
\ar[rrr]^{
~~~~~\otimes_{i=1}^N \prj{(\lambda_1^i,\lambda_2^i),\lambda^i,k^i}{}
}
& & &
H_{\ul\lambda}
\ar[rrr]^{\prj{\ul\lambda,\lambda,l}{}}
& & &
H_\lambda
\,,
}
\eeq
where 
$\ul\lambda = (\lambda^1,\dots,\lambda^N)$ 
with 
$\lambda^i \in \widehat G^2_{(\lambda_1^i,\lambda_2^i)}$, 
$k^i = 1 , \dots , m_{(\lambda_1^i,\lambda_2^i)}(\lambda^i)$ 
and 
$\lambda \in \widehat G^N_{\ul\lambda}$, 
$l = 1 , \dots , m_{\ul\lambda}(\lambda)$. 
Composition of the injection corresponding to \eqref{eq: decomp21} with the projection \eqref{eq: decomp12} yields a unitary representation isomorphism of irreps $H_\lambda$. Hence, Schur's lemma implies that 
\al{\nonumber
&
\prj{(\lambda_1,\lambda_2),\lambda',k}{}
\circ
\left(\prj{\ul\lambda_1,\lambda_1,l_1}{} \otimes \prj{\ul\lambda_2,\lambda_2,l_2}{}\right)
\circ
\left(\otimes_{i=1}^N \inj{(\lambda_1^i,\lambda_2^i),\lambda^i,k^i}{}\right)
\circ
\inj{\ul\lambda,\lambda,l}{}
\\ \label{eq: UnitaryCoeff-1}
& \hspace{8cm}
=
\delta_{\lambda}^{\lambda'}
\,
U
^{\ul\lambda_1,\lambda_1,l_1;\ul\lambda_2,\lambda_2,l_2;k}
_{\ul\lambda,\lambda,l;\ul k}
 \,
\id_{H_\lambda}
 } 
with certain coefficients 
$
U
^{\ul\lambda_1,\lambda_1,l_1;\ul\lambda_2,\lambda_2,l_2;k}
_{\ul\lambda,\lambda,l;\ul k}
$. 
Here, we have denoted $\ul k = (k^1,\dots,k^N)$. Now, the multiplication law may be expressed in terms of these coefficients.

\btm\label{MultLaw-Gen}

In terms of the basis functions, the multiplication in $\mc R$ is given by 
 \ala{\nonumber
\dBF{\ul\lambda_1}{\lambda_1}{l_1}{l'_1} \cdot \dBF{\ul\lambda_2}{\lambda_2}{l_2}{l'_2}
& =
\sum_{\ul\lambda} 
 \,
\sum_{\ul k=(1,\dots,1)}^{m_{\ul\lambda_1,\ul\lambda_2}(\ul\lambda)}
 \,
\sum_\lambda
 \,
\sum_{l,l'=1}^{m_{\ul\lambda}(\lambda)}
 \,
\sum_{k=1}^{m_{(\lambda_1,\lambda_2)}(\lambda)}
 \!
 \,
U
^{\ul\lambda_1,\lambda_1,l_1;\ul\lambda_2,\lambda_2,l_2;k}
_{\ul\lambda,\lambda,l;\ul k}
\\
 & \hspace{4.5cm}
\cdot
\left(
U
^{\ul\lambda_1,\lambda_1,l'_1;\ul\lambda_2,\lambda_2,l'_2;k}
_{\ul\lambda,\lambda,l';\ul k}
\right)^\ast
 ~
\dBF{\ul\lambda}{\lambda}{l}{l'}
\,.
 }
The same formula holds true for the basis functions $\dBFC{\ul\lambda}{\lambda}{l}{l'}$ on $G_\CC^N$.

\etm

\bbw

See Proposition 3.10 in \cite{FJRS}.
\ebw

\bbm\label{Bem-unitaries}

Note that the coefficients $U$ in Theorem \ref{MultLaw-Gen} depend on the unitary $G$-representation isomorphisms \eqref{G-D-vp}. In the next section, we will see that they are given by appropriate recoupling coefficients.
\qeb

\ebm

\bbm\label{Bem-N=1}

For completeness, we also recall the case $N=1$. Here, the quasicharacters are ordinary characters 
$$
\chi_\lambda(a) = \tr\big(D^\lambda(a)\big)
\,,\qquad
\lambda \in \widehat G
\,,
$$
and the multiplication law can be obtained directly from the Clebsch-Gordan series,
$$
D^{\lambda_1} \otimes D^{\lambda_2}
=
\bigoplus_{\lambda \in \widehat G} m_{(\lambda_1,\lambda_2)}(\lambda) \, D^\lambda
\,.
$$
Indeed, we have
$$
\tr\big(D^{\lambda_1}(a)\big) \tr\big(D^{\lambda_2}(a)\big)
=
\tr\big(D^{\lambda_1}(a) \otimes D^{\lambda_2}(a)\big)
=
\sum_{\lambda \in \widehat G} m_{(\lambda_1,\lambda_2)}(\lambda) 
\tr\big(D^\lambda(a)\big)
$$
and hence
\beq\label{G-Mult-N=1-2}
\chi_{\lambda_1} \cdot \chi_{\lambda_2}
=
\sum_{\lambda \in \widehat G} m_{(\lambda_1,\lambda_2)}(\lambda) \,\, \chi_{\lambda}
\,.
\eeq
This generalizes to
\beq\label{G-Mult-N=1}
\chi_{\lambda_1} \cdot \, \cdots \, \cdot \chi_{\lambda_r}
=
\sum_{\lambda \in \widehat G} 
m_{(\lambda_1,\dots,\lambda_r)}(\lambda) \,\, \chi_{\lambda}
\,,
\eeq
where $m_{(\lambda_1,\dots,\lambda_r)}(\lambda)$ denotes the multiplicity of the irrep $D^\lambda$ in $D^{\lambda_1} \otimes \cdots \otimes D^{\lambda_r}$.
\qeb

\ebm


\section{Recoupling calculus}
\label{A-RC}


As observed in the preceding section, to fix concrete basis functions $\BF{\ul\lambda}{\lambda}{i}{j}$, we have to fix the unitary $G$-representation isomorphisms \eqref{G-D-vp} entering their definition. As a consequence, we obtain concrete formulae for the unitary operators in the multiplication law of the algebra $\mc R$, expressed in terms of $G$-recoupling coefficients. This relates the algebra structure to the combinatorics of recoupling theory for $G$-respresentations, see \cite{BL1,BL2,Louck,Yutsis,J1,J2}.


\subsection{Reduction schemes, binary trees, and recoupling coefficients}
\label{subsec:BinaryTrees}


Given an element $\lambda \in \widehat G$, recall that $(H_\lambda,D^\lambda)$ denotes the standard irrep of highest weight $\lambda$. Let $\hat\weight(\lambda)$ denote the weight system of this representation and let $\weight(\lambda)$ denote the set of pairs $\mu=(\hat\mu,\check\mu)$, where $\hat\mu \in \hat\weight(\lambda)$ and $\check\mu$ is a multiplicity counter for $\hat\mu$. We refer to the pairs $\mu$ as weight labels. All results carry over to arbitrary orthonormal bases by interpreting $\mu$ as a label without an inner structure and by ignoring statements about weights. The representation space $H_\lambda$ is spanned by an orthonormal weight vector basis $\{\ket{\lambda \mu} : \mu \in \weight(\lambda)\}$, where $\ket{\lambda \, \mu}$ denotes the normalized common eigenvector of the image under $D^\lambda$ of the Cartan subalgebra chosen in the Lie algebra $\mf g$ of $G$ corresponding to the eigenvalue functional $\mr i \hat\mu$ and the multiplicity counter $\check\mu$. Here, $\mr i$ denotes the imaginary unit. The matrix elements of $D^\lambda(a)$, $a \in G$, in that basis are
$$
D^\lambda_{\mu \mu'}(a)
= 
\bra{\lambda \mu} D^\lambda(a) \ket{\lambda \mu'}
\,,\qquad 
\mu,\mu'\in\weight(\lambda)\,.
$$
Accordingly, the matrix elements of
$$
D^{\ul\lambda}(\ul a)
= 
D^{\lambda^1}(a_1)\otimes D^{\lambda^2}(a_2) \otimes \cdots \otimes D^{\lambda^N}(a_N)
$$ 
with respect to the tensor product weight basis 
$$
\ket{\ul\lambda \, \ul\mu}
= 
\ket{\lambda^1 \mu^1} \otimes \cdots \otimes \ket{\lambda^N \mu^N}
\,,\qquad
\mu^n \in \weight(\lambda^n)
\,,\qquad
n=1,\dots, N
\,,
$$
are given by
$$
\bra{\ul\lambda \, \ul\mu} D^{\ul\lambda}(\ul a) \ket{\ul\lambda \, \ul\mu'}
=
D^{\lambda^1}_{\mu_1 \mu'_1}(a_1) 
\cdots 
D^{\lambda^N}_{\mu_N \mu'_N}(a_N)
\,.
$$
We write \smash{$D_d^{\ul\lambda}$} for the induced diagonal representation of $G$ and $\ul\mu \in \weight(\ul\lambda)$ for the condition that $\mu^n \in \weight(\lambda^n)$ for all $n=1,\dots,N$. Moreover, we put $\fnlsum{\ul{\hat\mu}}:=\sum_{n=1}^N \hat\mu^n$ (sum of linear functionals). 

Recall that for highest weights $\lambda^1,\lambda^2 \in \widehat G$, it may happen that in the decomposition of $H_{(\lambda^1,\lambda^2)} = H_{\lambda_1} \otimes H_{\lambda_2}$ into irreps, given by the Clebsch-Gordan series
\beq\label{G-TePr-Zlg-1}
H_{\lambda_1} \otimes H_{\lambda_2}
\cong 
\bigoplus_{\lambda \in \widehat G}
m_{(\lambda_1,\lambda_2)}(\lambda) H_\lambda
\,,
\eeq
the multiplicity $m_{(\lambda_1,\lambda_2)}(\lambda) > 1$. We assume that in each such case, a concrete orthogonal decomposition of each isotypical subspace $H_{(\lambda_1,\lambda_2),\lambda}$ into $m_{(\lambda_1,\lambda_2)}(\lambda)$ irreducible subspaces has been chosen. By identifying each of them with $H_\lambda$, we obtain a unitary representation isomorphism
\beq\label{G-TePr-Zlg}
H_{(\lambda_1,\lambda_2),\lambda}
\cong
\bigoplus_{k=1}^{m_{(\lambda_1,\lambda_2)}(\lambda)}
H_\lambda
\eeq
for each isotypical subspace. Let us introduce the bracket sets
\ala{
\langle \lambda^1 , \lambda^2 \rangle
& :=
\{
\lambda \in \widehat G 
: 
m_{(\lambda_1,\lambda_2)}(\lambda) \neq 0
\}
\,,
\\
\llangle \lambda^1 , \lambda^2 \rrangle
& :=
\{
(\lambda,k) \in \widehat G \times \NN
: 
\lambda \in \langle \lambda^1 , \lambda^2 \rangle
\,,\, 
k=1,\dots,m_{(\lambda_1,\lambda_2)}(\lambda)
\}
\,.
}
The integer $k$ plays the role of a multiplicity counter. For the first bracket set, there is an iterated version like $\langle \lambda^1 , \langle \lambda^2 , \lambda^3 \rangle \rangle$ defined by taking the union of $\langle \lambda^1 , \lambda \rangle$ over $\lambda \in \langle \lambda^2 , \lambda^3 \rangle$. Note that 
this bracketing is associative,
$$
\langle \lambda^1 , \langle \lambda^2 , \lambda^3 \rangle \rangle
=
\langle \langle \lambda^1 , \lambda^2 \rangle , \lambda^3 \rangle
\,,
$$
and so for any given set $\ul\lambda$ of highest weights we may write $\langle \ul\lambda \rangle$. The elements $\lambda$ of this set label the (uniquely determined) isotypical subspaces $H_{\ul\lambda,\lambda}$ of $(H_{\ul\lambda},D^{\ul\lambda}_d)$. Concrete irreducible subspaces for the diagonal $G$-representation $(H_{\ul\lambda},D_d^{\ul\lambda})$ can be obtained by choosing a reduction scheme for $N$-fold tensor products of $G$-irreps, which breaks the reduction into iterated reductions of twofold tensor products. Such reduction schemes are enumerated combinatorially by specifying complete bracketing schemes, or equivalently, by rooted binary trees \cite{BL1,BL2,Louck,J1,J2}.

\bbm

A rooted tree is an undirected connected graph which does not contain cycles and which has a distinguished vertex, called the root. The parent of a vertex $x$ is the vertex connected to $x$ on the unique path to the root. Any vertex on that path is called an ascendant of $x$. A child of a vertex $x$ is a vertex of which $x$ is the parent. A descendant of $x$ is a vertex of which $x$ is an ascendant. We write $x < y$ if $x$ is an ascendant of $y$ (or equivalently $y$ is a descendant of $x$). The number of edges attached to a vertex is called the valence of that vertex. A vertex of valence $1$ is called a leaf. All other vertices are called nodes. A rooted binary tree is a rooted tree whose root has valence $2$ and whose other nodes have valence $3$. For convenience, we will view the root as a node of valence 3, with an additional pendant root edge. The nodes different from the root are called internal.

A labelling of a rooted binary tree $T$ is an assignment $\alpha : x \mapsto \alpha^x$ of a label to every vertex of $T$. The pair $(T,\alpha)$ is called a labelled, rooted binary tree.

It is convenient to view a rooted binary tree as being embedded in the plane. Then, the child vertices of every node can be ordered from left to right. Conversely, an ordering of the child vertices of every node defines a unique planar embedding. Given such an ordering, or equivalently an embedding in the plane, one speaks of an ordered tree. In particular, in such a tree, the leaves are ordered and thus their labels are given by a sequence.
\qeb

\ebm

\bbs
\label{B-5}

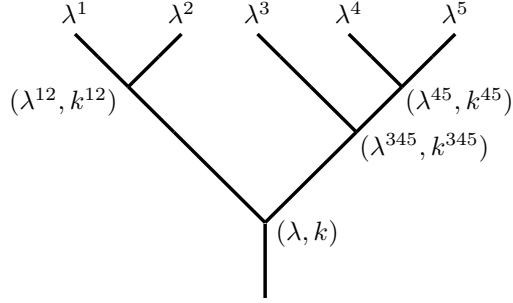
\begin{figure}

\begin{center}

\begin{tikzpicture}

\btreeset{grow=up, scale=0.6, line width=1.3pt}

\BinaryTree[local bounding box=INIT2,line width=1.3pt,bottom padding = 10pt]{%
:!
l!l!l!l,
l!l!r!r,
r!l!l!l,
r!r!l!l,
r!r!r!r
}{4}

\draw[thick] (btree-l-l-l-l) node [above]{\small $\lambda^1$};
\draw[thick] (btree-l-l-r-r) node [above]{\small $\lambda^2$};
\draw[thick] (btree-r-l-l-l) node [above]{\small $\lambda^3$};
\draw[thick] (btree-r-r-l-l) node [above]{\small $\lambda^4$};
\draw[thick] (btree-r-r-r-r) node [above]{\small $\lambda^5$};

\draw[thick] (btree-l-l)+(0,-0.2) node [left]{\small $(\lambda^{12},k^{12})$};
\draw[thick] (btree-r)+(-0.1,-0.2) node [right]{\small $(\lambda^{345},k^{345})$};
\draw[thick] (btree-r-r)+(-0.1,-0.2) node [right]{\small $(\lambda^{45},k^{45})$};

\draw[line width=1.3pt] (btree-root) -- ($(btree-root)+(0,-1)$);
\draw[thick] (btree-root)+(0,-.15) node [right]{\small $(\lambda,k)$};

\end{tikzpicture}

\end{center}

\caption{\label{fig:Ex5Leaf}
A reduction scheme for a tensor product of $N=5$ irreps.
}

\end{figure}

Consider the specific case of the reduction of a tensor product of $N=5$ irreps according to the bracketing 
\[
((H_{\lambda^1}\otimes H_{\lambda^2})\otimes (H_{\lambda^3}\otimes (H_{\lambda^4}\otimes H_{\lambda^5})))
\,,
\]
as shown in Figure \ref{fig:Ex5Leaf}. 
The rooted binary tree corresponding to this reduction scheme is as follows. There are $5$ leaves representing the tensor factors, $3$ internal nodes, representing intermediate stages of the reduction procedure, and the root representing the final irreducible subspaces. The leaves are labelled by the highest weights $\lambda^1, \cdots, \lambda^5$ of the respective irreducible representations. The $3$ internal nodes are labelled by $(\lambda^{12},k^{12})$, $(\lambda^{45},k^{45})$, $(\lambda^{345},k^{345})$ enumerating the admissible weight values occurring in the pairwise tensor products indicated by their child nodes and the root is labelled by $(\lambda^{12345},k^{12345}) \equiv (\lambda,k)$. We have
$
(\lambda^{12},k^{12}) \in \llangle \lambda^1,\lambda^2 \rrangle
$,
$
(\lambda^{45},k^{45}) \in \llangle \lambda^4,\lambda^5 \rrangle
$,
$
(\lambda^{345},k^{345}) \in \llangle \lambda^3,\lambda^{45} \rrangle
$,
and
$
(\lambda,k) \in \llangle \lambda^{12},\lambda^{345} \rrangle
$.
\qeb

\ebs

\bbs[Standard coupling]
\label{B-Raupe}

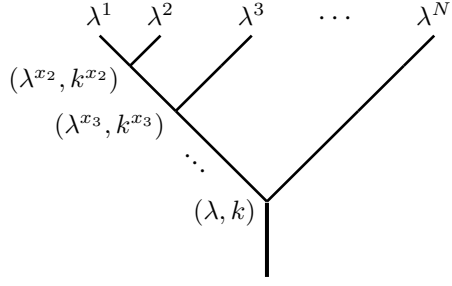
\begin{figure}

\begin{center}

\begin{tikzpicture}

\btreeset{grow=up, scale=0.6, line width=1.5pt}

\BinaryTree[local bounding box=INIT2,line width=1pt,bottom padding = 10pt]{%
:!
l!l!l,
l!l!r,
l!r!r,
r!r!r     
}{3}

\draw[thick] (btree-l-l-l) node [above]{\small $\lambda^1$};
\draw[thick] (btree-l-l-r) node [above]{\small $\lambda^2$};
\draw[thick] (btree-l-r-r) node [above]{\small $\lambda^3$};
\draw[thick] (btree-l-r-r) node [above]{\hspace{2.25cm}$\cdots $};
\draw[thick] (btree-r-r-r) node [above]{\small $\lambda^N$};

\draw[thick] (btree-l-l)+(0,-0.2) node [left]{\small $(\lambda^{x_2},k^{x_2})$};
\draw[thick] (btree-l)+(0,-0.2) node [left]{\small $(\lambda^{x_3},k^{x_3})$};
\draw[thick] (btree-l)+(0.35,-0.6) node [left]{$\cdot$};
\draw[thick] (btree-l)+(0.45,-0.7) node [left]{$\cdot$};
\draw[thick] (btree-l)+(0.55,-0.8) node [left]{$\cdot$};
\draw[thick] (btree-root)+(0,-0.15) node [left]{\small $(\lambda,k)$};

\draw[line width=1.5] (btree-root) -- ($(btree-root)+(0,-1)$);

\end{tikzpicture}

\end{center}

\caption{\label{fig:StdTree} The standard coupling tree.}

\end{figure}

Given $N$ highest weights $\lambda^1 , \dots , \lambda^N$, we may start with decomposing $H_{\lambda^1}\otimes H_{\lambda^2}$ into the unique irreducible subspaces $(H_{\lambda^1}\otimes H_{\lambda^2})_{(\lambda^{12},k^{12})}$ with $(\lambda^{12},k^{12}) \in \llangle \lambda^1 , \lambda^2 \rrangle$. Then, we decompose the invariant subspaces 
$$
(H_{\lambda^1}\otimes H_{\lambda^2})_{(\lambda^{12},k^{12})} \otimes H_{\lambda^3}
 \subset 
H_{\lambda^1} \otimes H_{\lambda^2} \otimes H_{\lambda^3}
$$
into the unique irreducible subspaces 
$$
(
(H_{\lambda^1}\otimes H_{\lambda^2})_{(\lambda^{12},k^{12})} 
\otimes 
H_{\lambda^3}
)_{(\lambda^{123},k^{123})}
 \,,\quad
(\lambda^{123},k^{123}) \in \llangle \lambda^{12} , \lambda^3 \rrangle
\,.
$$
Iterating this, we end up with a decomposition of $H_{\ul\lambda}$ into unique irreducible subspaces 
$$
(
 \cdots
(
(H_{\lambda^1}\otimes H_{\lambda^2})_{(\lambda^{12},k^{12})}
\otimes 
H_{\lambda^3})_{(\lambda^{123},k^{123})} 
\cdots \otimes 
H_{\lambda^N})_{(\lambda^{1\dots N},k^{1\dots N})}
\,,
$$
where $(\lambda^{1\dots n},k^{1\dots n}) \in \llangle \lambda^{1 \dots n-1} , \lambda^n \rrangle$ for $n = 3 , 4 , \dots , N$. We may number the nodes of $T$ by assigning the number $n$ to the parent of leaf $n$, where $n=2,\dots,N$. Then, $x_2,\dots,x_{N-1}$ are the internal nodes and $x_N$ is the root. Accordingly, we may write $(\lambda^{x_n},k^{x_n}) = (\lambda^{1\dots n},k^{1\dots n})$ for $n=2,\dots,N-1$ and $(\lambda,k) = (\lambda^{1\dots N},k^{1\dots N})$. The corresponding bracketing is $(\cdots((,),)\cdots,)$ and the corresponding coupling tree is the caterpillar tree shown in Figure \rref{fig:StdTree}, also referred to as the standard coupling tree.
\qeb

\ebs

In general, for an $N$-fold tensor product of $G$-irreps, the reduction schemes (bracketings) correspond 1-1 to rooted binary trees with $N$ leaves, referred to as coupling trees. Such a tree has $N-2$ internal nodes. The leaves of a coupling tree correspond to the irreps entering the tensor product and the root corresponds to an irreducible subspace of the tensor product. The root will usually be denoted by $\rt$. 

A given labelling $\alpha$ of a coupling tree $T$ assigns to every leaf $y$ of $T$ a highest weight $\alpha^y=\lambda^y \in \widehat G$ and to every node $x$ of $T$ a pair $\alpha^x = (\lambda^x,k^x) \in \widehat G \times \NN$. We say that $\alpha$ is admissible if $\alpha^x \in \llangle\lambda^{x'},\lambda^{x''}\rrangle$ for every node $x$ of $T$, where $x'$ and $x''$ denote the child vertices of $x$. In what follows, we will assume all labellings to be admissible without explicitly stating that. By forgetting about the multiplicity counters, $\alpha$ induces a labelling of $T$ by highest weights, referred to as the highest weight labelling underlying $\alpha$. By forgetting about the highest weights, $\alpha$ induces a labelling of $T$ by multipicity counters, referred to as the multiplicity counter labelling underlying $\alpha$. Given a coupling tree $T$ with leaf labelling $\ul\lambda$, let $\mc L^T(\ul\lambda)$ denote the set of admissible labellings of $T$ having $\ul\lambda$ as their leaf labelling. Given, in addition, a highest weight $\lambda \in \langle \ul\lambda \rangle$, let $\mc L^T(\ul\lambda,\lambda)$ denote the set of labellings of $T$ having $\ul\lambda$ as their leaf labelling and $\lambda$ as the highest weight label of the root. These subsets establish a disjoint decomposition of the totality of all (admissible) labellings of $T$. We will say that two given labellings of $T$ are combinable if they belong to the same such subset, that is, if they share the same leaf labelling and the same highest weight of the root. To be combinable is an equivalence relation.

\bbm

Each internal node $x$ is the root of a unique subtree made up by $x$ and all its descendants. This subtree represents a reduction scheme for a tensor product of $l$ $G$-irreps, where $l$ is the number of leaves in the subtree. In case the leaves are numbered by $1 , \dots , N$, the subtrees associated with the nodes are in one-to-one correspondence with the subsets $S \subseteq \{1,\cdots,N\}$ made up by their leaves, and one may use these subsets to name the nodes, as in Examples \rref{B-5} and \rref{B-Raupe}. Thus, the leaf labelling reads $\ul\lambda = (\lambda^1 , \dots , \lambda^N)$ and the node labels read $\alpha^S = (\lambda^S,k^S)$. The subset $S=\{1,\dots,N\}$ corresponds to the root, so that $\alpha^{\{1,\dots,N\}} = (\lambda^{\{1,\dots,N\}},k^{\{1,\dots,N\}}) = (\lambda,k)$. Admissibility ensures that 
$$
\lambda^S \in \langle \lambda^n : n \in S \rangle
\,.
$$
The subsets $S$ form a hierarchy, that is, for any two members $S$ and $S'$, either $S\cap S' = \emptyset$\,, $S\cap S' = S$\,, or $S\cap S' = S'$. 
\qeb

\ebm

Now, for given leaf labelling $\ul\lambda$, the diagonal representation $\big(H_{\ul\lambda} , D_d^{\ul\lambda}\big)$ decomposes into uniquely determined isotypical subspaces $H_{\ul\lambda,\lambda}$ labelled by an admissible highest weight label $\lambda \in \langle \ul\lambda \rangle$ of the root. According to the coupling tree $T$ chosen, the isotypical subspaces $H_{\ul\lambda,\lambda}$ can be further decomposed in a non-unique way into irreducible subspaces $H^T_\alpha$ enumerated by the tree labellings $\alpha \in \mc L^T(\ul\lambda,\lambda)$. Thus,
$$
H_{\ul\lambda}
\cong
\bigoplus_{\lambda \in \langle \ul\lambda \rangle}
~
\bigoplus_{\alpha \in \mc L^T(\ul\lambda,\lambda)} 
H^T_\alpha
$$
as $G$-representations, and it is the labels $\alpha^x$ of the internal nodes $x$ and the multiplicity counter $k$ of the root that are in 1-1 correspondence with the irreducible subspaces obtained by means of the reduction scheme related to $T$. Accordingly, the multiplicity of the irrep $(H_\lambda,D^\lambda)$ of highest weight $\lambda$ in the diagonal representation on $H_{\ul\lambda}$ is 
$$
m_{\ul\lambda}(\lambda) = \big|\mc L^T(\ul\lambda,\lambda)\big|
\,,
$$ 
and the isomorphism \eqref{G-D-vp} reads
\beq\label{G-D-vpj}
H_{\ul\lambda}
\cong
\bigoplus_{\lambda \in \langle \ul\lambda \rangle} 
~ 
\bigoplus_{\alpha \in \mc L^T(\ul\lambda,\lambda)} H_\lambda
\,.
\eeq
That is, the latter is obtained by identifying each irreducible subspace $H^T_\alpha$ with a copy of $H_\lambda$. By Schur's Lemma, each of these identifications is unique up to a phase. In what follows, we assume that a phase has been chosen. To summarize, the invisible choices made in the definition of the isomorphism \eqref{G-D-vp} via $T$ amount to these phases and to the choice of a unitary $G$-representation isomorphism \eqref{G-TePr-Zlg} for every pair $(\lambda_1,\lambda_2) \in \widehat G \times \widehat G$ (affecting the definition of the subspaces $H^T_{\alpha}$). Note that we may absorb the direct sum over the final highest weights $\lambda$ into the sum over the labellings and thus write 
\beq\label{G-D-vpj-a}
H_{\ul\lambda}
\cong
\bigoplus_{\alpha \in \mc L^T(\ul\lambda)} H_\lambda
\,.
\eeq
We use the $\alpha$ as labels for the copies of $H_\lambda$ in the direct sum on the right hand side. This has the following consequences. 

Firstly, the projections and injections \eqref{G-D-p-i} obtained by composing this isomorphism with the natural projections and injections of the direct sum read
$$
H_{\ul\lambda} \overset{\prj \alpha T}{\longrightarrow} H_\lambda
\,,\qquad
H_\lambda \overset{\inj{\alpha}{T}}{\longrightarrow} H_{\ul\lambda}
\,.
$$
By construction, each $\prj \alpha T$ is obtained by composing the elementary projections associated with twofold tensor products according to $T$. Thus, $\prj \alpha T$ is uniquely determined by the choice of a reduction tree $T$, a labelling $\alpha$, and the choice of an isomorphism \eqref{G-TePr-Zlg} for every combination of $\lambda_1,\lambda_2 \in \widehat G$ and every $\lambda \in \langle \lambda_1,\lambda_2 \rangle$. An analogous statement holds true for $\inj \alpha T$. 

Secondly, the quasicharacters read $\cq{T}{\alpha}{\alpha'}$ and the representation homomorphisms entering their definition read $A(T)^{\alpha}_{\alpha'}$, for any combinable $\alpha$, $\alpha'$. Using the injections $\inj \alpha T$ and the normalized weight bases $\{\ket{\lambda \mu} : \mu \in \weight(\lambda)\}$ chosen in $H_\lambda$ for every $\lambda \in \widehat G$, we can define elements of $\mc H_{\ul\lambda}$ by
\beq
\label{basis-H-ul-lambda}
\ket{T;\alpha,\mu} 
:= 
\inj \alpha T \big(\ket{\lambda \mu}\big)
\,,\qquad
\alpha \in \mc L^T(\ul\lambda,\lambda)
\,, \quad
\lambda \in \langle \ul\lambda \rangle
\lambda \in \langle \ul\lambda \rangle
\,,~
\mu \in \weight(\lambda)
\,.
\eeq
These elements form an orthonormal basis in $H_{\ul\lambda}$,
$$
\braket{T;\alpha,\mu}{T;\alpha',\mu'} 
=
\delta_{\alpha \alpha'} \, \delta_{\mu \mu'}
\,,
$$
and they are common eigenvectors of the Cartan subalgebra with eigenvalue functional $\mr i\hat\mu$. For fixed $\alpha$, the elements $\ket{T;\alpha,\mu}$, $\mu \in \weight(\lambda)$, form an orthonormal weight basis in the irreducible subspace $H^T_\alpha$. Using \eqref{A-T} and the relation $\prj \alpha T \circ \inj {\alpha'} T = \delta_{\alpha\alpha'} \id_{H_\lambda}$ holding for $\alpha,\alpha' \in \mc L^T(\ul\lambda,\lambda)$, we compute
$$
A(T)^{\alpha'}_{\alpha} (\ket{T;\alpha'',\mu})
=
\frac{1}{\sqrt{\dim(H_\lambda)}}
\inj {\alpha} T \circ \prj{\alpha'}T (\ket{T;\alpha'',\mu})
=
\frac{\delta_{\alpha',\alpha''}}{\sqrt{\dim(H_\lambda)}} \, \ket{T;\alpha,\mu}
$$
for all $\alpha, \alpha', \alpha'' \in \mc L^T(\ul\lambda,\lambda)$. This implies 
\beq\label{Form-A-G}
A(T)^{\alpha'}_\alpha
 =
\frac{1}{\sqrt{\dim(H_\lambda)}}
\sum_{\mu\in\weight(\lambda)} 
\ket{T;\alpha,\mu} \bra{T;\alpha',\mu}
 \,,\qquad
\alpha,\alpha' \in \mc L^T(\ul\lambda,\lambda) 
\,,
\eeq
and
\beq\label{Form-chi}
\dcq{T}{\alpha}{\alpha'}(\ul a)
 =
\sum_{\mu\in\weight(\lambda)}
\bra{T;\alpha',\mu} D^{\ul\lambda}(\ul a) \ket{T;\alpha,\mu}
 \,,\qquad
\alpha,\alpha' \in \mc L^T(\ul\lambda,\lambda)
\,.
\eeq

It will turn out that the key to unravelling the algebraic structure of the invariants in terms of the quasicharacters is the dependence on the coupling tree and the transformation law between different such trees. For $i=1,2$, let $T_i$ be a coupling tree, $\ul\lambda_i$ a leaf labelling of $T_i$, $\lambda_i \in \langle \ul\lambda_i \rangle$ and $\alpha_i \in \mc L^{T_i}(\ul\lambda_i,\lambda_i)$. If $\ul\lambda_2$ is obtained from $\ul\lambda_1$ by a permutation $\sigma$ of the entries, there exists a unitary transformation $\Pi : H_{\ul\lambda_1} \to H_{\ul\lambda_2}$ permuting the tensor factors according to $\sigma$, that is,
$$
\Pi \ket{\ul\lambda_1\,\ul\mu}
= 
\ket{\ul\lambda_2\,\sigma(\ul\mu)} 
\,,\qquad
\ul\mu\in\weight(\ul\lambda_1)
\,.
$$
Then, by orthogonality of isotypical subspaces, the overlap of the basis vectors $\ket{T_1;\alpha_1,\mu_1}$ and $\ket{T_2;\alpha_2,\mu_2}$ vanishes unless $\lambda_1 =\lambda_2$. In that case, $\prj{\alpha_2}{T_2} \circ \Pi \circ \inj{\alpha_1}{T_1}$ is a unitary representation automorphism of the unitary irrep $(H_{\lambda_1},D^{\lambda_1})$ and hence it is given by multiplication by a complex factor of modulus one. Denoting this factor by $R\big(T_2|T_1\big)^{\alpha_2}_{\alpha_1}$, we obtain
\beq\label{G-D-Racah}
\bra{T_2;\alpha_2,\mu_2} \Pi \ket{T_1;\alpha_1,\mu_1}
=
\delta_{\lambda_1 \lambda_2} 
\,
\delta_{\mu_1 \mu_2} 
\,
R\big(T_2|T_1\big)^{\alpha_2}_{\alpha_1}
\,.
\eeq
We will refer to $R\big(T_2|T_1\big)^{\alpha_2}_{\alpha_1}$ as a recoupling coefficient. In case $G=\SU(2)$, the recoupling coefficients coincide with the angular momentum recoupling coefficients (Racah coefficients) \cite{BL2}.

\bbm\label{Bem-Racah}

Explicitly, we have 
\beq\label{G-Racah-expl}
R\big(T_2|T_1\big)^{\alpha_2}_{\alpha_1}
=
\bra{T_2;\alpha_2,\mu} \Pi \ket{T_1;\alpha_1,\mu}
\,,
\eeq
independently of $\mu$. Note that \eqref{G-Racah-expl} yields the symmetry
\beq\label{G-Racah-sym}
R\big(T_2|T_1\big)^{\alpha_2}_{\alpha_1}
=
\left(R\big(T_1|T_2\big)^{\alpha_1}_{\alpha_2}\right)^\ast
\,.
\eeq
Consider a further coupling tree $T_3$ with labelling $\alpha_3$ whose leaf labelling $\ul\lambda_3$ is obtained from $\ul\lambda_2$ by some further permutation. Denote the respective induced unitary transformation permuting the tensor factors by $\Pi_{12} : H_{\ul\lambda_1} \to H_{\ul\lambda_2}$ and $\Pi_{23} : H_{\ul\lambda_2} \to H_{\ul\lambda_3}$. Then, $\Pi_{23} \circ \Pi_{12}$ permutes the tensor factors according to the composite permutation. Assume further that $\alpha_1$ and $\alpha_3$ assign the same weight label $\lambda$ to the root. Then, by inserting an appropriate unit into \eqref{G-Racah-expl}, with index $2$ replaced by $3$ and $\Pi$ replaced by $\Pi_{23} \circ \Pi_{12}$, we obtain the cycle relation
\beq\label{G-Racah-Zykl}
R\big(T_3|T_1\big)^{\alpha_3}_{\alpha_1}
=
\sum_{\alpha_2 \in \mc L^{T_2}(\ul\lambda_2,\lambda)}
R\big(T_3|T_2\big)^{\alpha_3}_{\alpha_2}
R\big(T_2|T_1\big)^{\alpha_2}_{\alpha_1}
\,.
\eeq
\qeb

\ebm

\bsz[Change of Coupling Tree] \label{prop:ChangeOfTree}

Let $T_1$ and $T_2$ be coupling trees with the same number of leaves, let $\ul\lambda$ be a leaf labelling for both of them and let $\lambda \in \langle \ul\lambda \rangle$. Then, for labellings $\alpha_1 , \alpha_1' \in \mc L^{T_1}(\ul\lambda,\lambda)$ and $\alpha_2 , \alpha_2' \in \mc L^{T_2}(\ul\lambda,\lambda)$, the transformation rule between quasicharacters coupled according to $T_1$ and $T_2$ is
\begin{align*}
\dcq{T_1}{\alpha_1}{\alpha_1'}
= &
\sum_{\alpha_2 , \alpha_2' \in \mc L^{T_2}(\ul\lambda,\lambda)}  
R\big(T_1|T_2\big)^{\alpha_1'}_{\alpha_2'} 
\,\,
\dcq{T_2}{\alpha_2}{\alpha_2'}
\,\,
R\big(T_2|T_1\big)^{\alpha_2}_{\alpha_1}
\,.
\end{align*}

\esz

\bbw

Inserting complete sets of coupled states for the coupling tree $T_2$,
$$
\unit 
= 
\sum_{\alpha_2 \in \mc L^{T_2}(\ul\lambda)}
\sum_{\mu_2}
\ket{T_2;\alpha_2,\mu_2}\bra{T_2;\alpha_2,\mu_2}
\,,
$$
into \eqref{Form-chi}, we obtain
\begin{align*}
\dcq{T_1}{\alpha_1}{\alpha_1'}(\ul a)
=&
\sum_{\mu}
~
\sum_{\alpha_2 , \alpha_2' \in \mc L^{T_2}(\ul\lambda)}
~
\sum_{\mu_2,\mu_2'}
\braket{T_1;\alpha_1',\mu}{T_2;\alpha_2',\mu_2'}
\\
& \hspace{2.5cm}
\cdot
\bra{T_2;\alpha_2',\mu_2'}
D^{\ul\lambda}(\ul a) 
\ket{T_2;\alpha_2,\mu_2}
\braket{T_2;\alpha_2,\mu_2}{T_1;\alpha_1,\mu}
\,.
\end{align*}
Application of \eqref{G-D-Racah} now yields the assertion. 
\ebw


\subsection{The multiplication law for quasicharacters}
\label{sec:ProductsSubSec}


In this section, we study the multiplicative structure of the algebra $\mc R$, that is, we analyze the general multiplication law given by Theorem \rref{MultLaw-Gen} in terms of the recoupling calculus developed above. To start with, we define two operations on trees.

\bdf[\textbf{Operations on coupling trees}]

Let $T_1$, $T_2$ and $T$ be coupling trees with, respectively, $N_1$, $N_2$ and $N$ leaves. 

\textbf{Tree join:}
The join $T_1 \!\cdot\! T_2$ is the coupling tree with $N_1+N_2$ leaves formed by glueing the pendant root edges of $T_1$ and $T_2$ to make a new root and add to this new root a pendant edge. 

\textbf{Leaf duplication:} 
The tree $T^\vee$ is the coupling tree with $2N$ leaves formed by replacing each leaf of $T$ by the root of a copy of the rooted binary tree with two leaves (a cherry), thereby identifying the leaf edge with the pendant root edge of that copy. 

\edf

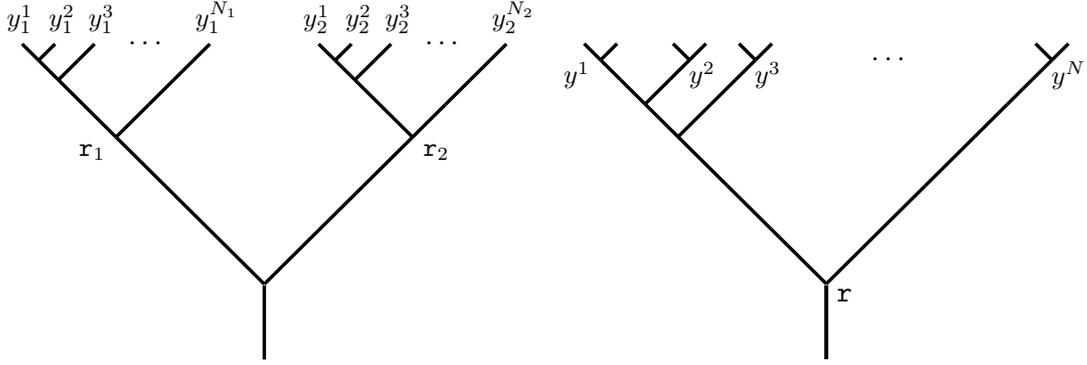
\begin{figure}

\begin{center}

\begin{tikzpicture}

\btreeset{grow=up, scale=0.65, line width=1.3pt}

\BinaryTree[local bounding box=INIT2,line width=1.3pt,bottom padding = 10pt]{%
:!
l!l!l!l!l!l,
l!l!l!l!l!r,
l!l!l!l!r!r,
l!l!r!r!r!r,
r!r!l!l!l!l,
r!r!l!l!l!r,
r!r!l!l!r!r,
r!r!r!r!r!r
}{6}

\draw[thick] (btree-l-l-l-l-l-l) node [above]{\small $y_1^1$\,};
\draw[thick] (btree-l-l-l-l-l-r) node [above]{\small \,\,\,$y_1^2$};
\draw[thick] (btree-l-l-l-l-r-r) node [above]{\small \,\,\,$y_1^3$};
\draw[thick] (btree-l-l-l-l-r-r) node [right]{\hspace{0.3cm}$\cdots$};
\draw[thick] (btree-l-l-r-r-r-r) node [above]{\small \,\,\,$y_1^{N_1}$};
\draw[thick] (btree-r-r-l-l-l-l) node [above]{\small $y_2^1$\,};
\draw[thick] (btree-r-r-l-l-l-r) node [above]{\small \,\,\,$y_2^2$};
\draw[thick] (btree-r-r-l-l-r-r) node [above]{\small \,\,\,$y_2^3$};
\draw[thick] (btree-r-r-l-l-r-r) node [right]{\hspace{0.3cm}$\cdots$};
\draw[thick] (btree-r-r-r-r-r-r) node [above]{\small \,\,\,$y_2^{N_2}$};

\draw[thick] (btree-l-l)+(0,-0.2) node [left]{\small $\rt_1$};
\draw[thick] (btree-r-r)+(0,-0.2) node [right]{\small $\rt_2$};

\draw[line width=1.3pt] (btree-root) -- ($(btree-root)+(0,-1)$);

\end{tikzpicture}
\begin{tikzpicture}

\btreeset{grow=up, scale=0.65, line width=1.3pt}

\BinaryTree[local bounding box=INIT2,line width=1.3pt,bottom padding = 10pt]{
:!
l!l!l!l!l!l,
l!l!l!l!l!r,
l!l!l!r!r!r,
l!l!l!r!r!l,
l!l!r!r!r!r,
l!l!r!r!r!l,
r!r!r!r!r!r,
r!r!r!r!r!l
}{6}

\draw[thick] (btree-l-l-l-l-l)+(0,-0.2) node [left]{\small $y^1$};
\draw[thick] (btree-l-l-l-r-r)+(-0.15,-0.2) node [right]{\small $y^2$};
\draw[thick] (btree-l-l-r-r-r)+(-0.15,-0.2) node [right]{\small $y^3$};
\draw[thick] (btree-l-l-r-r-r) node [right]{\hspace{1.25cm} $\cdots$};
\draw[thick] (btree-r-r-r-r-r)+(-0.15,-0.2) node [right]{\small $y^N$};

\draw[line width=1.5pt] (btree-root) -- ($(btree-root)+(0,-1)$);
\draw[thick] (btree-root)+(0,-0.15) node [right]{$\rt$};

\end{tikzpicture}

\end{center}
\caption{\label{Abb-TT-std} Join $T_1 \!\cdot\! T_2$ of two standard trees $T_i$ with leaves $y_i^1 , \dots , y_i^{N_i}$ and root $\rt_i$, $i=1,2$ (left) and leaf duplication $T^\vee$ of a standard tree $T$ with leaves $y^1 , \dots , y^N$ and root $\rt$ (right).}
\end{figure}

See Figure \ref{Abb-TT-std} for specific examples of tree join and leaf duplication. 

For $i=1,2$, let $\rt_i$ denote the root of $T_i$ and assume that admissible labellings $\alpha_i$ of $T_i$ with underlying highest weight labelling $x \mapsto \lambda_i^x$ are given. 

In case of the operation of join, one can obtain an admissible labelling of $T_1 \!\cdot\! T_2$ by retaining the labellings $\alpha_1$ for $T_1$ and $\alpha_2$ for $T_2$, and by assigning to the new root some value $(\lambda,k) \in \llangle \lambda_1^{\rt_1} , \lambda_2^{\rt_2} \rrangle$. The labelling of $T_1 \!\cdot\! T_2$ so arising will be denoted by $\join{\alpha_1}{\alpha_2}{(\lambda,k)}$. In addition, given leaf labellings $\ul\lambda_i$ of $T_i$, let $\ul\lambda_1 \!\cdot\! \ul\lambda_2$ denote the leaf labelling of $T_1 \!\cdot\! T_2$ obtained by retaining the leaf labellings $\ul\lambda_1$ for $T_1$ and $\ul\lambda_2$ for $T_2$. Thus, if $\ul\lambda_i$ is the leaf labelling belonging to $\alpha_i$, then $\ul\lambda_1 \!\cdot\! \ul\lambda_2$ is the leaf labelling belonging to $\join{\alpha_1}{\alpha_2}{(\lambda,k)}$. Having drawn $T_1$ and $T_2$ in a plane, with $T_1$ to the left and $T_2$ to the right, then the leaf labelling $\ul\lambda_1 \!\cdot\! \ul\lambda_2$ is given by
\beq\label{G-D-join}
\ul\lambda_1 \!\cdot\! \ul\lambda_2 
= 
(\lambda_1^1, \cdots , \lambda_1^{N_1}, \lambda_2^1, \cdots, \lambda_2^{N_2})
\eeq
and the corresponding tensor product by $H_{\lambda_1^1} \otimes \cdots \otimes H_{\lambda_1^{N_1}} \otimes H_{\lambda_2^1} \otimes \cdots \otimes H_{\lambda_2^{N_2}}$. If $N_1=N_2=N$, there corresponds a representation of $G^N$,
$$
D^{\ul\lambda_1 \!\cdot\! \ul\lambda_2}(\ul a)
=
D^{\ul\lambda_1}(\ul a) \otimes D^{\ul\lambda_2}(\ul a)
=
D^{\lambda_1^1}(a_1) \otimes \cdots \otimes D^{\lambda_1^N}(a_N)
\otimes
D^{\lambda_2^1}(a_1) \otimes \cdots \otimes D^{\lambda_2^N}(a_N)
$$
In case of the operation of leaf duplication, the situation is different. Here, given labellings $\alpha_1$ of $T_1$ and $\alpha_2$ of $T_2$, their only parts which can be assigned are their leaf labellings $\ul\lambda_1$ and $\ul\lambda_2$, respectively. They define a leaf labelling $\ul\lambda_1 \!\ast\! \ul\lambda_2$ of $T^\vee$ by assigning $\lambda_i^y$ to the $i$-th child vertex of the cherry replacing leaf $y$ of $T$. As the nodes of $T^\vee$ correspond 1-1 to the vertices of $T$, every node labelling of $T^\vee$ splits  into a labelling of $T$ and an assignment $\ul k$ of a multiplicity counter $k^y$ to every leaf $y$ of $T$. Conversely, assume that we are given a further labelling $\alpha_3$ of $T$ with leaf labelling $\ul\lambda_3$ and an assignment $\ul k_3$ of a positive integer to every leaf of $T$. If $(\lambda_3^y,k_3^y) \in \llangle \lambda_1^y,\lambda_2^y \rrangle$ for all leaves $y$ of $T$, then the node labelling of $T^\vee$ arising from $\alpha_3$ and $\ul k_3$ is compatible with the leaf labelling $\ul\lambda_1\!\ast\!\ul\lambda_2$, and so the two combine to an admissible labelling of $T^\vee$. This labelling will be denoted by $\ld{\alpha_1}{\alpha_2}{\alpha_3,\ul k_3}$. Finally, having drawn $T^\vee$ in a plane, with the first child vertex of every cherry to the left and the second one to the right, the corresponding sequence of leaf labels is given by
\beq\label{G-D-dupl}
\ul\lambda_1 \!\ast\! \ul\lambda_2 
= 
(\lambda_1^1,\lambda_2^1,\cdots,\lambda_1^N,\lambda_2^N)
\eeq
and the corresponding tensor product by $H_{\lambda_1^1} \otimes H_{\lambda_2^1} \otimes \cdots \otimes H_{\lambda_1^N}  \otimes H_{\lambda_2^N}$. There corresponds a representation of $G^N$, 
$$
D^{\ul\lambda_1 \!\ast\! \ul\lambda_2}(\ul a)
=
D^{\lambda_1^1}(a_1) \otimes D^{\lambda_2^1}(a_1) 
\otimes \cdots \otimes 
D^{\lambda_1^N}(a_N) \otimes D^{\lambda_2^N}(a_N)
\,.
$$

\bbm

Leaf duplication is a special case of (rooted) tree composition, $T_1 \!\ast\!  T_2 $, defined as the tree with $N_1 N_2$ leaves formed by replacing each leaf of $T_1$ by the root of a copy of $T_2$, thereby identifying the leaf edge with the pendant root edge of that copy. Thus, $T^\vee = T \!\ast\! \vee$. Given $N_2$ leaf labellings $\ul\lambda_{(1)} , \dots , \ul\lambda_{(N_2)}$ of $T_1$, one defines the leaf labelling of $T_1 \!\ast\! T_2$ by
$$
\ul\lambda_{(1)} \ast \cdots \ast \ul\lambda_{(N_2)}
:=
\big(
\lambda^1_{(1)} , \dots , \lambda^1_{(N_2)} , \dots , \lambda^{N_1}_{(1)} , \dots , \lambda^{N_1}_{(N_2)}
\big)
\,.
$$
It is immediate that tree join and tree composition are subject to the 'distributive law'
\beq
(T_1 \!\cdot\! T_2) \!\ast\! T_3 = (T_1 \!\ast\! T_3) \!\cdot\! (T_2 \!\ast\! T_3) 
\,.
\eqqeb
\eeq

\ebm

\noindent
In the sequel we adapt the notations $\ul\lambda_1 \!\cdot\! \ul\lambda_2$ and $\ul\lambda_1 \!\ast\! \ul\lambda_2$ given by \eqref{G-D-join} and \eqref{G-D-dupl} to weight labels to obtain weight labellings $\ul\mu_1 \!\cdot\! \ul\mu_2$ and $\ul\mu_1 \!\ast\! \ul\mu_2$ of the leaves of $T_1 \!\cdot\! T_2$ and $T^\vee$, respectively. There correspond tensor product states $\ket{\ul\lambda_1 \bcdot \ul\lambda_2 \ \ul\mu_1 \bcdot \ul\mu_2}$ and $\ket{\ul\lambda_1 \!\ast\! \ul\lambda_2 \ \ul\mu_1 \!\ast\! \ul\mu_2}$, respectively. In the case where $T_1$, $T_2$ and $T$ have the same number $N$ of leaves, there exists a permutation operator $\Pi$ such that
\beq
\label{G-D-U}
\ket{\ul\lambda_1 \bcdot \ul\lambda_2 \ \ul\mu_1 \bcdot \ul\mu_2}
=
\Pi \ket{\ul\lambda_1 \!\ast\! \ul\lambda_2 \ \ul\mu_1 \!\ast\! \ul\mu_2} 
\,,\qquad
\Pi^{-1} D^{\ul\lambda_1 \bcdot \ul\lambda_2} (\ul a) \Pi 
= 
D^{\ul\lambda_1 \!\ast\! \ul\lambda_2} (\ul a)
\,.
\eeq

With these preliminaries, we are able to express the multiplication law for quasicharacters in terms of the recoupling coefficients. For that purpose, let labellings $\alpha_1$, $\alpha_2$, $\alpha_3$ of $T$, an assignment $\ul k$ of a positive integer to every leaf of $T$ and a positive integer $k$ be given. Let $x \mapsto \lambda_i^x$ be the highest weight labelling underlying $\alpha_i$. We write $(\alpha_3,\ul k,k) \in \llangle \alpha_1,\alpha_2 \rrangle$ if $(\lambda_3^y,k^y) \in \llangle \lambda_1^y,\lambda_2^y \rrangle$ for all leaves $y$ of $T$ and $(\lambda_3^\rt,k) \in \llangle \lambda_1^\rt,\lambda_2^\rt \rrangle$, where $\rt$ denotes the root of $T$. We define 
$$
\rcneu{T}{\alpha_1}{\alpha_2}{\alpha_3}{\ul k}{k}
:=
\begin{cases}
R\big(T \bcdot T|T^\vee\big)
^{\join{\alpha_1}{\alpha_2}{(\lambda_3^\rt,k)}}
_{\ld{\alpha_1}{\alpha_2}{\alpha_3,\ul k}}
& |\abs 
(\alpha_3,\ul k,k) \in \llangle \alpha_1,\alpha_2 \rrangle
\,,
\\
0 & |\abs \text{otherwise}
\end{cases}
$$
and the adjoint
$$
\rctneu{T}{\alpha_1}{\alpha_2}{\alpha_3}{\ul k}{k}
:=
\begin{cases}
R\big(T^\vee|T \bcdot T\big)
^{\ld{\alpha_1}{\alpha_2}{\alpha_3,\ul k}}
_{\join{\alpha_1}{\alpha_2}{(\lambda_3^\rt,k)}}
& |\abs 
(\alpha_3,\ul k,k) \in \llangle \alpha_1,\alpha_2 \rrangle
\,,
\\
0 & |\abs \text{otherwise.}
\end{cases}
$$
By \eqref{G-Racah-sym}, 
$$
\rcneu{T}{\alpha_1}{\alpha_2}{\alpha_3}{\ul k}{k}
=
\big(\rctneu{T}{\alpha_1}{\alpha_2}{\alpha_3}{\ul k}{k}\big)^\ast
\,.
$$ 
It turns out that for our choice of the isomorphism \eqref{G-D-vp}, the coefficients $U$ in Formula \eqref{eq: UnitaryCoeff-1} are given by the recoupling coefficients $R(T)$:

\ble
\label{L-R-U}

For $i=1,2,3$, let $\ul\lambda_i$ be a leaf labelling of $T$, let $\lambda_i \in \langle \ul\lambda_i \rangle$ and let $\alpha_i \in \mc L^T(\ul\lambda_i,\lambda_i)$. Let $\ul k$ be an assignment of a positive integer to every leaf of $T$, let $k$ be a positive integer and assume that $(\alpha_3,\ul k,k) \in \llangle \alpha_1,\alpha_2 \rrangle$. Then, 
$$
\rcneu{T}{\alpha_1}{\alpha_2}{\alpha_3}{\ul k}{k}
=
U
^{\ul\lambda_1,\lambda_1,l_1;\ul\lambda_2,\lambda_2,l_2;k}
_{\ul\lambda_3,\lambda_3,l_3;\ul k}
\,,
$$
where $l_i$ is the positive integer corresponding to $\alpha_i$ relative to some chosen enumeration of the elements of $\mc L^T(\ul\lambda_i,\lambda_i)$, $i=1,2,3$. 

\ele

\bbw

For any $\mu \in \weight(\lambda_3)$, we have 
\ala{
\rcneu{T}{\alpha_1}{\alpha_2}{\alpha_3}{\ul k}{k}
& =
R(T \bcdot T|T^\vee)
^{\join{\alpha_1}{\alpha_2}{(\lambda_3,k)}}
_{\ld{\alpha_1}{\alpha_2}{\alpha_3,\ul k}}
\\
& =
\bigbra{T \bcdot T;\join{\alpha_1}{\alpha_2}{(\lambda_3,k)},\mu}
\,\Pi\,
\bigket{T^\vee;\ld{\alpha_1}{\alpha_2}{\alpha_3,\ul k},\mu}
\\
& =
\left\langle
\lambda_3 \ \mu
\left|
\prj{\join{\alpha_1}{\alpha_2}{(\lambda_3,k)}}{T \bcdot T}
\circ 
\Pi 
\circ
\inj{\ld{\alpha_1}{\alpha_2}{\alpha_3,\ul k}}{T^\vee}
\right|
\lambda_3 \ \mu
\right\rangle
\,,
}
where $\Pi$ denotes the unitary transformation given by \eqref{G-D-U}. We observe that 
$$
\Pi \circ \inj{\ld{\alpha_1}{\alpha_2}{\alpha_3,\ul k}}{T^\vee}
=
\left(\otimes_{i=1}^N \inj{(\lambda_1^i,\lambda_2^i),\lambda^i,k^i}{}\right)
\circ
\inj{\ul\lambda,\lambda,l}{}
$$
and that
$$
\prj{\join{\alpha_1}{\alpha_2}{(\lambda_3,k)}}{T \bcdot T}
=
\prj{(\lambda_1,\lambda_2),\lambda',k}{}
\circ
\left(
\prj{\ul\lambda_1,\lambda_1,l_1}{} \otimes \prj{\ul\lambda_2,\lambda_2,l_2}{}
\right)
\,.
$$
Hence, the assertion follows from \eqref{eq: UnitaryCoeff-1}.
\ebw

Using this lemma, we immediately obtain the following corollary to Theorem 
\ref{MultLaw-Gen}.

\bfg[\textbf{Multiplication law for quasicharacters}]
\label{TreeProductRule}\label{F-TPR}

For a given coupling tree $T$, the product of quasicharacters is given by
$$
\dcq{T}{\alpha_1^{\phantom{\prime}}}{\alpha_1'}
\,
\dcq{T}{\alpha_2^{\phantom{\prime}}}{\alpha_2'}
=
\sum_{\alpha,\alpha'}
\sum_{\ul k,k}
\,
\rcneu{T}{\alpha_1'}{\alpha_2'}{\alpha'}{\ul k}{k}
\,\,
\dcq{T}{\alpha^{\phantom{\prime}}}{\alpha'}
\,\,
\rctneu{T}{\alpha_1}{\alpha_2}{\alpha}{\ul k}{k}
\,,
$$
where the sum is over all combinable labellings $\alpha$, $\alpha'$ of $T$, all assignments $\ul k$ of a positive integer to every leaf of $T$ and all positive integers $k$ such that $(\alpha,\ul k,k) \in \llangle \alpha_1,\alpha_2 \rrangle$.
\qed

\efg

For a direct proof of this corollary using the Clebsch-Gordan calculus see the appendix.

\bbm
\label{Bem-T-TPR}
~
\ben
\item As a result, the structure constants of the algebra $\mc R$ with respect to the basis $\{\dcq{T}{\alpha}{\alpha'}\}$ are given by 
$
\sum_{\ul k,k}
\rcneu{T}{\alpha_1'}{\alpha_2'}{\alpha_3'}{\ul k}{k}
\rctneu{T}{\alpha_1}{\alpha_2}{\alpha_3}{\ul k}{k}
$
with free summation indices $\alpha_3$ and $\alpha_3'$.

\item For a threefold product, we obtain
\ala{
\dcq{T}{\alpha_1^{\phantom{\prime}}}{\alpha_1'}
\, &
\dcq{T}{\alpha_2^{\phantom{\prime}}}{\alpha_2'}
\,
\dcq{T}{\alpha_3^{\phantom{\prime}}}{\alpha_3'}
\\
= &
\sum_{\beta_2,\beta_2' \atop \beta_3,\beta_3'}
\sum_{\ul k_2,k_2 \atop \ul k_3,k_3}
\,
\rcneu{T}{\alpha_1'}{\alpha_2'}{\beta_2'}{\ul k_2}{k_2}
\,\,
\rcneu{T}{\beta_2'}{\alpha_3'}{\beta_3'}{\ul k_3}{k_3}
\,\,
\dcq{T}{\beta_3^{\phantom{\prime}}}{\beta_3'}
\,\,
\rctneu{T}{\beta_2}{\alpha_3}{\beta_3}{\ul k_3}{k_3}
\,\,
\rctneu{T}{\alpha_1}{\alpha_2}{\beta_2}{\ul k_2}{k_2}
\,.
}
This generalizes in an obvious way to $n$-fold products.
\qeb

\een

\ebm

\bbm
\label{TreeProductCorollary}

The above formula generalizes directly to the pointwise multiplication $\hat\chi(T_1)\bcdot \hat\chi(T_2)$ of modified quasicharacters for two different coupling trees $T_1$ and $T_2$ having the same number of leaves. The result is expressible in terms of the $\hat\chi(T_3)$-basis provided by an arbitrary third tree $T_3$ with the same number of leaves, with $T \bcdot T$ replaced by $T_1 \bcdot T_2$\,, and $T^\vee$ replaced by $T_3^\vee$,
$$
\dcq{T_1}{\alpha_1}{\alpha_1'}
\,
\dcq{T_2}{\alpha_2}{\alpha_2'}
=
\sum_{\alpha_3,\alpha_3'}
\sum_{\ul k,k}
R\big(T_1 \bcdot T_2 | T_3^\vee\big)
^{\join{\alpha_1'}{\alpha_2'}{(\lambda_3^\rt,k)}}
_{\ld{\alpha_1'}{\alpha_2'}{\alpha_3',\ul k}}
\,\,
\dcq{T_3}{\alpha_3}{\alpha_3'}
\,\,
R\big(T_3^\vee | T_1 \bcdot T_2\big)
^{\ld{\alpha_1}{\alpha_2}{\alpha_3,\ul k}}
_{\join{\alpha_1}{\alpha_2}{(\lambda_3^\rt,k)}}
$$
where the sum is over all combinable labellings $\alpha_3$, $\alpha_3'$ of $T_3$, all assignments $\ul k_3$ of a positive integer to every leaf of $T_3$ and all positive integers $k_3$ such that
$(\alpha_3,\ul k_3) \in \llangle \alpha_1,\alpha_2 \rrangle$ and $(\lambda_3,k) \in \llangle \lambda_1,\lambda_2 \rrangle$. In this case, the recoupling coefficients measure the overlap of two $2N$-leaf coupling trees, and so each of them entails $2N-2$ internal labels; including the final coupling to $\lambda_3$ and adding the $2N$ leaf labels $\ul\lambda_1, \ul\lambda_2$, they are thus of $(6N\!-\!3)j$-type, reflecting the count for structure constants arising from expanding the pointwise product of two $\chi$'s into a sum, $3(2N\!-\!1) = 6N\!-\!3$. The structure constants reflect the commutativity of the pointwise product because of their symmetry (under interchange of $\alpha_1$ with $\alpha_2$ and $\alpha'_1$ with $\alpha'_2$). Associativity on the other hand is not manifest, but is clearly a concomitant of the tensor-categorial origins of the recoupling calculus itself; there appears moreover to be a functorial association between the pointwise algebraic product, and the above combinatorial tree operations. In this light, it might be expected that the compound structure constants derived above should be replaced by single, but more elaborated, recouplings involving higher degree tree operations such as $\langle (T\!\cdot\!T)\!\cdot\!T|T\!*\!((\cdot,\cdot),\cdot)\rangle$\,, and their $n$-ary generalizations.
\qeb

\ebm


\subsection{Composite Clebsch-Gordan coefficients}
\label{A-RC-CCG}


In this subsection, we analyze the combinatorics of angular momentum theory phrased in 
terms of Clebsch-Gordan coefficients for the case of a compact group $G$. Using this 
calculus, we will be able to reduce the recoupling coefficients arising in our quasicharacter manipulations to products of more basic coefficients (see \S \ref{A-RC-Red}below).

Let $T$ be a coupling tree, $\ul\lambda$ a leaf labelling and $\lambda \in \langle\ul\lambda\rangle$. For $\alpha \in \mc L^T(\ul\lambda,\lambda)$, $\ul\mu\in\weight(\ul\lambda)$ and $\mu\in\weight(\lambda)$, we define
$$
\ccg{T}{\alpha}{\ul\mu}{\mu}
:= 
\braket{\ul\lambda \ \ul\mu}{T;\alpha,\mu}
\,.
$$
Then, for $\mu \in \weight(\lambda)$, we have
\beq\label{G-CCG-Entw}
\ket{T;\alpha,\mu} 
= 
\sum_{\ul\mu \in \weight(\ul\lambda)}
\ccg{T}{\alpha}{\ul\mu}{\mu} \, \ket{\ul\lambda\,\ul\mu}
\,.
\eeq
Since $\ket{\ul\lambda\,\ul\mu}$ and $\ket{T;\alpha,\mu}$ are common eigenvectors of a commutative algebra of skew-adjoint operators with eigenvalue functionals $\mr i \fnlsum{\ul{\hat\mu}}$ and $\mr i \hat\mu$, respectively, we have $\ccg{T}{\alpha}{\ul\mu}{\mu} = 0$ unless $\fnlsum{\ul{\hat\mu}}=\hat\mu$. Hence, the sum restricts automatically to $\ul\mu$ satisfying $\fnlsum{\ul{\hat\mu}}=\hat\mu$. In case $N=2$, the only coupling tree is the cherry $\boldsymbol\vee$, with labelling assigning $\lambda^1$ and $\lambda^2$ to the leaves and $(\lambda,k) \in \llangle \lambda^1,\lambda^2 \rrangle$ to the root. For $\mu^n \in \weight(\lambda^n)$ and $\mu \in \weight(\lambda)$, the coefficients 
$$
\ocg{\lambda^1}{\lambda^2}{\lambda}{k}{\mu^1}{\mu^2}{\mu}
:=
\ccg
{\boldsymbol\vee}
{\text{\small$($}\lambda^1,\lambda^2,(\lambda,k)\text{\small$)$}}
{(\mu^1,\mu^2)}
{\mu}
$$
are the analogues of the ordinary Clebsch-Gordan coefficients in the case $G=\SU(2)$ and hence will be referred to as the ordinary Clebsch-Gordan coefficients for $G$.
These cofficients can be chosen to be real \cite{AKHD}. Occasionally, we will make use of this. The coefficients $\ccg{T}{\alpha}{\ul\mu}{\mu}$ will turn out to be products of ordinary Clebsch-Gordan coefficients for $G$. Therefore, they will be referred to as composite Clebsch-Gordan coefficients for $G$. Quasicharacters and recoupling coefficients can be expressed in terms of composite Clebsch-Gordan coefficients as follows.

\bsz\label{S-CCG}
~
\ben

\sitem\label{i-S-CCG-qc}
For every coupling tree $T$, leaf labelling $\ul\lambda$, $\lambda \in \langle \ul\lambda \rangle$ and $\alpha,\alpha' \in \mc L^T(\ul\lambda,\lambda)$, one has
$$
\dcq{T}{\alpha}{\alpha'}(\ul a)
=
\sum_{\mu\in\weight(\lambda)} 
\,
\sum_{\ul\mu,\ul\mu'\in\weight(\ul\lambda)}
 \,
\ccg{T}{\alpha}{\ul\mu}{\mu}
\, 
\left(\ccg{T}{\alpha'}{\ul\mu'}{\mu}\right)^\ast
\,
D^{\lambda^1}_{\mu'^1 \mu^1}(a_1) \cdots D^{\lambda^N}_{\mu'^N \mu^N}(a_N)
\,.
$$

\sitem\label{i-S-CCG-rc} 
For $i=1,2$, let $T_i$ be a coupling tree, $\ul\lambda_i$ a leaf labelling, $\lambda_i \in \langle \ul\lambda_i \rangle$ and $\alpha_i \in \mc L^{T_i}(\ul\lambda_i,\lambda_i)$. If $\ul\lambda_2 = \sigma(\ul\lambda_1)$ for some permutation $\sigma$ and $\lambda_2=\lambda_1$, then
$$
R\big(T_2|T_1\big)^{\alpha_2}_{\alpha_1}
=
\sum_{\ul\mu\in\weight(\ul\lambda_1)}
\Big(\ccg{T_2}{\alpha_2}{\sigma(\ul\mu)}{\mu}\Big)^\ast
\,
\ccg{T_1}{\alpha_1}{\ul\mu}{\mu}
$$
for any $\mu\in\weight(\lambda_1)$.

\een

\esz

In both situations, the sum over $\ul\mu$ (and $\ul\mu'$) restricts automatically to contributions where $\fnlsum{\ul{\hat\mu}} = \hat\mu$.

\bbw

Point \rref{i-S-CCG-qc} follows by plugging \eqref{G-CCG-Entw} into \eqref{Form-chi}. Point \rref{i-S-CCG-rc} follows by plugging \eqref{G-CCG-Entw} into the definition of $R(T_1|T_2)$ and using that $\bra{\ul\lambda_2 \, \ul\mu_2} \Pi \ket{\ul\lambda_1 \, \ul\mu_1} = \delta_{\ul\mu_2,\sigma(\ul\mu_1)}$, where $\Pi$ denotes the unitary transformation permuting the tensor factors according to $\sigma$. 
\ebw

\comment{
To conclude, let us note the main properties of Clebsch-Gordan coefficients.

\ble[Properties of ordinary Clebsch-Gordan coefficients]\label{L-CG}

For $\lambda^1,\lambda^2 \in \widehat G$, $(\lambda,k) \in \llangle \lambda^1,\lambda^2 \rrangle$ and $\mu^1 \in \weight(\lambda^1)$, $\mu^2 \in \weight(\lambda^2)$, $\mu \in \weight(\lambda)$, the following holds.

\ben

\sitem\label{i-L-CG-sym}
Symmetry:\abs 
$
\ocg{\lambda^1}{\lambda^2}{\lambda}{k}{\mu^1}{\mu^2}{\mu}
=
\ocg{\lambda^2}{\lambda^1}{\lambda}{k}{\mu^2}{\mu^1}{\mu}
$.

\sitem\label{i-L-CG-ogon}
Orthogonality:\abs 
$
\ocg{\lambda^1}{\lambda^2}{\lambda}{k}{\mu^1}{\mu^2}{\mu}
=
0
$
unless $\hat\mu^1 + \hat\mu^2 = \hat\mu$.

\een

\ele

\bbw

\rref{i-L-CG-sym}.\abs
The transformation defined by exchanging the tensor factors is a unitary representation isomorphism. 

\rref{i-L-CG-ogon}.\abs
The vectors $\bigket{(\lambda^1,\lambda^2) \ (\mu^1,\mu^2)}$ and 
$
\ket{
\boldsymbol\vee
;
\text{\small$($}\lambda^1,\lambda^2,(\lambda,k)\text{\small$)$}
,
\mu
}
$
are common eigenvectors of a commutative algebra of skew-adjoint operators with eigenvalue functionals $\mr i(\hat\mu_1+\hat\mu_2)$ and $\mr i \hat\mu$, respectively.
\ebw
}

Let us introduce the following terminology. Given a coupling tree $T$, a leaf labelling $\ul\lambda$ and $\ul\mu\in\weight(\ul\lambda)$, for every vertex $x$ of $T$, let $\hat\mu^x$ be the sum of $\hat\mu^y$ over all leaves $y$ among the descendants of $x$. We refer to the assignment $x \mapsto \hat\mu^x$ as the weight labelling of $T$ generated by $\ul\mu$. Given, in addition, $\alpha\in\mc L^T(\ul\lambda)$ with underlying highest weight labelling $x \mapsto \lambda^x$, we say that $\ul\mu$ and $\alpha$ are compatible if $\hat\mu^x\in\hat\weight(\lambda^x)$ for all nodes $x$ (for the leaves this holds true by construction).

\ble[Product formula for composite Clebsch-Gordan coefficients]\label{L-CCG}

Let $T$ be a coupling tree, $\ul\lambda$ a leaf labelling of $T$, $\lambda \in \langle \ul\lambda \rangle$, $\alpha \in \mc L^T(\ul\lambda,\lambda)$, $\ul\mu\in\weight(\ul\lambda)$ and $\mu\in\weight(\lambda)$. Let $x \mapsto \lambda^x$ be the highest weight labelling and $x \mapsto k^x$ the multiplicity counter labelling underlying $\alpha$. Let $x \mapsto \hat\mu^x$ be the weight labelling generated by $\ul\mu$. Then, $\ccg{T}{\alpha}{\ul\mu}{\mu}=0$ unless $\ul\mu$ and $\alpha$ are compatible. In that case,
\beq
\label{eq:CompositeCG}
\ccg{T}{\alpha}{\ul\mu}{\mu}
 =
\sum_{\check\mu}
\,
\prod_{x \text{ node of } T} 
\left(
C
^{\lambda^{x'}\lambda^{x''}\lambda^x,k^x}
_{(\hat\mu^{x'},\check\mu^{x'})(\hat\mu^{x''},\check\mu^{x''})(\hat\mu^x,\check\mu^x)}
\right)
\,,
\eeq
where $x'$, $x''$ denote the child vertices of $x$ and where the sum runs over all assignments $x \mapsto \check\mu^x = 1 , \dots , m_{\lambda^x}(\hat\mu^x)$ of a weight multiplicity counter to every internal node $x$. 

\ele

\bbw

The proof is by induction on the number $N$ of leaves. For $N=2$, the assertion holds by definition of the ordinary $G$-Clebsch-Gordan coefficients. Thus, assume that it holds for $N$ and that $T$ has $N+1$ leaves.  Let $\rt$ be the root of $T$. For clarity, we write $\lambda^\rt$ for $\lambda$. The child nodes $\rt'$ and $\rt''$ of $\rt$ are the roots of subtrees $S'$ and $S''$, respectively. Accordingly, $\ul\lambda$ splits into leaf labellings $\ul\lambda'$ of $S'$ and $\ul\lambda''$ of $S''$, $\ul\mu$ splits into $\ul\mu' \in \weight(\ul\lambda')$ and $\ul\mu'' \in \weight(\ul\lambda'')$, and $\alpha$ splits into labellings $\alpha'$ of $S'$, $\alpha''$ of $S''$ and $\alpha^\rt=(\lambda^\rt,k^\rt)$ of the root. For convenience, let us introduce the notation $\tilde\alpha$ for the induced labelling of the cherry $\boldsymbol\vee$ made up by the nodes $\rt$, $\rt'$ and $\rt''$. Thus, $\tilde\alpha$ consists of the leaf labels $\lambda^{\rt'}$, $\lambda^{\rt''}$ and the root label $\alpha^\rt$.

Since $H_{\ul\lambda} = H_{\ul\lambda'} \otimes H_{\ul\lambda''}$ under the bracketing associated with $T$, we may expand $\ket{T;\alpha,\mu}$ with respect to the orthonormal basis vectors $\ket{S';\beta',\nu'} \otimes \ket{S'';\beta'',\nu''}$. To compute the expansion coefficients, we observe that the diagram 
$$
\xymatrix{
{H_{\ul\lambda'} \otimes H_{\ul\lambda''}} 
\ar[rrrr]^{=} 
\ar[d]^{\prj{\alpha'}{S'} \otimes \prj{\alpha''}{S''}}  
& & & &
{H_{\ul\lambda}} 
\ar[d]^{\prj{\alpha}{T}} 
\\
{
H_{\lambda^{\rt'}} \otimes H_{\lambda^{\rt''}}
} 
\ar[rrrr]^{\prj{\tilde{\alpha}}{\boldsymbol\vee}}
& & & &
H_{\lambda^\rt}
}
$$
commutes, by construction of the projections involved. Thus, for the corresponding injections we have 
$$
\inj{\alpha}{T}
=
\left(\inj{\alpha'}{S'} \otimes \inj{\alpha''}{S''}\right)
\circ 
\inj{\tilde\alpha}{\boldsymbol\vee}
\,.
$$
Hence,
\al{\nonumber
\ket{T;\alpha,\mu}
& =
\left(\inj{\alpha'}{S'} \otimes \inj{\alpha''}{S''}\right)
\circ 
\inj{\tilde\alpha}{\boldsymbol\vee}
\big(\ket{\lambda^\rt \mu}\big)
\\ \nonumber
& =
\sum_{\mu'\in\weight(\lambda^{\rt'})}
\,
\sum_{\mu''\in\weight(\lambda^{\rt''})}
\,
\ocg{\lambda^{\rt'}}{\lambda^{\rt''}}{\lambda^\rt}{k^\rt}{\mu'}{\mu''}{\mu}
\left(\inj{\alpha'}{S'} \otimes \inj{\alpha''}{S''}\right)
\big(\ket{\lambda^{\rt'} \mu'} \otimes \ket{\lambda^{\rt''} \mu''}\big)
\\
& =
\sum_{\mu'\in\weight(\lambda^{\rt'})}
\,
\sum_{\mu''\in\weight(\lambda^{\rt''})}
\,
\ocg{\lambda^{\rt'}}{\lambda^{\rt''}}{\lambda^\rt}{k^\rt}{\mu'}{\mu''}{\mu}
\,
\ket{S';\alpha',\mu'} \otimes \ket{S'';\alpha'',\mu''}
\,.
}
Using this, we find
\ala{
\ccg{T}{\alpha}{\ul\mu}{\mu}
& =
\bigbraket
{\ul\lambda \, \ul\mu}
{T;\alpha,\mu}
=
\sum_{\mu'\in\weight(\lambda^{\rt'}) \atop \mu''\in\weight(\lambda^{\rt''})}
\,
\ocg{\lambda^{\rt'}}{\lambda^{\rt''}}{\lambda^\rt}{k^\rt}{\mu'}{\mu''}{\mu}
\,
\bigbraket
{\ul\lambda' \, \ul\mu'}
{S';\alpha',\mu'}
\bigbraket
{\ul\lambda'' \, \ul\mu''}
{S'';\alpha'',\mu''}
.
}
We have $\bigbraket{\ul\lambda' \, \ul\mu'}{S';\alpha',\mu'}=0$ unless $\hat\mu' = \fnlsum{\ul{\hat\mu'}}$ and $\bigbraket{\ul\lambda'' \, \ul\mu''}{S'';\alpha'',\mu''}=0$ unless $\hat\mu'' = \fnlsum{\ul{\hat\mu''}}$. Moreover, $\fnlsum{\ul{\hat\mu'}} = \hat\mu^{\rt'}$ and $\fnlsum{\ul{\hat\mu''}} = \hat\mu^{\rt''}$. It follows that $\ccg{T}{\alpha}{\ul\mu}{\mu}$ vanishes unless $\hat\mu^{\rt'} \in \hat\weight(\lambda^{\rt'})$ and $\hat\mu^{\rt''} \in \hat\weight(\lambda^{\rt''})$. In that case,
$$
\ccg{T}{\alpha}{\ul\mu}{\mu}
=
\sum_{\check\mu^{\rt'}=1}^{m_{\lambda^{\rt'}}(\hat\mu^{\rt'})}
\,
\sum_{\check\mu^{\rt''}=1}^{m_{\lambda^{\rt''}}(\hat\mu^{\rt''})}
\,
C
^{\lambda^{\rt'} \lambda^{\rt''} \lambda^\rt , k^\rt}
_{
(\hat\mu^{\rt'},\check\mu^{\rt'})
(\hat\mu^{\rt''},\check\mu^{\rt''})
\,
\mu
}
\,
\ccg{S'}{\alpha'}{\ul\mu'}{(\hat\mu^{\rt'},\check\mu^{\rt'})}
\,
\ccg{S''}{\alpha''}{\ul\mu''}{(\hat\mu^{\rt''},\check\mu^{\rt''})}
\,.
$$
Since $S'$ and $S''$ have at most $N$ leaves, the induction assumption holds for the two composite Clebsch-Gordan coefficients on the right hand side. Since the nodes of $S'$ and the nodes of $S''$ make up the internal nodes of $T$, the assertion follows.
\ebw


\subsection{Reduction of recoupling coefficients}
\label{A-RC-Red}


In this subsection, we are going to express the recoupling coefficients $R(T)$ appearing in the multiplication law for quasicharacters in terms of the elementary recoupling coefficients $R(\boldsymbol\vee)$ for a cherry $\boldsymbol\vee$. The latter correspond to the recoupling between $\boldsymbol\vee \!\cdot\! \boldsymbol\vee$ and $\boldsymbol\vee^\vee$, see Figure \rref{Abb-RV}. 
\begin{figure}

~\hfill
\begin{tikzpicture}

\btreeset{grow=up, scale=0.6, line width=1.5pt}

\BinaryTree[local bounding box=INIT2,line width=1pt,bottom padding = 10pt]{%
:!
l!l,
l!r,
r!l,
r!r     
}{2}

\draw[thick] (btree-l-l) node [above]{\small $\lambda_1^1$};
\draw[thick] (btree-l-r) node [above]{\small $\lambda_1^2$};
\draw[thick] (btree-r-l) node [above]{\small $\lambda_2^1$};
\draw[thick] (btree-r-r) node [above]{\small $\lambda_2^2$};

\draw[thick] (btree-l)+(0,-0.2) node [left]{\small $(\lambda_1,k_1)$};
\draw[thick] (btree-r)+(0,-0.2) node [right]{\small $(\lambda_2,k_2)$};
\draw[thick] (btree-root)+(0,-0.15) node [right]{\small $(\lambda_3,k)$};

\draw[thick] (btree-root)+(0,-0.85) node [below]{(a)};

\draw[line width=1.5] (btree-root) -- ($(btree-root)+(0,-0.5)$);

\end{tikzpicture}
\hfill
\begin{tikzpicture}

\btreeset{grow=up, scale=0.6, line width=1.5pt}

\BinaryTree[local bounding box=INIT2,line width=1pt,bottom padding = 10pt]{%
:!
l!l,
l!r,
r!l,
r!r     
}{2}

\draw[thick] (btree-l-l) node [above]{\small $\lambda_1^1$};
\draw[thick] (btree-l-r) node [above]{\small $\lambda_2^1$};
\draw[thick] (btree-r-l) node [above]{\small $\lambda_1^2$};
\draw[thick] (btree-r-r) node [above]{\small $\lambda_2^2$};

\draw[thick] (btree-l)+(0,-0.2) node [left]{\small $(\lambda_3^1,k^1)$};
\draw[thick] (btree-r)+(0,-0.2) node [right]{\small $(\lambda_3^2,k^2)$};
\draw[thick] (btree-root)+(0,-0.15) node [right]{\small $(\lambda_3,k_3)$};

\draw[thick] (btree-root)+(0,-0.85) node [below]{(b)};

\draw[line width=1.5] (btree-root) -- ($(btree-root)+(0,-0.5)$);

\end{tikzpicture}
\hfill~

\caption{\label{Abb-RV} The elementary recoupling coefficient $\rcneu{\boldsymbol\vee}{\alpha_1}{\alpha_2}{\alpha_3}{\ul k}{k}$ relates the coupling tree $\boldsymbol\vee \!\cdot\! \boldsymbol\vee$ with labelling $\join{\alpha_1}{\alpha_2}{(\lambda_3,k)}$  (left) with the coupling tree $\boldsymbol\vee^\vee$ with labelling $\ld{\alpha_1}{\alpha_2}{\alpha_3,\ul k}$ (right). Combinatorially both $\boldsymbol\vee\!\cdot\!\boldsymbol\vee$ and $\boldsymbol\vee^\vee$ are identical with the balanced 4 leaf tree.}

\end{figure}
One may organize the parameters entering this coefficient into a matrix symbol as follows. For $i=1,2,3$, let $\alpha_i$ be a labelling of $\boldsymbol\vee$ assigning $\lambda_i^1$ and $\lambda_i^2$ to the leaves and $(\lambda_i,k_i)$ to the root. Given an assignment $\ul k = (k^1,k^2)$ of positive integers to the leaves of $\boldsymbol\vee$ and a positive integer $k$, one defines
$$
\bpma
\lambda_1^1 & \lambda_1^2 & \lambda_1 & k_1
\\
\lambda_2^1 & \lambda_2^2 & \lambda_2 & k_2
\\
\lambda_3^1 & \lambda_3^2 & \lambda_3 & k_3
\\
k^1 & k^2 & k
\epma
:=
\rcneu{\boldsymbol\vee}{\alpha_1}{\alpha_2}{\alpha_3}{\ul k}{k}
\,.
$$
In case $G=\SU(2)$, where the multiplicity labels may be omitted, these matrix symbols reduce to the ordinary Wigner $9j$ symbols, up to a dimension factor. Therefore, we will refer to them as $9\lambda$ symbols. By construction, the entries $\#1,\dots,\#4$ of any full row satisfy $(\#3,\#4) \in \llangle \#1,\#2 \rrangle$, because the $\alpha_i$ are admissible labellings of $\boldsymbol\vee$. We refer to this statement as the full row property. We may extend the definition of the $9\lambda$ symbol to arbitrary weights in the upper left $(3 \times 3)$-block and arbitrary positive integers as the remaining entries by putting it to $0$ whenever the full row property does not hold. Moreover, by definition of the recoupling coefficient on the right hand side, the $9\lambda$ symbol vanishes unless $(\alpha_3,\ul k,k) \in \llangle \alpha_1,\alpha_2 \rrangle$. Thus, it vanishes unless the entries $\#1,\dots,\#4$ of any full column satisfy $(\#3,\#4) \in \llangle \#1,\#2 \rrangle$. We refer to this statement as the full column property.

\bbm\label{Bem-CCG}

By point \rref{i-S-CCG-rc} of Proposition \rref{S-CCG} and \eqref{eq:CompositeCG}, one finds the following expression of the $9\lambda$ symbol in terms of ordinary Clebsch-Gordan coefficients,
\ala{
\bpma
\lambda_1^1 & \lambda_1^2 & \lambda_1\phantom{^1} & k_1
\\
\lambda_2^1 & \lambda_2^2 & \lambda_2 & k_2
\\
\lambda_3^1 & \lambda_3^2 & \lambda_3 & k_3
\\
k^1 & k^2 & k
\epma
= &
\sum_{\ul\mu_1,\ul\mu_2}
\,
\sum_{\check\mu_1=1}^{m_{\lambda_1}(\hat\mu_1)}
\,
\sum_{\check\mu_2=1}^{m_{\lambda_2}(\hat\mu_2)}
\,
\sum_{\check\mu_3^1=1}^{m_{\lambda_3^1}(\hat\mu_3^1)}
\,
\sum_{\check\mu_3^2=1}^{m_{\lambda_3^2}(\hat\mu_3^2)}
\Big(
C
^{\lambda_1^1 \lambda_1^2 \lambda_1 , k_1}
_{\mu_1^1 \ \mu_1^2 \ \mu_1}
\Big)^\ast
\\
& \hspace{0.5cm}
\cdot
\Big(
C
^{\lambda_2^1 \lambda_2^2 \lambda_2 , k_2}
_{\mu_2^1 \ \mu_2^2 \ \mu_2}
\Big)^\ast
\Big(
C
^{\lambda_1 \lambda_2 \lambda_3 , k}
_{\mu_1 \ \mu_2 \ \mu_3}
\Big)^\ast
\, 
C
^{\lambda_1^1 \lambda_2^1 \lambda_3^1 , k^1}
_{\mu_1^1 \ \mu_2^1 \ \mu_3^1}
\,
C
^{\lambda_1^2 \lambda_2^2 \lambda_3^2 , k^2}
_{\mu_1^2 \ \mu_2^2 \ \mu_3^2}
\,
C
^{\lambda_3^1 \lambda_3^2 \lambda_3 , k_3}
_{\mu_3^1 \ \mu_3^2 \ \mu_3}
}
for any $\mu_3 \in \weight(\lambda_3)$. Here, the outer sum runs over $\ul\mu_1 \in \weight(\ul\lambda_1)$ and $\ul\mu_2 \in \weight(\ul\lambda_2)$ such that $\hat\mu_i:=\hat\mu_i^1+\hat\mu_i^2 \in \hat\weight(\lambda_i)$, $i=1,2$ and $\hat\mu_3^n:=\hat\mu_1^n+\hat\mu_2^n \in \hat\weight(\lambda_3^n)$, $n=1,2$, and we have denoted $\mu_i:=(\hat\mu_i,\check\mu_i)$, $i=1,2$, and $\mu_3^n:=(\hat\mu_3^n,\check\mu_3^n)$, $n=1,2$. Clearly, the sum over $\ul\mu_1$, $\ul\mu_2$ may be further restricted to pairs satisfying $\hat\mu_1+\hat\mu_2=\hat\mu_3$. On the other hand, by orthogonality, in the summation, any  relation between the weights may be omitted, so that the sum runs over independent $\ul\mu_i \in \weight(\ul\lambda_i)$, $i=1,2,3$, and $\mu_i \in \weight(\lambda_i)$, $i=1,2$.
\qeb

\ebm

We need the following two special formulae for the recoupling coefficients of tree joins.

\ble\label{L-R}

For $i=1,\dots,4$, let $T_i$ be a coupling tree, $\ul\lambda_i$ a leaf labelling, $\lambda_i \in \langle\ul\lambda_i\rangle$ and $\alpha_i \in \mc L^T(\ul\lambda_i,\lambda_i)$. For $i=1,2$, assume that $T_i$ and $T_{i+2}$ have the same number of leaves and that $\ul\lambda_{i+2}$ can be obtained from $\ul\lambda_i$ by a permutation of entries.
\ben
\sitem\label{i-L-R-2}
For all $\lambda \in \langle \lambda_1,\lambda_2 \rangle \cap \langle \lambda_3,\lambda_4 \rangle$ and $k_{12} = 1 , \dots , m_{(\lambda_1,\lambda_2)}(\lambda)$, $k_{34} = 1 , \dots , m_{(\lambda_3,\lambda_4)}(\lambda)$, we have
$$
R
\big(T_1\!\cdot\! T_2|T_3\!\cdot\! T_4\big)
^{\join{\alpha_1}{\alpha_2}{(\lambda,k_{12})}}
_{\join{\alpha_3}{\alpha_4}{(\lambda,k_{34})}}
=
\delta_{\lambda_1 \lambda_3} 
\,
\delta_{\lambda_2 \lambda_4}
\,
\delta_{k_{12} k_{34}}
\,
R\big(T_1|T_3\big)^{\alpha_1}_{\alpha_3}
\,
R
\big(T_2|T_4\big)^{\alpha_2}_{\alpha_4}
.
$$
\sitem\label{i-L-R-4}
Given, in addition, $\lambda'_i \in \langle \ul\lambda_i \rangle$ and $\alpha_i' \in \mc L^T(\ul\lambda_i,\lambda_i)$, $i=1,\dots,4$, then for all 
\ala{
(\lambda_{12},k_{12}) & \in \llangle \lambda_1,\lambda_2 \rrangle
\,, & 
(\lambda_{34},k_{34}) & \in \llangle \lambda_3,\lambda_4 \rrangle
\,, 
\\
(\lambda_{13},k_{13}) & \in \llangle \lambda'_1,\lambda'_3 \rrangle
\,, &
(\lambda_{24},k_{24}) & \in \llangle \lambda'_2,\lambda'_4 \rrangle
\,,
}
all 
$
\lambda 
\in 
\langle \lambda_{12},\lambda_{34} \rangle
\cap 
\langle \lambda_{13},\lambda_{24} \rangle
$
and all 
$$
k = 1 , \dots , m_{(\lambda_{12},\lambda_{34})}(\lambda)
\,,\qquad
k' = 1 , \dots , m_{(\lambda_{13},\lambda_{24})}(\lambda)
\,,
$$
we have 
\ala{
&
R
\big(
(T_1\!\cdot\! T_2)\!\cdot\!(T_3\!\cdot\! T_4)
\big|
(T_1\!\cdot\! T_3)\!\cdot\!(T_2\!\cdot\! T_4)
\big)
^{
\join{
\join{\alpha_1}{\alpha_2}{(\lambda_{12},k_{12})}
}{
\join{\alpha_3}{\alpha_4}{(\lambda_{34},k_{34})}
}{
(\lambda,k)
}
}
_{
\join{
\join{\alpha_1'}{\alpha_3'}{(\lambda_{13},k_{13})}
}{
\join{\alpha_2'}{\alpha_4'}{(\lambda_{24},k_{24})}
}{
(\lambda,k')
}
}
\\
& \hspace{6cm}
=
\delta_{\alpha_1,\alpha_1'} \cdots \delta_{\alpha_4,\alpha_4'}
\bpma
\lambda_1 & \lambda_2 & \lambda_{12} & k_{12}
\\
\lambda_3 & \lambda_4 & \lambda_{34} & k_{34}
\\
\lambda_{13} & \lambda_{24} & \lambda & k'
\\
k_{13} & k_{24} & k
\epma
.
}
\een
\ele

\bbw

\rref{i-L-R-2}.\abs
In the proof, we use the shorthand notations
$$
T_{ij} = T_i \bcdot T_j
\,,\qquad
T_{ijkl}
= 
(T_i \bcdot T_j) \bcdot (T_k \bcdot T_l)
\,.
$$
Let $\sigma_{13}$ and $\sigma_{24}$ be the permutations turning $\ul\lambda_1$ into $\ul\lambda_3$ and $\ul\lambda_2$ into $\ul\lambda_4$, respectively. They define a permutation $\sigma$ turning $\ul\lambda_1\!\cdot\!\ul\lambda_2$ into $\ul\lambda_3\!\cdot\!\ul\lambda_4$. By point \rref{i-S-CCG-rc} of Proposition \rref{S-CCG}, for any $\mu\in\weight(\lambda)$, 
\beq\label{G-L-R-1}
R(T_{12}|T_{34})
^{\join{\alpha_1}{\alpha_2}{(\lambda,k_{12})}}
_{\join{\alpha_3}{\alpha_4}{(\lambda,k_{34})}}
=
\sum_{\ul\mu_1 \in \weight(\ul\lambda_1) \atop \ul\mu_2 \in \weight(\ul\lambda_2)}
\Big(
\ccg
{T_{12}}
{\join{\alpha_1}{\alpha_2}{(\lambda,k_{12})}}
{\ul\mu_1\!\cdot\!\ul\mu_2}
{\mu}
\Big)^\ast
\,
\ccg
{T_{34}}
{\join{\alpha_3}{\alpha_4}{(\lambda,k_{34})}}
{\sigma(\ul\mu_1\!\cdot\!\ul\mu_2)}
{\mu}
\,.
\eeq
Clearly, $\ul\mu_i$ and $\alpha_i$ are compatible for $i=1,2$ iff $\ul\mu_1 \bcdot \ul\mu_2$ and $\join{\alpha_1}{\alpha_2}{(\lambda,k_{12})}$ are compatible. If this condition is violated, then Proposition \rref{L-CCG} yields that 
$
\ccg
{T_{12}}
{\join{\alpha_1}{\alpha_2}{(\lambda,k_{12})}}
{\ul\mu_1\!\cdot\!\ul\mu_2}
{\mu}
=
0
$.
Hence, for every nonzero contribution to the right hand side of \eqref{G-L-R-1}, compatibility holds true. Then, Proposition \rref{L-CCG} implies that 
$
\ccg{T_{12}}{\join{\alpha_1}{\alpha_2}{(\lambda,k_{12})}}{\ul\mu_1\!\cdot\!\ul\mu_2}{\mu}
$
and 
$
\ccg{T_i}{\alpha_i}{\ul\mu_i}{(\fnlsum{\ul{\hat\mu}_i},\check\mu_i)}
$
for any $\check\mu_i=1,\dots,m_{\lambda_i}(\fnlsum{\ul{\hat\mu}_i})$, $i=1,2$, can be decomposed according to \eqref{eq:CompositeCG}. It follows that
$$
\ccg
{T_{12}}
{\join{\alpha_1}{\alpha_2}{(\lambda,k_{12})}}
{\ul\mu_1\!\cdot\!\ul\mu_2}
{\mu}
=
\sum_{\check\mu_1=1}^{m_{\lambda_1}(\hat\mu_1)}
\,
\sum_{\check\mu_2=1}^{m_{\lambda_2}(\hat\mu_2)}
\,
\ccg{T_1}{\alpha_1}{\ul\mu_1}{\mu_1}
\,
\ccg{T_2}{\alpha_2}{\ul\mu_2}{\mu_2}
\,
C
^{\lambda_1 \lambda_2 \lambda , k_{12}}
_{\mu_1 \, \mu_2 \, \mu}
\,,
$$
where we have denoted $\hat\mu_i:=\fnlsum{\ul{\hat\mu}_i}$ and $\mu_i:=(\hat\mu_i,\check\mu_i)$. Since $\fnlsum{\ul{\hat\mu}_1}=\fnlsum{\sigma_{13}(\ul{\hat\mu}_1)}$ and $\fnlsum{\ul{\hat\mu}_2}=|\fnlsum{\sigma_{24}(\ul{\hat\mu}_2)}$, an analogous argument yields that for every nonzero contribution to the right hand side of \eqref{G-L-R-1}, we have $\hat\mu_1 \in \hat\weight(\lambda_3)$ and $\hat\mu_2 \in \hat\weight(\lambda_4)$, and 
$$
\ccg
{T_{34}}
{\join{\alpha_3}{\alpha_4}{(\lambda,k_{34})}}
{\sigma(\ul\mu_1\!\cdot\!\ul\mu_2)}
{\mu}
=
\sum_{\check\mu_3=1}^{m_{\lambda_3}(\hat\mu_1)}
\,
\sum_{\check\mu_4=1}^{m_{\lambda_4}(\hat\mu_2)}
\,
\ccg{T_3}{\alpha_3}{\sigma_{13}(\ul\mu_1)}{\mu_3}
\,
\ccg{T_4}{\alpha_4}{\sigma_{24}(\ul\mu_2)}{\mu_4}
\,
C
^{\lambda_3 \lambda_4 \lambda,k_{34}}
_{\mu_3 \, \mu_4 \, \mu}
\,,
$$
where we have denoted $\mu_3:=(\hat\mu_1,\check\mu_3)$ and $\mu_4:=(\hat\mu_2,\check\mu_4)$. Grouping the factors and decomposing the sum accordingly, we obtain
\al{\nonumber
&
R(T_1\!\cdot\! T_2|T_3\!\cdot\! T_4)
^{\join{\alpha_1}{\alpha_2}{(\lambda,k_{12})}}
_{\join{\alpha_3}{\alpha_4}{(\lambda,k_{34})}}
\\ \nonumber
& \hspace{0.5cm}
=
\sum_{
\hat\mu_1\in\hat\weight(\lambda_1)\cap\hat\weight(\lambda_3)
\atop
\hat\mu_2\in\hat\weight(\lambda_2)\cap\hat\weight(\lambda_4)
}
\,
\sum_{\check\mu_1=1}^{m_{\lambda_1}(\hat\mu_1)}
\,
\sum_{\check\mu_2=1}^{m_{\lambda_2}(\hat\mu_2)}
\,
\sum_{\check\mu_3=1}^{m_{\lambda_3}(\hat\mu_1)}
\,
\sum_{\check\mu_4=1}^{m_{\lambda_4}(\hat\mu_2)}
\,
\left(
\sum_{\ul\mu_1\in\weight(\lambda_1)}
\big(\ccg{T_1}{\alpha_1}{\ul\mu_1}{\mu_1}\big)^\ast
\,
\ccg{T_3}{\alpha_3}{\sigma_{13}(\ul\mu_1)}{\mu_3}
\right)
\\ \label{G-L-R-CCCC}
& \hspace{1.5cm}
\cdot 
\left(
\sum_{\ul\mu_2\in\weight(\lambda_2)}
\big(\ccg{T_2}{\alpha_2}{\ul\mu_2}{\mu_2}\big)^\ast
\,
\ccg{T_4}{\alpha_4}{\sigma_{24}(\ul\mu_2)}{\mu_4}
\right)
\big(
C
^{\lambda_1 \lambda_2 \lambda , k_{12}}
_{\mu_1 \, \mu_2 \, \mu}
\big)^\ast
C
^{\lambda_3 \lambda_4 \lambda,k_{34}}
_{\mu_3 \, \mu_4 \, \mu}
\,.
}
Using $\ket{\ul\lambda_3,\sigma_{13}(\ul\mu_1)} = \Pi_{13} \ket{\ul\lambda_1,\ul\mu_1}$, where $\Pi_{13}$ is the unitary transformation permuting the tensor factors according to $\sigma_{13}$, and \eqref{G-D-Racah}, we find
\ala{
\sum_{\ul\mu_1\in\weight(\lambda_1)}
\big(\ccg{T_1}{\alpha_1}{\ul\mu_1}{\mu_1}\big)^\ast
\,
\ccg{T_3}{\alpha_3}{\sigma_{13}(\ul\mu_1)}{\mu_3}
& =
\sum_{\ul\mu_1\in\weight(\lambda_1)}
\braket{T_1;\alpha_1,\mu_1}{\ul\lambda_1,\ul\mu_1}
\bra{\ul\lambda_1,\ul\mu_1}
\Pi_{13} 
\ket{T_3;\alpha_3,\mu_1}
\\
& =
\bra{T_1;\alpha_1,\mu_1}\Pi_{13}\ket{T_3;\alpha_3,\mu_3}
\\
& =
\delta_{\lambda_1 \lambda_3} 
\,
\delta_{\check\mu_1 \check\mu_3} 
\,
R\big(T_1|T_3\big)^{\alpha_1}_{\alpha_3}
\,.
}
By analogy, the second expression in parentheses equals 
$
\delta_{\lambda_2 \lambda_4}
\,
\delta_{\check\mu_2 \check\mu_4} 
\,
R\big(T_2|T_4\big)^{\alpha_2}_{\alpha_4}
$. 
Then, by unitarity of the isomorphisms \eqref{G-TePr-Zlg} chosen, the remaining sums yield 
$$
\sum_{\mu_1\in\weight(\lambda_1)}
\,
\sum_{\mu_2\in\weight(\lambda_2)}
\big(
C
^{\lambda_1 \lambda_2 \lambda , k_{12}}
_{\mu_1 \, \mu_2 \, \mu}
\big)^\ast
C
^{\lambda_1 \lambda_2 \lambda,k_{34}}
_{\mu_1 \, \mu_2 \, \mu}
=
\delta_{k_{12} k_{34}}
\,.
$$

\begin{figure}

~\hfill
\begin{tikzpicture}

\btreeset{grow=up, scale=0.8, line width=1.5pt}

\BinaryTree[local bounding box=INIT2,line width=1pt,bottom padding = 10pt]{%
:!
l!l,
l!r,
r!l,
r!r,
l!l,
l!r,
r!l,     
r!r
}{2}

\draw[thick] ($(btree-l-l)+(-0.75,0)$) -- ($(btree-l-l)+(0.75,0)$);
\draw[thick] ($(btree-l-l)+(0.75,0)$) -- ($(btree-l-l)+(0.75,0.75)$);
\draw[thick] ($(btree-l-l)+(0.75,0.75)$) -- ($(btree-l-l)+(-0.75,0.75)$);
\draw[thick] ($(btree-l-l)+(-0.75,0.75)$) -- ($(btree-l-l)+(-0.75,0)$);

\draw[thick] (btree-l-l) node [above]{\small $T_1 \,,\, \alpha'_1$};

\draw[thick] ($(btree-l-r)+(-0.75,0)$) -- ($(btree-l-r)+(0.75,0)$);
\draw[thick] ($(btree-l-r)+(0.75,0)$) -- ($(btree-l-r)+(0.75,0.75)$);
\draw[thick] ($(btree-l-r)+(0.75,0.75)$) -- ($(btree-l-r)+(-0.75,0.75)$);
\draw[thick] ($(btree-l-r)+(-0.75,0.75)$) -- ($(btree-l-r)+(-0.75,0)$);

\draw[thick] (btree-l-r) node [above]{\small $T_2 \,,\, \alpha'_2$};

\draw[thick] ($(btree-r-l)+(-0.75,0)$) -- ($(btree-r-l)+(0.75,0)$);
\draw[thick] ($(btree-r-l)+(0.75,0)$) -- ($(btree-r-l)+(0.75,0.75)$);
\draw[thick] ($(btree-r-l)+(0.75,0.75)$) -- ($(btree-r-l)+(-0.75,0.75)$);
\draw[thick] ($(btree-r-l)+(-0.75,0.75)$) -- ($(btree-r-l)+(-0.75,0)$);

\draw[thick] (btree-r-l) node [above]{\small $T_3 \,,\, \alpha'_3$};

\draw[thick] ($(btree-r-r)+(-0.75,0)$) -- ($(btree-r-r)+(0.75,0)$);
\draw[thick] ($(btree-r-r)+(0.75,0)$) -- ($(btree-r-r)+(0.75,0.75)$);
\draw[thick] ($(btree-r-r)+(0.75,0.75)$) -- ($(btree-r-r)+(-0.75,0.75)$);
\draw[thick] ($(btree-r-r)+(-0.75,0.75)$) -- ($(btree-r-r)+(-0.75,0)$);

\draw[thick] (btree-r-r) node [above]{\small $T_4 \,,\, \alpha'_4$};

\draw[thick] (btree-l)+(0,-0.2) node [left]{\small $(\lambda_{12},k_{12})$};
\draw[thick] (btree-r)+(0,-0.2) node [right]{\small $(\lambda_{34},k_{34})$};
\draw[thick] (btree-root)+(0,-0.15) node [right]{\small $(\lambda,k)$};

\draw[line width=1.5] (btree-root) -- ($(btree-root)+(0,-0.5)$);


\end{tikzpicture}
\hfill
\begin{tikzpicture}

\btreeset{grow=up, scale=0.8, line width=1.5pt}

\BinaryTree[local bounding box=INIT2,line width=1pt,bottom padding = 10pt]{%
:!
l!l,
l!r,
r!l,
r!r,
l!l,
l!r,
r!l,     
r!r
}{2}

\draw[thick] ($(btree-l-l)+(-0.75,0)$) -- ($(btree-l-l)+(0.75,0)$);
\draw[thick] ($(btree-l-l)+(0.75,0)$) -- ($(btree-l-l)+(0.75,0.75)$);
\draw[thick] ($(btree-l-l)+(0.75,0.75)$) -- ($(btree-l-l)+(-0.75,0.75)$);
\draw[thick] ($(btree-l-l)+(-0.75,0.75)$) -- ($(btree-l-l)+(-0.75,0)$);

\draw[thick] (btree-l-l) node [above]{\small $T_1 \,,\, \alpha'_1$};

\draw[thick] ($(btree-l-r)+(-0.75,0)$) -- ($(btree-l-r)+(0.75,0)$);
\draw[thick] ($(btree-l-r)+(0.75,0)$) -- ($(btree-l-r)+(0.75,0.75)$);
\draw[thick] ($(btree-l-r)+(0.75,0.75)$) -- ($(btree-l-r)+(-0.75,0.75)$);
\draw[thick] ($(btree-l-r)+(-0.75,0.75)$) -- ($(btree-l-r)+(-0.75,0)$);

\draw[thick] (btree-l-r) node [above]{\small $T_3 \,,\, \alpha'_3$};

\draw[thick] ($(btree-r-l)+(-0.75,0)$) -- ($(btree-r-l)+(0.75,0)$);
\draw[thick] ($(btree-r-l)+(0.75,0)$) -- ($(btree-r-l)+(0.75,0.75)$);
\draw[thick] ($(btree-r-l)+(0.75,0.75)$) -- ($(btree-r-l)+(-0.75,0.75)$);
\draw[thick] ($(btree-r-l)+(-0.75,0.75)$) -- ($(btree-r-l)+(-0.75,0)$);

\draw[thick] (btree-r-l) node [above]{\small $T_2 \,,\, \alpha'_2$};

\draw[thick] ($(btree-r-r)+(-0.75,0)$) -- ($(btree-r-r)+(0.75,0)$);
\draw[thick] ($(btree-r-r)+(0.75,0)$) -- ($(btree-r-r)+(0.75,0.75)$);
\draw[thick] ($(btree-r-r)+(0.75,0.75)$) -- ($(btree-r-r)+(-0.75,0.75)$);
\draw[thick] ($(btree-r-r)+(-0.75,0.75)$) -- ($(btree-r-r)+(-0.75,0)$);

\draw[thick] (btree-r-r) node [above]{\small $T_4 \,,\, \alpha'_4$};

\draw[thick] (btree-l)+(0,-0.2) node [left]{\small $(\lambda_{13},k_{13})$};
\draw[thick] (btree-r)+(0,-0.2) node [right]{\small $(\lambda_{24},k_{24})$};
\draw[thick] (btree-root)+(0,-0.15) node [right]{\small $(\lambda,k')$};

\draw[line width=1.5] (btree-root) -- ($(btree-root)+(0,-0.5)$);

\end{tikzpicture}
\hfill~

\caption{\label{Abb-L-R} The coupling trees $(T_1 \!\cdot\! T_2) \!\cdot\! (T_3 \!\cdot\! T_4)$ and $(T_1 \!\cdot\! T_2) \!\cdot\! (T_3 \!\cdot\! T_4)$ of point \rref{i-L-R-4} of Lemma \rref{L-R}. The root labels $\alpha_i^{\rt_i}$ and $\alpha_i'^{\rt_i}$ of $T_i$ contain the weight labels $\lambda_i$ and $\lambda'_i$, respectively.}

\end{figure}
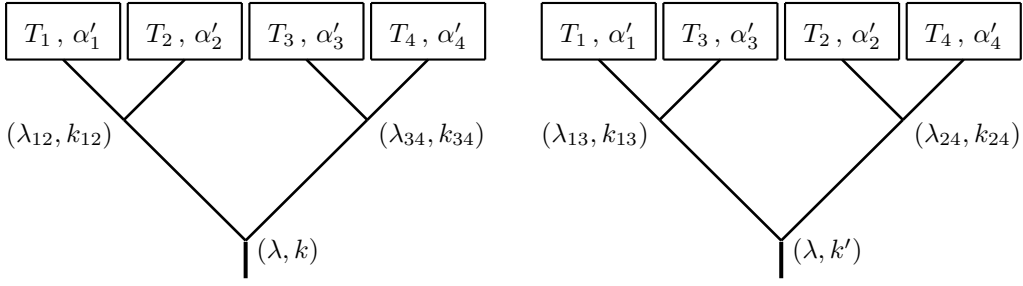

\rref{i-L-R-4}.\abs We proceed by analogy. First, we use point \rref{i-S-CCG-rc} of Proposition \rref{S-CCG} to rewrite the recoupling coefficient under consideration as 
\al{\nonumber
&
\sum_{\ul\mu_1\in\weight(\ul\lambda_1)}
\cdots
\sum_{\ul\mu_4\in\weight(\ul\lambda_4)}
\Big(
\ccg
{T_{1234}}
{
\join{
\join{\alpha_1}{\alpha_2}{(\lambda_{12},k_{12})}
}{
\join{\alpha_3}{\alpha_4}{(\lambda_{34},k_{34})}
}{(\lambda,k)}
}
{(\ul\mu_1 \!\cdot\! \ul\mu_2) \!\cdot\! (\ul\mu_3 \!\cdot\! \ul\mu_4)}
{\mu}
\Big)^\ast
\\ \label{G-L-R-4-CC}
& \hspace{4.5cm}
\cdot\,
\ccg
{T_{1324}}
{
\join{
\join{\alpha_1'}{\alpha_3'}{(\lambda_{13},k_{13})}
}{
\join{\alpha_2'}{\alpha_4'}{(\lambda_{24},k_{24})}
}{
(\lambda,k')
}
}
{(\ul\mu_1 \!\cdot\! \ul\mu_3) \!\cdot\! (\ul\mu_2 \!\cdot\! \ul\mu_4)}
{\mu}
}
for some chosen $\mu \in \weight(\lambda)$. For $i=1,2\dots,4$, put $\hat\mu_i:=\fnlsum{\ul{\hat\mu}_i}$. Then, we define cherry labellings $\alpha_{12}$, $\alpha_{34}$, $\alpha$ by, respectively,
$$
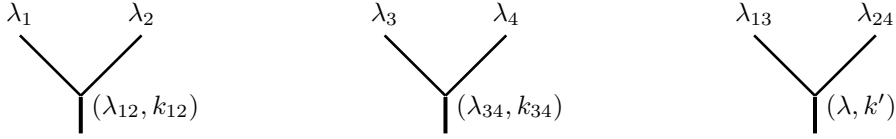
\begin{tikzpicture}
\btreeset{grow=up, scale=0.4, line width=1.5pt}
\BinaryTree[local bounding box=INIT2,line width=1pt,bottom padding = 10pt]{%
:!l,r}{1}
\draw[thick] (btree-l) node [above]{\small $\lambda_1$};
\draw[thick] (btree-r) node [above]{\small $\lambda_2$};
\draw[line width=1.5] (btree-root) -- ($(btree-root)+(0,-0.5)$);
\draw[thick] (btree-root)+(0,-0.15) node [right]{\small $(\lambda_{12},k_{12})$};

\end{tikzpicture}
\hspace{2cm}
\begin{tikzpicture}

\btreeset{grow=up, scale=0.4, line width=1.5pt}
\BinaryTree[local bounding box=INIT2,line width=1pt,bottom padding = 10pt]{%
:!l,r}{1}
\draw[thick] (btree-l) node [above]{\small $\lambda_3$};
\draw[thick] (btree-r) node [above]{\small $\lambda_4$};
\draw[line width=1.5] (btree-root) -- ($(btree-root)+(0,-0.5)$);
\draw[thick] (btree-root)+(0,-0.15) node [right]{\small $(\lambda_{34},k_{34})$};

\end{tikzpicture}
\hspace{2cm}
\begin{tikzpicture}

\btreeset{grow=up, scale=0.4, line width=1.5pt}
\BinaryTree[local bounding box=INIT2,line width=1pt,bottom padding = 10pt]{%
:!l,r}{1}
\draw[thick] (btree-l) node [above]{\small $\lambda_{13}$};
\draw[thick] (btree-r) node [above]{\small $\lambda_{24}$};
\draw[line width=1.5] (btree-root) -- ($(btree-root)+(0,-0.5)$);
\draw[thick] (btree-root)+(0,-0.15) node [right]{\small $(\lambda,k')$};

\end{tikzpicture}
$$
and the multiplicity pair $\ul k := (k_{13},k_{24})$, check compatibility and use \eqref{eq:CompositeCG} to factorize 
\ala{
&
\ccg
{T_{1234}}
{
\join{
\join{\alpha_1}{\alpha_2}{(\lambda_{12},k_{12})}
}{
\join{\alpha_3}{\alpha_4}{(\lambda_{34},k_{34})}
}{
(\lambda,k)
}
}
{(\ul\mu_1 \!\cdot\! \ul\mu_2) \!\cdot\! (\ul\mu_3 \!\cdot\! \ul\mu_4)}
{\mu}
\\
& \hspace{5cm}
=
\sum_{\check\mu_1=1}^{m_{\lambda_1}(\hat\mu_1)}
\cdots
\sum_{\check\mu_4=1}^{m_{\lambda_4}(\hat\mu_4)}
\left(
\prod_{i=1}^4
\ccg{T_i}{\alpha_i}{\ul\mu_i}{\mu_i}
\right)
\,
\ccg
{\boldsymbol\vee\!\cdot\!\boldsymbol\vee}
{\join{\alpha_{12}}{\alpha_{34}}{(\lambda,k)}}
{\ul\mu_{12} \bcdot \ul\mu_{34}}
{\mu}
\,,
\\
&
\ccg
{T_{1324}}
{
\join{
\join{\alpha_1'}{\alpha_3'}{(\lambda_{13},k_{13})}
}{
\join{\alpha_2'}{\alpha_4'}{(\lambda_{24},k_{24})}
}{
(\lambda,k')
}
}
{(\ul\mu_1 \!\cdot\! \ul\mu_3) \!\cdot\! (\ul\mu_2 \!\cdot\! \ul\mu_4)}
{\mu}
\\
& \hspace{5cm}
=
\sum_{\check\mu_1'=1}^{m_{\lambda_1}(\hat\mu_1)}
\cdots
\sum_{\check\mu_4'=1}^{m_{\lambda_4}(\hat\mu_4)}
\left(
\prod_{i=1}^4
\ccg{T_i}{\alpha_i'}{\ul\mu_i}{\mu_i'}
\right)
\ccg
{\boldsymbol\vee^\vee}
{\ld{\alpha_{12}}{\alpha_{34}}{\alpha,\ul k}}
{\ul\mu_{12}' \!\ast\! \ul\mu_{34}'}
{\mu}
\,,
}
see Figure \rref{Abb-L-R}. Here, we have denoted $\mu_i:=(\hat\mu_i,\check\mu_i)$, $\mu_i':=(\hat\mu_i,\check\mu_i')$ and $\ul{\mu}_{ij} := (\mu_i,\mu_j)$, $\ul{\mu}_{ij}' := (\mu_i',\mu_j')$, $i,j=1,\dots,4$. By grouping the factors and decomposing the sum, we find that \eqref{G-L-R-4-CC} equals 
\al{\nonumber
&
\sum_{\hat\mu_1\in\hat\weight(\lambda_1)}
\sum_{\check\mu_1,\check\mu_1'=1}^{m_{\lambda_1}(\hat\mu_1)}
\cdots
\sum_{\hat\mu_4\in\hat\weight(\lambda_4)}
\sum_{\check\mu_4,\check\mu_4'=1}^{m_{\lambda_4}(\hat\mu_4)}
\left(
\prod_{i=1}^4 
\left(
\sum_{\ul\mu_i\in\weight(\ul\lambda_i)} 
\Big(\ccg{T_i}{\alpha_i}{\ul\mu_i}{\mu_i}\Big)^\ast
\,
\ccg{T_i}{\alpha_i'}{\ul\mu_i}{\mu_i'}
\right)
\right)
\\ \label{G-L-R-3}
& \hspace{5.75cm}
\cdot
\Big(
\ccg
{\boldsymbol\vee\!\cdot\!\boldsymbol\vee}
{\join{\alpha_{12}}{\alpha_{34}}{(\lambda,k)}}
{\ul\mu_{12} \bcdot \ul\mu_{34}}
{\mu}
\Big)^\ast
\,
\ccg
{\boldsymbol\vee^\vee}
{\ld{\alpha_{12}}{\alpha_{34}}{\alpha,\ul k}}
{\ul\mu_{12}' \!\ast\! \ul\mu_{34}'}
{\mu}
\,.
}
Since 
$$
\sum_{\ul\mu_i\in\weight(\ul\lambda_i)} 
\Big(\ccg{T_i}{\alpha_i}{\ul\mu_i}{\mu_i}\Big)^\ast
\,
\ccg{T_i}{\alpha_i'}{\ul\mu_i}{\mu_i'}
=
\sum_{\ul\mu_i} 
\braket{T_i;\alpha_i,\mu_i}{\ul\lambda_i \ul\mu_i}
\braket{\ul\lambda_i \ul\mu_i}{T_i;\alpha_i',\mu_i'}
=
\delta_{\alpha_i,\alpha_i'}
\,
\delta_{\check\mu_i,\check\mu_i'}
\,,
$$
\eqref{G-L-R-3} boils down to
\ala{
&
\left(
\prod_{i=1}^4 
\delta_{\alpha_i,\alpha_i'}
\right)
\sum_{\mu_1\in\weight(\lambda_1)}
\cdots
\sum_{\mu_4\in\weight(\lambda_4)}
\Big(
\ccg
{\boldsymbol\vee\!\cdot\!\boldsymbol\vee}
{\join{\alpha_{12}}{\alpha_{34}}{(\lambda,k)}}
{\ul\mu_{12} \bcdot \ul\mu_{34}}
{\mu}
\Big)^\ast
\,
\ccg
{\boldsymbol\vee^\vee}
{\ld{\alpha_{12}}{\alpha_{34}}{\alpha,\ul k}}
{\ul\mu_{12} \!\ast\! \ul\mu_{34}}
{\mu}
\\
& \hspace{9cm}
=
\left(
\prod_{i=1}^4 
\delta_{\alpha_i,\alpha_i'}
\right)
\rcneu{\boldsymbol\vee}{\alpha_{12}}{\alpha_{34}}{\alpha}{\ul k}{k}
\,,
}
where we have used by point \rref{i-S-CCG-rc} of Proposition \rref{S-CCG}.
\ebw

\btm\label{T-R}

Let $T$ be a coupling tree. For $i=1,2,3$, let $\alpha_i$ be a labelling  of $T$ with underlying highest weight labelling $x \mapsto \lambda_i^x$ and underlying multiplicity counter labelling $x \mapsto k_i^x$. Let $\ul k$ be an assignment of a positive integer to every leaf of $T$ and let $k$ be a positive integer. Then,
$$
\rcneu{T}{\alpha_1}{\alpha_2}{\alpha_3}{\ul k}{k}
=
\sum
\prod_{x \text{ node of } T}
\bpma
\lambda_1^{x'} & \lambda_1^{x''} & \lambda_1^x & k_1^x
\\
\lambda_2^{x'} & \lambda_2^{x''} & \lambda_2^x & k_2^x
\\
\lambda_3^{x'} & \lambda_3^{x''} & \lambda_3^x & k_3^x
\\
k^{x'} & k^{x''} & k^x
\epma
\,,
$$
where $x'$ and $x''$ denote the child vertices of $x$ and $k^x=k$ in case $x$ is the root of $T$. The sum is over all assignments of an integer $k^x = 1,\dots,m_{(\lambda_1^x,\lambda_2^x)}(\lambda_3^x)$ to every internal node $x$ of $T$.

\etm

In each of the multiplicity counter assignments, the sum runs through complements $\ul k$ and $k$ to an assignment of a multiplicity counter to every vertex of $T$, and so in the formula given, $k^x$ denotes the multiplicity counter assigned to vertex $x$ irrespective of whether $x$ is a leaf (where $k^x$ is prescribed by $\ul k$), an internal node (where $k^x$ is a summation variable) or the root (where $k^x=k$).

\bbw

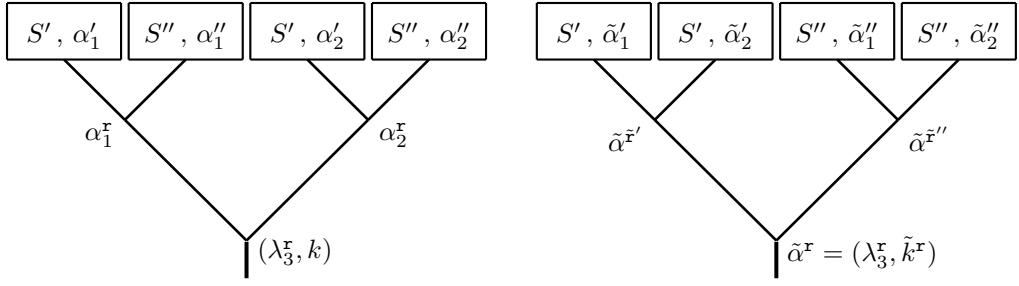
\begin{figure}

~\hfill
\begin{tikzpicture}

\btreeset{grow=up, scale=0.8, line width=1.5pt}

\BinaryTree[local bounding box=INIT2,line width=1pt,bottom padding = 10pt]{%
:!
l!l,
l!r,
r!l,
r!r,
l!l,
l!r,
r!l,     
r!r
}{2}

\draw[thick] ($(btree-l-l)+(-0.75,0)$) -- ($(btree-l-l)+(0.75,0)$);
\draw[thick] ($(btree-l-l)+(0.75,0)$) -- ($(btree-l-l)+(0.75,0.75)$);
\draw[thick] ($(btree-l-l)+(0.75,0.75)$) -- ($(btree-l-l)+(-0.75,0.75)$);
\draw[thick] ($(btree-l-l)+(-0.75,0.75)$) -- ($(btree-l-l)+(-0.75,0)$);

\draw[thick] (btree-l-l) node [above]{\small $S' \,,\, \alpha'_1$};

\draw[thick] ($(btree-l-r)+(-0.75,0)$) -- ($(btree-l-r)+(0.75,0)$);
\draw[thick] ($(btree-l-r)+(0.75,0)$) -- ($(btree-l-r)+(0.75,0.75)$);
\draw[thick] ($(btree-l-r)+(0.75,0.75)$) -- ($(btree-l-r)+(-0.75,0.75)$);
\draw[thick] ($(btree-l-r)+(-0.75,0.75)$) -- ($(btree-l-r)+(-0.75,0)$);

\draw[thick] (btree-l-r) node [above]{\small $S'' \,,\, \alpha''_1$};

\draw[thick] ($(btree-r-l)+(-0.75,0)$) -- ($(btree-r-l)+(0.75,0)$);
\draw[thick] ($(btree-r-l)+(0.75,0)$) -- ($(btree-r-l)+(0.75,0.75)$);
\draw[thick] ($(btree-r-l)+(0.75,0.75)$) -- ($(btree-r-l)+(-0.75,0.75)$);
\draw[thick] ($(btree-r-l)+(-0.75,0.75)$) -- ($(btree-r-l)+(-0.75,0)$);

\draw[thick] (btree-r-l) node [above]{\small $S' \,,\, \alpha'_2$};

\draw[thick] ($(btree-r-r)+(-0.75,0)$) -- ($(btree-r-r)+(0.75,0)$);
\draw[thick] ($(btree-r-r)+(0.75,0)$) -- ($(btree-r-r)+(0.75,0.75)$);
\draw[thick] ($(btree-r-r)+(0.75,0.75)$) -- ($(btree-r-r)+(-0.75,0.75)$);
\draw[thick] ($(btree-r-r)+(-0.75,0.75)$) -- ($(btree-r-r)+(-0.75,0)$);

\draw[thick] (btree-r-r) node [above]{\small $S'' \,,\, \alpha''_2$};

\draw[thick] (btree-l)+(0,-0.2) node [left]{\small $\alpha_1^\rt$};
\draw[thick] (btree-r)+(0,-0.2) node [right]{\small $\alpha_2^\rt$};
\draw[thick] (btree-root)+(0,-0.15) node [right]{\small $(\lambda_3^\rt,k)$};

\draw[line width=1.5] (btree-root) -- ($(btree-root)+(0,-0.5)$);


\end{tikzpicture}
\hfill
\begin{tikzpicture}

\btreeset{grow=up, scale=0.8, line width=1.5pt}

\BinaryTree[local bounding box=INIT2,line width=1pt,bottom padding = 10pt]{%
:!
l!l,
l!r,
r!l,
r!r,
l!l,
l!r,
r!l,     
r!r
}{2}

\draw[thick] ($(btree-l-l)+(-0.75,0)$) -- ($(btree-l-l)+(0.75,0)$);
\draw[thick] ($(btree-l-l)+(0.75,0)$) -- ($(btree-l-l)+(0.75,0.75)$);
\draw[thick] ($(btree-l-l)+(0.75,0.75)$) -- ($(btree-l-l)+(-0.75,0.75)$);
\draw[thick] ($(btree-l-l)+(-0.75,0.75)$) -- ($(btree-l-l)+(-0.75,0)$);

\draw[thick] (btree-l-l) node [above]{\small $S' \,,\, \tilde\alpha'_1$};

\draw[thick] ($(btree-l-r)+(-0.75,0)$) -- ($(btree-l-r)+(0.75,0)$);
\draw[thick] ($(btree-l-r)+(0.75,0)$) -- ($(btree-l-r)+(0.75,0.75)$);
\draw[thick] ($(btree-l-r)+(0.75,0.75)$) -- ($(btree-l-r)+(-0.75,0.75)$);
\draw[thick] ($(btree-l-r)+(-0.75,0.75)$) -- ($(btree-l-r)+(-0.75,0)$);

\draw[thick] (btree-l-r) node [above]{\small $S' \,,\, \tilde\alpha'_2$};

\draw[thick] ($(btree-r-l)+(-0.75,0)$) -- ($(btree-r-l)+(0.75,0)$);
\draw[thick] ($(btree-r-l)+(0.75,0)$) -- ($(btree-r-l)+(0.75,0.75)$);
\draw[thick] ($(btree-r-l)+(0.75,0.75)$) -- ($(btree-r-l)+(-0.75,0.75)$);
\draw[thick] ($(btree-r-l)+(-0.75,0.75)$) -- ($(btree-r-l)+(-0.75,0)$);

\draw[thick] (btree-r-l) node [above]{\small $S'' \,,\, \tilde\alpha''_1$};

\draw[thick] ($(btree-r-r)+(-0.75,0)$) -- ($(btree-r-r)+(0.75,0)$);
\draw[thick] ($(btree-r-r)+(0.75,0)$) -- ($(btree-r-r)+(0.75,0.75)$);
\draw[thick] ($(btree-r-r)+(0.75,0.75)$) -- ($(btree-r-r)+(-0.75,0.75)$);
\draw[thick] ($(btree-r-r)+(-0.75,0.75)$) -- ($(btree-r-r)+(-0.75,0)$);

\draw[thick] (btree-r-r) node [above]{\small $S'' \,,\, \tilde\alpha''_2$};

\draw[thick] (btree-l)+(0,-0.25) node [left]{\small $\tilde\alpha^{\tilde \rt'}$};
\draw[thick] (btree-r)+(0,-0.25) node [right]{\small $\tilde\alpha^{\tilde \rt''}$};
\draw[thick] (btree-root)+(0,-0.15) node [right]{\small $\tilde\alpha^\rt=(\lambda_3^\rt,\tilde k^\rt)$};

\draw[line width=1.5] (btree-root) -- ($(btree-root)+(0,-0.5)$);

\end{tikzpicture}
\hfill~

\caption{\label{Abb-TT-Tt} The coupling trees $T \!\cdot\! T$ with labelling $\join{\alpha_1}{\alpha_2}{(\lambda_3^\rt,k)}$ (left) and $\tilde T$ with labelling $\tilde\alpha$ (right) used in the proof of Theorem \rref{T-R}.}

\end{figure}

Let $\rt$ denote the root of $T$. By definition of $\rcneu{T}{\alpha_1}{\alpha_2}{\alpha_3}{\ul k}{k}$, we have to show that the right hand side of the asserted formula vanishes unless 
\beq\label{G-T-R-Bed}
(\alpha_3,\ul k,k) \in \llangle \alpha_1,\alpha_2 \rrangle
\eeq
and that, in that case, it coincides with 
$
R\big(T \!\cdot\! T|T^\vee\big)
^{\join{\alpha_1}{\alpha_2}{(\lambda_3^\rt,k)}}
_{\ld{\alpha_1}{\alpha_2}{\alpha_3,\ul k}}
$.
The first statement follows from the observation that, otherwise, we have either $(\lambda_3^\rt,k) \notin \llangle \lambda_2^\rt,\lambda_1^\rt \rrangle$ or there is a leaf $y$ of $T$ such that $(\lambda_3^y,k^y) \notin \llangle \lambda_1^y,\lambda_2^y \rrangle$. In the first case, the factor contributed by $\rt$ to any of the summands on the right hand side is a $9\lambda$ symbol with a column having the entries $\lambda_1^\rt$, $\lambda_2^\rt$, $\lambda_3^\rt$, $k$. Therefore, this factor vanishes by the full column property. In the second case, the factor contributed by the parent of $y$ to any of the summands on the right hand side vanishes by the full column property as well.

Thus, in what follows, we may assume that \eqref{G-T-R-Bed} holds. The main step in the proof will consist in splitting 
$
R\big(T \!\cdot\! T|T^\vee\big)
^{\join{\alpha_1}{\alpha_2}{(\lambda_3^\rt,k)}}
_{\ld{\alpha_1}{\alpha_2}{\alpha_3,\ul k}}
$
into a product of analogous coefficients with $T$ replaced by 
subtrees as follows. The child nodes $\rt'$ and $\rt''$ of $\rt$ are the roots of subtrees $S'$ and $S''$, respectively. By restriction, for $i=1,2,3$, $\alpha_i$ induces labellings $\alpha_i'$ of $S'$ and $\alpha_i''$ of $S''$, and $\ul\lambda_i$ splits into leaf labellings $\ul\lambda_i'$ of $S'$ and $\ul\lambda_i''$ of $S''$. Accordingly, $\ul k$ splits into $\ul k'$ and $\ul k''$. We have
$$
T = S' \bcdot S''
\,,\qquad
\alpha_i = \join{\alpha_i'}{\alpha_i''}{\alpha_i^\rt}
\,,\qquad
i=1,2,3
\,.
$$
Consequently,
\beq\label{G-T-R-TT}
T \bcdot T = (S' \bcdot S'') \bcdot (S' \bcdot S'')
\,,\quad
\join{\alpha_1}{\alpha_2}{(\lambda_3^\rt,k)}
=
\join
{\join{\alpha_1'}{\alpha_1''}{\alpha_1^\rt}}
{\join{\alpha_2'}{\alpha_2''}{\alpha_2^\rt}}
{(\lambda_3^\rt,k)}
\,,
\eeq
see Figure \rref{Abb-TT-Tt}, and 
\beq\label{G-T-R-Tv}
T^\vee = S'^\vee \bcdot S''^\vee
\,,\qquad
\ld{\alpha_1}{\alpha_2}{\alpha_3,\ul k}
=
\join
{\ld{\alpha_1'}{\alpha_2'}{\alpha_3',\ul k'}}
{\ld{\alpha_1''}{\alpha_2''}{\alpha_3'',\ul k''}}
{\alpha_3^\rt}
\,.
\eeq
Consider the intermediate rooted binary tree
\beq\label{G-T-R-Tt}
\tilde T := (S' \bcdot S') \bcdot (S'' \bcdot S'')
\eeq
with leaf labelling $\tilde{\ul\lambda} := (\ul\lambda_1'\bcdot \ul\lambda_2') \bcdot (\ul\lambda_1'' \bcdot \ul\lambda_2'')$. By \eqref{G-Racah-Zykl}, we have the decomposition
\beq\label{G-T-R-RR}
R\big(T \bcdot T|T^\vee\big)
^{\join{\alpha_1}{\alpha_2}{(\lambda_3^\rt,k)}}
_{\ld{\alpha_1}{\alpha_2}{\alpha_3,\ul k}}
=
\sum_{\tilde\alpha \in \mc L^{\tilde T}(\tilde{\ul\lambda},\lambda_3^\rt)}
R(T\!\cdot\!T|\tilde T)
^{\join{\alpha_1}{\alpha_2}{(\lambda_3^\rt,k)}}
_{\tilde\alpha}
\,\,
R(\tilde T|T^\vee)
^{\tilde\alpha}
_{\ld{\alpha_1}{\alpha_2}{\alpha_3,\ul k}}
\,.
\eeq
According to the decomposition \eqref{G-T-R-Tt}, the labelling $\tilde\alpha$ of $\tilde T$ induces labellings $\tilde\alpha_i'$ of the $i$-th copy of $S'$ and $\tilde\alpha_i''$ of the $i$-th copy of $S''$. Denoting the roots of $S' \!\cdot\! S'$ and $S'' \!\cdot\! S''$ by $\tilde \rt'$ and $\tilde \rt''$, respectively, we can write
\beq\label{G-tl}
\tilde\alpha
= 
\join
{\join{\tilde\alpha_1'}{\tilde\alpha_2'}{\tilde\alpha^{\tilde \rt'}}}
{\join{\tilde\alpha_1''}{\tilde\alpha_2''}{\tilde\alpha^{\tilde \rt''}}}
{(\lambda_3^\rt,\tilde k^\rt)}
\,,
\eeq
where $\tilde k^\rt$ is the multiplicity counter assigned by $\tilde\alpha$ to $\rt$. See Figure \rref{Abb-TT-Tt}. In view of \eqref{G-tl} and \eqref{G-T-R-TT}, we can apply point \rref{i-L-R-4} of Lemma \rref{L-R} by identifying $T_1$ with the first copy of $S'$, $T_2$ with the first copy of $S''$, $T_3$ with the second copy of $S'$, and $T_4$ with the second copy of $S''$ (as shown in Figures \rref{Abb-L-R} and \rref{Abb-TT-Tt}) to get
\beq\label{G-TT-Tt}
R(T\!\cdot\!T|\tilde T)
^{\join{\alpha_1}{\alpha_2}{(\lambda_3^\rt,k)}}
_{\tilde\alpha}
=
\left(
\prod_{i=1,2}
\delta_{\alpha_i',\tilde\alpha_i'}
\,
\delta_{\alpha_i'',\tilde\alpha_i''}
\right)
\bpma
\lambda_1^{\rt'} & \lambda_1^{\rt''} & \lambda_1^\rt & k_1^\rt
\\
\lambda_2^{\rt'} & \lambda_2^{\rt''} & \lambda_2^\rt & k_2^\rt
\\
\tilde\lambda^{\tilde \rt'} & \tilde\lambda^{\tilde \rt''} & \lambda_3^\rt & \tilde k^\rt
\\
\tilde k^{\tilde \rt'} & \tilde k^{\tilde \rt''} & k
\epma
.
\eeq
Here, we have written out $\tilde\alpha^{\tilde \rt'}$ as $(\tilde\lambda^{\tilde \rt'},\tilde k^{\tilde \rt'})$ and $\tilde\alpha^{\tilde \rt''}$ as $(\tilde\lambda^{\tilde \rt''},\tilde k^{\tilde \rt''})$. In view of \eqref{G-tl} and \eqref{G-T-R-Tv}, we can apply point \rref{i-L-R-2} of Lemma \rref{L-R} to $T_1 = S' \!\cdot\! S'$, $T_2 = S'' \!\cdot\! S''$, $T_3=S'^\vee$ and $T_4=S''^\vee$ to get 
\al{\nonumber
&
R(\tilde T|T^\vee)
^{\tilde\alpha}
_{\ld{\alpha_1}{\alpha_2}{\alpha_3,\ul k}}
\\ \label{G-TT-Tt-2}
& \hspace{1cm}
=
\delta_{\tilde\lambda^{\tilde \rt'} \lambda_3^{\rt'}}
\,
\delta_{\tilde\lambda^{\tilde \rt''} \lambda_3^{\rt''}}
\,
\delta_{\tilde k^\rt k_3^\rt}
\,
R(S' \bcdot S'|S'^\vee)
^{\join{\tilde\alpha_1'}{\tilde\alpha_2'}{\tilde\alpha^{\tilde \rt'}}}
_{\ld{\alpha_1'}{\alpha_2'}{\alpha_3',\ul k'}}
\,
R(S'' \bcdot S''|S''^\vee)
^{\join{\tilde\alpha_1''}{\tilde\alpha_2''}{\tilde\alpha^{\tilde \rt''}}}
_{\ld{\alpha_1''}{\alpha_2''}{\alpha_3'',\ul k''}}
\,.
}
Here, we have used that $(\lambda_3')^{\rt'} = \lambda_3^{\rt'}$ and $(\lambda_3'')^{\rt''} = \lambda_3^{\rt''}$. Substituting \eqref{G-TT-Tt} and \eqref{G-TT-Tt-2} into \eqref{G-T-R-RR}, renaming $\kappa^{\rt'} = \tilde k^{\tilde \rt'}$, $\kappa^{\rt''} = \tilde k^{\tilde \rt''}$ and taking the sum over $\tilde\alpha$, we obtain
\ala{
R\big(T \!\cdot\! T|T^\vee\big)
^{\join{\alpha_1}{\alpha_2}{(\lambda_3^\rt,k)}}
_{\ld{\alpha_1}{\alpha_2}{\alpha_3,\ul k}}
& =
\sum
_{\kappa^{\rt'}=1}
^{m_{(\lambda_1^{\rt'},\lambda_2^{\rt'})}(\lambda_3^{\rt'})}
\,
\sum
_{\kappa^{\rt''}=1}
^{m_{\lambda_1^{\rt''},\lambda_2^{\rt''}}(\lambda_3^{\rt''})}
\bpma
\lambda_1^{\rt'} & \lambda_1^{\rt''} & \lambda_1^\rt & k_1^\rt
\\
\lambda_2^{\rt'} & \lambda_2^{\rt''} & \lambda_2^\rt & k_2^\rt
\\
\lambda_3^{\rt'} & \lambda_3^{\rt''} & \lambda_3^\rt & k_3^\rt
\\
\kappa^{\rt'} & \kappa^{\rt''} & k
\epma
\\
& \hspace{1cm}
\cdot
R(S' \bcdot S'|S'^\vee)
^{\join{\alpha_1'}{\alpha_2'}{(\lambda_3^{\rt'},k^{\rt'})}}
_{\ld{\alpha_1'}{\alpha_2'}{\alpha_3',\ul k'}}
\,
R(S'' \bcdot S''|S''^\vee)
^{\join{\tilde\alpha_1''}{\tilde\alpha_2''}{(\lambda_3^{\rt''},k^{\rt''})}}
_{\ld{\alpha_1''}{\alpha_2''}{\alpha_3'',\ul k''}}
\,.
}
Now, we can iterate this formula by replacing $T$ by $S'$, $S''$, thereby replacing successively $\rt$ by the root $x$ of a subtree and $k$ by $\kappa^x$, until we arrive at the situation where one of the child nodes of $x$, $x'$ say, is a leaf of $T$. In that case, the subtree $S'$ with root $x'$ is the trivial coupling tree consisting of $x'$ alone, so that both $S' \bcdot S'$ and $S'^\vee$ are cherries, and we have $\alpha_i' = \lambda_i^{x'}$ and $\ul k' = k^{x'}$, so that 
$\join{\alpha_1'}{\alpha_2'}{(\lambda_3^{x'},\kappa^{x'})}$ and $\ld{\alpha_1'}{\alpha_2'}{\alpha_3',\ul k'}$ differ only in their multiplicity counters $\kappa^{x'}$, $k^{x'}$. Thus, the recoupling coefficient is in fact a coupling coefficient, so that unitarity of the representation isomorphisms \eqref{G-TePr-Zlg} implies
$$
R(S' \bcdot S'|S'^\vee)
^{\join{\alpha_1'}{\alpha_2'}{(\lambda_3^{x'},k^{x'})}}
_{\ld{\alpha_1'}{\alpha_2'}{\alpha_3',\ul k'}}
=
\delta_{\kappa^{x'} k^{x'}}
\,.
$$
Summation over $\kappa^{x'}$ then yields that in the $9\lambda$ symbols, $\kappa^{x'}$ gets replaced by $k^{x'}$, with the remaining summation being over those $\kappa^x$ where $x$ is an internal node. This completes the proof. 
\ebw

\bbs[Structure constants for standard coupling]

\begin{figure}

\begin{center}

\begin{tikzpicture}

\btreeset{grow=up, scale=0.65, line width=1.3pt}

\BinaryTree[local bounding box=INIT2,line width=1.3pt,bottom padding = 10pt]{%
:
!l!l!l!l!l!l,
l!l!l!l!l!r,
l!l!l!r!r!r,
l!l!r!r!r!r,
r!r!r!r!r!r,
r!r!l!r!r!r,
r!r!l!l!l!r,
r!r!l!l!l!l
}{6}

\draw[thick] (btree-l-l-l-l-l-l) node [above]{\small \,$\lambda_1^1$};
\draw[thick] (btree-l-l-l-l-l-r) node [above]{\small \,$\lambda_1^2$};
\draw[thick] (btree-l-l-l-l-l-r) node [right]{\hspace{0.15cm}$\cdots$};
\draw[thick] (btree-l-l-l-r-r-r) node [above]{\small $\lambda_1^{N-1}$};
\draw[thick] (btree-l-l-r-r-r-r) node [above]{\small $\lambda_1^N$};

\draw[thick] (btree-l-l-l)+(-0.1,-0.05) node [left]{\small $\rt'$};
\draw[thick] (btree-l-l-r-r-r-r) node [below]{\small ~$\rt''$};
\draw[thick] (btree-l-l)+(0,-0.2) node [left]{\small $\rt$};

\draw[thick] (btree-r-r-l-l-l-l) node [above]{\small \,$\lambda_2^1$};
\draw[thick] (btree-r-r-l-l-l-r) node [above]{\small \,$\lambda_2^2$};
\draw[thick] (btree-r-r-l-l-l-r) node [right]{\hspace{0.15cm}$\cdots$};
\draw[thick] (btree-r-r-l-r-r-r) node [above]{\small $\lambda_2^{N-1}$};
\draw[thick] (btree-r-r-r-r-r-r) node [above]{\small $\lambda_2^N$};

\draw[thick] (btree-r-r-l)+(-0.1,-0.05) node [left]{\small $\rt'$};
\draw[thick] (btree-r-r-r-r-r-r) node [below]{\small ~$\rt''$};
\draw[thick] (btree-r-r)+(0,-0.2) node [right]{\small \!$\rt$};

\draw[line width=1.3pt] (btree-root) -- ($(btree-root)+(0,-1)$);

\draw (-2.65,1.7) -- (-1.1,3.25);
\draw (-2.65,1.7) -- (-3.4,2.45);
\draw (-3.4,2.45) -- (-3.4,4);
\draw (-1.1,3.25) -- (-1.1,4);
\draw (-3.4,4) -- (-1.1,4);

\draw (-2.25,4) node [above] {\small $S'$};

\draw (-1.8,2.55) -- (-1.25,2);
\draw (-1.25,2) -- (-0.35,2.9);
\draw (-0.35,2.9) -- (-0.35,4);
\draw (-1.1,4) -- (-0.35,4);

\draw (-0.725,4) node [above] {\small $S''$};

\draw (1.25,1.7) -- (2.8,3.25);
\draw (1.25,1.7) -- (0.5,2.45);
\draw (0.5,2.45) -- (0.5,4);
\draw (2.8,3.25) -- (2.8,4);
\draw (0.5,4) -- (2.8,4);

\draw (1.65,4) node [above] {\small $S'$};

\draw (2.1,2.55) -- (2.65,2);
\draw (2.65,2) -- (3.55,2.9);
\draw (3.55,2.9) -- (3.55,4);
\draw (2.8,4) -- (3.55,4);

\draw (3.175,4) node [above] {\small $S''$};

\draw (1.5,-1) node [above] {$T \!\cdot\! T$};

\end{tikzpicture}
\begin{tikzpicture}

\btreeset{grow=up, scale=0.65, line width=1.3pt}

\BinaryTree[local bounding box=INIT2,line width=1.3pt,bottom padding = 10pt]{
:!
l!l!l!l!l!l,
l!l!l!l!l!r,
l!l!l!r!r!r,
l!r!r!r!r!r,
l!r!r!l!l!l,
l!r!r!l!l!r,
r!r!r!r!r!r,
r!r!r!r!l!l
}{6}

\draw[thick] (btree-l-l-l-l-l-l) node [above]{\small \,$\lambda_1^1$};
\draw[thick] (btree-l-l-l-l-l-r) node [above]{\small \,\,$\lambda_1^2$};
\draw[thick] (btree-l-l-l-l-l-r) node [right]{\hspace{0.15cm}$\cdots$};
\draw[thick] (btree-l-l-l-r-r-r) node [above]{\small $\lambda_1^{N-1}$\,\,\,};

\draw[thick] (btree-l-l-l)+(-0.1,-0.05) node [left]{\small $\rt'$};
\draw[thick] (btree-l-r-r)+(-0.1,-0.05) node [left]{\small $\rt'$};
\draw[thick] (btree-l)+(0,-0.2) node [left]{\small $\tilde \rt'$};

\draw[thick] (btree-l-r-r-l-l-l) node [above]{\small \,\,$\lambda_2^1$};
\draw[thick] (btree-l-r-r-l-l-r) node [above]{\small \,\,\,$\lambda_2^2$};
\draw[thick] (btree-l-r-r-l-l-r) node [right]{\hspace{0.15cm}$\cdots$};
\draw[thick] (btree-l-r-r-r-r-r) node [above]{\small \,\,$\lambda_2^{N-1}$\,\,\,};
\draw[thick] (btree-r-r-r-r-l-l) node [above]{\small \,\,$\lambda_1^N$};
\draw[thick] (btree-r-r-r-r-r-r) node [above]{\small \,\,$\lambda_2^N$};

\draw[thick] (btree-r-r-r-r-l-l)+(0.1,-0.2) node [left]{\small $\rt''$};
\draw[thick] (btree-r-r-r-r-r-r)+(-0.15,-0.2) node [right]{\small $\rt''$};
\draw[thick] (btree-r-r-r-r)+(-0.1,-0.2) node [right]{\small $\tilde \rt''$};

\draw (-2.65,1.7) -- (-1.2,3.15);
\draw (-2.65,1.7) -- (-3.4,2.45);
\draw (-3.4,2.45) -- (-3.4,4);
\draw (-1.2,3.15) -- (-1.2,4);
\draw (-3.4,4) -- (-1.2,4);

\draw (-2.3,4) node [above] {\small $S'$};

\draw (-0.45,1.7) -- (1,3.175);
\draw (-0.45,1.7) -- (-1.2,2.45);
\draw (-1.2,2.45) -- (-1.2,4);
\draw (1,3.15) -- (1,4);
\draw (-1.2,4) -- (1,4);

\draw (-0.1,4) node [above] {\small $S'$};

\draw (2.725,3.15) -- (2.075,2.5);
\draw (2.075,2.5) -- (1.725,2.8);
\draw (1.725,2.8) -- (1.725,4);
\draw (2.725,3.15) -- (2.725,4);
\draw (1.725,4) -- (2.725,4);

\draw (2.225,4) node [above] {\small $S''$};

\draw (3.35,2.525) -- (3.725,2.9);
\draw (3.35,2.525) -- (2.725,3.15);
\draw (3.725,2.9) -- (3.725,4);
\draw (2.725,3.15) -- (2.725,4);
\draw (2.725,4) -- (3.725,4);

\draw (3.225,4) node [above] {\small $S''$};

\draw[line width=1.5pt] (btree-root) -- ($(btree-root)+(0,-1)$);

\draw (1.5,-1) node [above] {$\tilde T$};

\end{tikzpicture}

\end{center}
\caption{\label{Abb-Tt-T-std} The tree join $T \cdot T$, the subtrees $S'$ and $S''$, and the intermediate coupling tree $\tilde T$ constructed in the proof of Theorem \rref{T-R}, for the standard coupling tree $T$. Here, $\rt$ is the root of $T$ and $\rt'$, $\rt''$ are its child nodes.}
\end{figure}
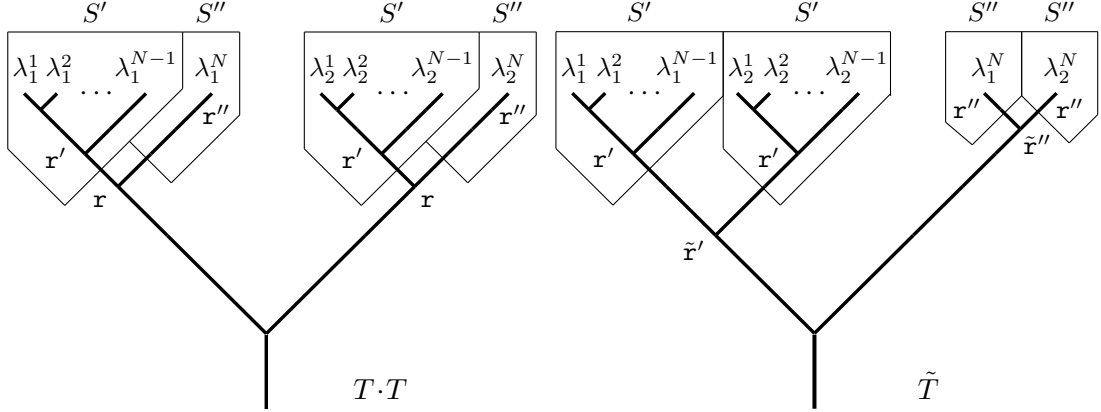

Consider the standard tree (caterpillar tree) of $N$ leaves and let labellings $\alpha_i$ with leaf labellings $\ul\lambda_i$, $i=1,2,3$, be given. As in Example \ref{B-Raupe}, we number the nodes of $T$ by assigning, for $n=2,\dots,N$, the number $n$ to the parent of the $n$-th leaf, see Figure \rref{fig:StdTree}. Then, the nodes numbered $2,\dots,N-1$ are internal and node number $N$ is the root. Accordingly, the labels of the internal nodes are $\alpha_i^{x_n}=(\lambda_i^{x_n},k_i^{x_n})$ and the multiplicity counters one has to sum over are $k^{x_n} = 1 , \dots , m_{(\lambda_1^{x_n},\lambda_2^{x_n})}(\lambda_3^{x_n})$ for $n=2,\dots,N-1$. In this notation, the formula given in Theorem \rref{T-R} reads
$$
\rcneu{T}{\alpha_1}{\alpha_2}{\alpha_3}{\ul k}{k}
=
\sum_{k^{x^2},\dots,k^{x^{N-1}}}
\bpma
\lambda_1^1 & \lambda_1^2 & \lambda_1^{x_2} & k_1^{x_2}
\\
\lambda_2^1 & \lambda_2^2 & \lambda_2^{x_2} & k_2^{x_2}
\\
\lambda_3^1 & \lambda_3^2 & \lambda_3^{x_2} & k_3^{x_2}
\\
k^1 & k^2 & k^{x_2} & 
\epma
\left(
\prod_{n=2}^{N-1}
\bpma
\lambda_1^{x_n} & \lambda_1^{n+1} & \lambda_1^{x_{n+1}} & k_1^{x_{n+1}}
\\
\lambda_2^{x_n} & \lambda_2^{n+1} & \lambda_2^{x_{n+1}} & k_2^{x_{n+1}}
\\
\lambda_3^{x_n} & \lambda_3^{n+1} & \lambda_3^{x_{n+1}} & k_3^{x_{n+1}}
\\
k^{x_n} & k^{n+1} & k^{x_{n+1}} & 
\epma
\right)
,
$$
where $k^{x_N} = k$ and where the summation range is given above. The subtree $S'$ used in the proof of that theorem consists of the leaves $1,\dots,N-1$ and the nodes $2,\dots,N-1$ and has the shape of the standard tree on $N-1$ leaves. The subtree $S''$, on the other hand, consists of the leaf $N$ only. The join $T \bcdot T$ and the intermediate coupling tree $\tilde T$, together with the subtrees $S'$ and $S'$, are shown in Figure \rref{Abb-Tt-T-std}. 
\qeb

\ebs


\subsection{The special case $G = \SU(2)$}
\label{alg-SU2}


In the case of $G=\SU(2)$, the highest weights $\lambda$ of irreps correspond to spins $j = 0 , \frac 1 2 , 1 , \frac 3 2 , \dots$. The elements of the weight system of the highest weight corresponding to spin $j$ are nondegenerate, and correspond to spin projections $m = -j , -j+1 , \dots , j$. Therefore, we write $j$ for $\lambda$, $\ul j$ for $\ul\lambda$, $m$ for $\mu$ and $\ul m$ for $\ul\mu$, and we omit the weight multiplicity counters. Thus, $(H_j,D^j)$ is the standard $\SU(2)$-irrep of spin $j$, spanned by the orthonormal ladder basis $\{\ket{j m} : m = -j , -j+1 , \dots , j\}$ which is unique up to a phase. Let us denote
$$
d_j := \dim(H_j) = 2 j + 1
\,,\qquad
d_{\ul j} := \dim(H_{\ul j}) = 2 (j_1 + \cdots + j_N) + N
\,.
$$
In contrast to the general case, in the decomposition of the tensor product of two irreps into irreps the multiplicities are equal to $1$. Thus, we need only one type of bracket, $\langle \cdot , \cdot \rangle$, given by 
$$
\langle j^1 , j^2 \rangle
 = 
\{|j^1-j^2| , |j^1-j^2|+1 , |j^1-j^2|+2 , \dots , j^1+j^2\}
\,,
$$
for any two spins $j^1$, $j^2$, which is known as the triangle rule. The unitary representation isomorphism \eqref{G-TePr-Zlg-1} reads
$$
H_{j^1} \otimes H_{j^2} \to \bigoplus_{j\in \langle j^1 , j^2 \rangle}H_j\,.
$$
Accordingly, the bracketing schemes and their description in terms of 
coupling trees simplify: a labelling $\alpha$ of a coupling tree $T$ assigns to every vertex $x$ of $T$ a spin $j^x$. It is admissible if for every node $x$ one has $j^x \in \langle j^{x'},j^{x''} \rangle$, where $x'$ and $x''$ denote the child nodes of $x$. The modified quasicharacters read 
\beq\label{Form-chi-SU2}
\dcq{T}{\alpha}{\alpha'}(\ul a)
 =
\sum_{m=-j}^j 
\bra{T;\alpha',m} D^{\ul j}(\ul a) \ket{T;\alpha,m}
 \,,\qquad
\alpha,\alpha' \in \mc L^T(\ul j,j)
\,.
\eeq
Computation of their norms \eqref{G-norm} yields \cite{HRS}
\beq\label{G-SU2-Norm}
\left\|\dcq{T}{\alpha}{\alpha'}(\ul a)\right\|^2
=
(\hbar\beta)^{3N/2}
\,
\frac{d_j}{d_{\ul j}}
\,
\mr e^{\hbar \beta^2 (d_{j^1}^2 + \cdots + d_{j^N}^2)}
\,,
\eeq
where $\beta$ is a scaling factor for the invariant scalar product $\braket{~}{~}$ on $\mf g$ defined by 
$$
\braket{X}{Y} = - \frac{1}{2\beta^2} \, \tr(XY)
\,.
$$
Formula \eqref{G-D-Racah} giving the recoupling coefficients reads 
\beq\label{G-D-Racah-SU2}
\bra{T_2;\alpha_2,m_2} \Pi \ket{T_1;\alpha_1,m_1}
=
\delta_{j_1 j_2} \delta_{m_1 m_2} 
R\big(T_2|T_1\big)^{\alpha_2}_{\alpha_1}
\,.
\eeq
The latter are the recoupling coefficients (Racah coefficients) of angular momentum theory \cite{BL2}. Explicitly,
\beq\label{G-Racah-expl-SU2}
R\big(T_2|T_1\big)^{\alpha_2}_{\alpha_1}
=
\bra{T_2;\alpha_2,m} \Pi \ket{T_1;\alpha_1,m}
\,,
\eeq
independently of $m$. The recoupling coefficients are real. Hence, we have the symmetry relation
\beq
\label{G-Racah-sym-SU2}
R\big(T_2|T_1\big)^{\alpha_2}_{\alpha_1}
=
R\big(T_1|T_2\big)^{\alpha_1}_{\alpha_2}
\,.
\eeq
The labelling of join and leaf duplication simplifies as follows. For $i=1,2,3$, let $\ul j_i$ be a leaf labelling of $T$, let $j_i \in \langle \ul j_i \rangle$ and let $\alpha_i\in \mc L^T(\ul j_i,j_i)$. For $T \bcdot T$, the label of the new root reduces to $j \in \langle j_1,j_2 \rangle$. Hence, labellings of $T \bcdot T$ read $\join{\alpha_1}{\alpha_2}{j}$. For $T^\vee$, node labellings compatible with the leaf labelling $\ul j_1 \!\ast\! \ul j_2$ are given by labellings $\alpha_3$ of $T$ satisfying the condition that for all leaves $y$ one has 
$$
j_3^y \in \langle j_1^y,j_2^y \rangle
\,,
$$
that is, $j_1^y$, $j_2^y$ and $j_3^y$ satisfy the triangle condition. The corresponding labellings of $T^\vee$ read $\ld{\alpha_1}{\alpha_2}{\alpha_3}$. If, in addition, also $j_1$, $j_2$ and $j_3$ satisfy the triangle condition, we will say that $\alpha_1$, $\alpha_2$ and $\alpha_3$ satisfy the triangle condition. Then, the definitions of the recoupling coefficient $R(T)$ and of its transpose read
$$
\rc{T}{\alpha_1}{\alpha_2}{\alpha_3}
:=
\begin{cases}
R\big(T \bcdot T|T^\vee\big)
^{\join{\alpha_1}{\alpha_2}{j_3}}
_{\ld{\alpha_1}{\alpha_2}{\alpha_3}}
& |\abs \alpha_1 , \alpha_2 , \alpha_3 \text{ satisfy the triangle condition,}
\\
0 & |\abs \text{otherwise}
\end{cases}
$$
and
$$
\rct{T}{\alpha_1}{\alpha_2}{\alpha_3}
:=
\begin{cases}
R\big(T^\vee|T \bcdot T\big)
^{\ld{\alpha_1}{\alpha_2}{\alpha_3}}
_{\join{\alpha_1}{\alpha_2}{j_3}}
& |\abs \alpha_1 , \alpha_2 , \alpha_3 \text{ satisfy the triangle condition,}
\\
0 & |\abs \text{otherwise,}
\end{cases}
$$
respectively. Since the recoupling coefficients can be chosen to be real, we have 
$$
\rc{T}{\alpha_1}{\alpha_2}{\alpha_3} = \rct{T}{\alpha_1}{\alpha_2}{\alpha_3}
\,.
$$
In this notation, the multiplication law for quasicharacters given by Corollary \rref{TreeProductRule} reads
\beq
\label{MultLaw-SU2}
\dcq{T}{\alpha_1^{\phantom{\prime}}}{\alpha_1'}
\,
\dcq{T}{\alpha_2^{\phantom{\prime}}}{\alpha_2'}
=
\sum_{\alpha_3,\alpha_3'}
\,
\rc{T}{\alpha_1'}{\alpha_2'}{\alpha_3'}
\,\,
\dcq{T}{\alpha_3^{\phantom{\prime}}}{\alpha_3'}
\,\,
\rct{T}{\alpha_1}{\alpha_2}{\alpha_3}
\,,
\eeq
where the sum is over all combinable labellings $\alpha_3$, $\alpha_3'$ of $T$ such that $\alpha_1$, $\alpha_2$ and $\alpha_3$ satisfy the triangle condition. 

Finally, we observe that the combinatorics based upon Clebsch-Gordan coefficients boils down to the combinatorics of angular momentum theory. According to Lemma \rref{L-CCG}, for a given leaf labelling $\ul j$, $j \in \langle \ul j \rangle$ and $\alpha \in \mc L^T(\ul j,j)$, the composite Clebsch-Gordan coefficient 
$$
\ccg{T}{\alpha}{\ul m}{m} = \bigbraket{\ul j\,\ul m}{T;\alpha,m}
$$
vanishes unless $|m^x|\leq j^x$ for every node $x$, where $m^x$ denotes the sum of $m^y$ over all leaves $y$ descending from $x$. In that case, 
\beq
\label{C-comp-dec}
\ccg{T}{\alpha}{\ul m}{m}
 =
\prod_{\text{nodes } x} 
C^{j^{x'} j^{x''} j^x}_{m^{x'}m^{x''}m^x}
\,,
\eeq
where $x'$, $x''$ denote the child vertices of $x$. Here, $C^{j_1 \, j_2 \, j_3}_{m_1 \, m_2 \, m_3}$ are the ordinary Clebsch-Gordan coefficients of angular momentum theory. It follows that the composite Clebsch-Gordan coefficients are real. The formula expressing the recoupling coefficients in terms of composite Clebsch-Gordan coefficients given in point \rref{i-S-CCG-rc} of Proposition \rref{S-CCG} reads 
\beq
\label{R-C-SU(2)}
R\big(T_2|T_1\big)^{\alpha_2}_{\alpha_1}
=
\sum_{\ul m}
C\big(T_2;\alpha_2,\sigma(\ul m)\big)
\,
C\big(T_1;\alpha_1,\ul m\big) 
\,,
\eeq
where $\sigma$ is the permutation turning $\ul j_1$ into $\ul j_2$ and the sum is over all spin projections $\ul m$ of $\ul j_1$ summing to $ m$ for some fixed spin projection $m$ of $j$. 

Finally, the decomposition of recoupling coefficients into 
a product of primitive $9\lambda$ symbols given by Theorem \ref{T-R} boils down to 
the decomposition of Racah-recoupling coefficients in terms of $9j$ symbols of angular momentum theory. In more detail, the $9\lambda$ symbols become $(3\times 3)$-arrays,
$$
\bpma
j_1^1 & j_1^2 & j_1
\\
j_2^1 & j_2^2 & j_2
\\
j_3^1 & j_3^2 & j_3
\epma
:=
\begin{cases}
{\rc{\boldsymbol\vee}{\alpha_1}{\alpha_2}{\alpha_3}}
& | \abs 
j_i \in \langle j_i^1,j_i^2 \rangle \text{ for all } i=1,2,3
\,,
\\
0
& | \abs \text{otherwise.}
\end{cases}
$$
Here, $\alpha_i$ assigns $j_i^1$, $j_i^2$ to the leaves of $\boldsymbol\vee$ and $j_i$ to its root. By definition of ${\rc{\boldsymbol\vee}{\alpha_1}{\alpha_2}{\alpha_3}}$, these symbols vanish unless all rows satisfy the triangle condition. They are related to Wigner's $9j$ symbols $\{\cdot\}$ via a dimension factor,
\beq\label{G-9j}
\bpma
j_1^1 & j_1^2 & j_1
\\
j_2^1 & j_2^2 & j_2
\\
j_3^1 & j_3^2 & j_3
\epma
=
\sqrt{d_{j_1} d_{j_2} d_{j_3^1} d_{j_3^2}}
\bBma
j_1^1 & j_1^2 & j_1
\\
j_2^1 & j_2^2 & j_2
\\
j_3^1 & j_3^2 & j_3
\eBma
.
\eeq
By Theorem \rref{T-R}, for every coupling tree $T$ and all labellings $\alpha_1$, $\alpha_2$, $\alpha_3$, we have 
\beq
\label{DecLaw-R-SU2}
\rc{T}{\alpha_1}{\alpha_2}{\alpha_3}
=
\prod_{x \text{ node of } T}
\bpma
j_1^{x'} & j_1^{x''} & j_1^x
\\
j_2^{x'} & j_2^{x''} & j_2^x
\\
j_3^{x'} & j_3^{x''} & j_3^x
\epma
\,,
\eeq
where $x'$ and $x''$ denote the child vertices of $x$. For the standard coupling tree, this decomposition formula was already found in \cite{FJRS}.


\section{Hamiltonian lattice quantum gauge theory}
\label{OT-LGT}



\subsection{Background}
\label{ClassPic}


Let us consider a (finite) lattice gauge theory with compact gauge group $G$ in the Hamiltonian approach. For details, see \cite{qcd2,qcd3,KS}. Let $\Lambda$ be a finite regular three-dimensional spatial lattice and let $\Lambda^0$, $\Lambda^1$ and $\Lambda^2$ denote, respectively, the sets of lattice sites, lattice links and lattice plaquettes. For the links and plaquettes, let there be chosen an arbitrary orientation. Gauge potentials (the variables) are approximated by their parallel transporters along links and gauge transformations (the symmetries) are approximated by their values at the lattice sites. Thus, the classical configuration space is the space $G^{\Lambda^1}$ of mappings $\Lambda^1 \to G$, the classical symmetry group is the group $G^{\Lambda^0}$ of mappings $\Lambda^0 \to G$ with pointwise multiplication and the action of $g \in G^{\Lambda^0}$ on $a \in G^{\Lambda^1}$ is given by  
\beq
\label{G-Wir-voll}
(g \!\cdot\! a)(\ell) := g(x) a(\ell) g(y)^{-1}\,,
\eeq
where $(x,y) = \ell \in \Lambda^1$ with $x$ and $y$ the source and target of $\ell$, respectively. The classical phase space is given by the associated Hamiltonian $G$-manifold and the reduced classical phase space is obtained from that by symplectic reduction, as developed in \cite{OrtegaRatiu,Buch,SjamaarLerman}. 
Dynamics is ruled by the classical counterpart of the Kogut-Susskind lattice Hamiltonian \cite{KS}.

In gauge theory, it is convenient to perform the symplectic reduction by two stages. First, one perfoms regular symplectic reduction with respect to the free action of the normal subgroup 
\beq
\label{G-ptgautrf}
\{g \in G^{\Lambda^0} : g(x_0) = \II\}
\eeq
of pointed gauge transformations. Here, $\II$ denotes the unit element of $G$. The quotient symplectic manifold carries the action of the residual group of local gauge transformations which is naturally isomorphic to $G$.  By choosing a maximal tree $\mc T$ in the graph $\Lambda^1$ one finds that the subset defined by the tree gauge, 
$$
\{a \in G^{\Lambda^1} 
: 
a(\ell) = \II \text{ for all } \ell \in \mc T\} \subset G^{\Lambda^1}
\,,
$$
intersects every orbit of the subgroup \eqref{G-ptgautrf} exactly once. Hence, the quotient manifold may be identified with the direct product $G^N$ of $N$ copies of $G$, where $N$ is the number of off-tree links. Then, the action of the residual gauge group turns into the action of $G$ on $G^N$ by diagonal conjugation,
\beq\label{G-Wir-Q}
g \!\cdot\! (a_1,\dots,a_N) = (g a_1 g^{-1} , \dots , g a_N g^{-1})
\,.
\eeq
To this action there corresponds a Hamiltonian $G$-action on the cotangent bundle $\ctg G^N$ and the second stage of the reduction procedure consists in zero level symplectic reduction of that Hamiltonian $G$-action. If $G$ is nonabelian, this action has non-trivial orbit types and thus requires singular symplectic reduction. The resulting reduced phase space $\pha$ then carries the structure of a stratified symplectic space, where the strata are given by the connected components of the orbit type subsets of $\pha$. It is well known that these strata are invariant under the dynamics generated by the classical counterpart of the Kogut-Suskind Hamiltonian. 

The quantum theory of the above model is constructed via canonical quantization. We refer 
to \cite{qcd2,qcd3,GR} for the details, including the discussion of the field algebra and 
the observable algebra of the quantum system. The (uniqe up to isomorphism) representation spaces for the field and the observable 
algebra are $L^2(G^N)$ and $L^2(G^N)^G$, respectively. It turns out that the quantum model so obtained may be, equivalently, derived by K\"ahler quantization and, in this context, the classical gauge orbit stratification may be implemented at the quantum level. As regular reduction commutes with quantization \cite[Thm.\ 5.2]{qcd3}, we may start with the partially reduced phase space $\ctg G^N$. The latter is endowed with a natural K\"ahler structure as follows.
Trivialization by left-invariant vector fields and polar decomposition define a $G$-equivariant diffemorphism between the cotangent bundle $\ctg G^N$ and the complexified Lie group $G_\CC^N$. On the one hand, according to \cite{Hall:cptype}, this diffeomorphism allows for combining the symplectic structure of $\ctg G^N$ with the complex structure of $G_\CC^N$ to a K\"ahler structure. K\"ahler quantization with subsequent reduction leads to the Hilbert space 
$$
{\cal H} = HL^2(G_\CC^N)^G
$$ 
of square-integrable $G$-invariant holomorphic functions on $G_\CC^N$. Via the Segal-Bargmann transformation, this Hilbert space is unitarily isomorphic to $L^2(G^N)^G$ and thus may be taken as the Hilbert space of the quantum lattice gauge theory under consideration.

 To study the question whether the strata produce quantum effects, we follow the idea to model the classical strata in quantum theory by appropriate subspaces of the corresponding Hilbert space, forming a costratification in the sense of Huebschmann \cite{Hue:Quantization,HRS}. This will be briefly explained in the next subsection.


\subsection{Calculus for bi-invariant quantum operators}
\label{QuantumOp}


Consider the following classes of bi-invariant operators on $\cal H$.
\begin{enumerate}
\item 
The elements of the algebra $\mc R$ of $G$-invariant representative functions acting as multiplication operators on $\cal H$.
\item
The  bi-invariant linear differential operators on $\cal H$. 
\item
Any linear combination of operators of the above type. 
\end{enumerate}

\noindent
Let us discuss the first class. As the normalized $G$-invariant representative functions 
\beq
\label{G-Bs}
\frac{\dcqc{T}{\alpha}{\alpha'}}{\|\dcqc{T}{\alpha}{\alpha'}\|}
\eeq
associated with a chosen reduction tree $T$ form an orthonormal basis 
in $\cal H$, any element $r$ of $\mc R$ may be expressed in terms of them and,  thus, the matrix elements of the corresponding 
multiplication operator $r^\CC$ may be computed explicitly. In more detail, 
these matrix elements are given by 
\beq\label{G-ME}
\hat r^\CC(T)^{\alpha'\beta}_{\alpha\beta'}
=
\frac
{\left\|\dcqc{T}{\alpha}{\alpha'}\right\|}
{\left\|\dcqc{T}{\beta}{\beta'}\right\|}
\,
M(T)^{\alpha'\beta}_{\alpha\beta'}
\,,
\eeq
where the coefficients $M(T)^{\alpha\beta'}_{\alpha'\beta}$ are defined by
\beq\label{G-MEK}
r^\CC \cdot \dcqc{T}{\alpha}{\alpha'}
=
\sum_{\beta,\beta'}
M(T)^{\alpha'\beta}_{\alpha\beta'}
\dcqc{T}{\beta}{\beta'}
\,.
\eeq
Now, the matrix elements are computed by first expanding $r^\CC$ with respect to the basis elements \eqref{G-Bs} and then applying the multiplication law for quasicharacters given in Corollary \rref{F-TPR}.

Concerning the second class, recall the following well known facts. A differential operator on $G_\CC^N$ is called left
(or right) invariant by G if it commutes with the action of G by
left (or right) translations. The algebra of left invariant linear
differential operators on $G_\CC^N$ can be identified with the complexified universal enveloping algebra $U(\mf g)$ of the Lie algebra $\mf g$ 
of G. Bi-invariant operators, that is,  operators which are both left and right invariant then correspond to the elements of the centre 
$Z(\mf g)$ of $U(\mf g)$. The centre is spanned by the so-called Casimir elements. By Schur's Lemma, in any irrep of $\mf g$, the Casimir operators are proportional to the identity and, thus, the constants of proportionality can be used to classify the irreps of $G$. On the other hand, the irreps are spanned by the quasicharacters given by \eqref{G-Bs}. Thus, these quasicharacters are eigenfunctions of 
the Casimir operators and, thus, are eigenfunctions of the bi-invariant differential operators. This implies that their matrix elements 
are given in terms of the corresponding eigenvalues. 

We may of course build linear combinations of bi-invariant operators of the first and of the second type. The Hamiltonian 
of quantum lattice gauge theory is of that type, see the next subsection for the case $G= \SU(2)$.\footnote{Clearly, 
a priori, we rather view the Hamiltonian as an operator on the Hilbert space $L^2(G^N)^G$. But, via the Segal-Bargmann 
transformation, we can transport it to $\cal H$. Thereby, the Casimir is extended in an obvious way to a differential operator 
on $G_{\CC}^N$ and the multiplication operator becomes a multiplication operator on $\cal H$..
}
One meets Hamiltonians in many other branches of physics having also that structure, see e.g. some models of nuclear physics. We conclude that the 
calculus developed in this paper reduces the study of the spectral problem of a Hamiltonian of the above type to a problem in 
linear algebra, see the next subsections for more details. 

Next, let us explain how this calculus may be used for the implementation of the classical stratification on quantum level:
one can show that the $G$-equivariant diffeomorphism $\ctg G^N \cong G_\CC^N$ descends to a homeomorphism between $\pha$ and the quotient $G_\CC^N//G_\CC$ of $G_\CC^N$ with respect to $G_\CC$-orbit closure equivalence. Then, given a stratum $\mc S \subset \pha$, let $r_1, \dots , r_n \in \mc R$ be chosen in such a way that

\ben

\item the zero locus of the complex analytic extensions $r_1^\CC, \dots , r_n^\CC$ to $G_\CC^N$, viewed as functions on $G_\CC^N//G_\CC$, coincides with the image of $\mc S$ under the homeomorphism $G_\CC^N//G_\CC \cong \pha$ (zero locus condition),

\item the ideal generated by $r_1, \dots , r_n$ in $\mc R$ is a radical ideal (radical ideal condition).

\een

\noindent
By Hilbert's Basissatz, such finite sets of relations exist. The subspace of $\cal H$ associated with $\mc S$ consists of those elements which are localized at $\mc S$ in the sense that they are orthogonal to all elements which vanish on $\mc S$. Since $\mc S$ is the zero locus of $r_1^\CC, \dots , r_n^\CC$, one has to evaluate the multiplication operator $\hat r_i^\CC$ on $\cal H$ defined by $r_i^\CC$ for each $i = 1 , \dots , n$. Now, this can be done by computing their matrix elements with respect to the basis \eqref{G-Bs} as explained above.  

In the remainder, we will present the quantum Hamiltonian for $G= \SU(2)$ and we will show how to compute the matrix elements 
for two instances of relations $r_i$, one for $G=\SU(2)$ and one for $G=\SU(3)$.


\subsection{Examples}


\subsubsection{The quantum Hamiltonian for $G=\SU(2)$}
\label{Q-Ham}


The pure gauge part of the quantum Hamiltonian\footnote{Here rather viewed as an operator on $L^2(G^N)^G$} for the case $G=\SU(2)$ reads as follows, see \cite{FJRS} for the details:
\beq
\label{Hamiltonian}
H = \frac{g^2}{2\delta} \, \mf C - \frac{1}{g^2 \delta} \, \mf W
\,.
\eeq
Here,
$$
\mf C := \sum_{\ell \in \Lambda^1} E_{ij}(\ell) E_{ji}(\ell)
$$
is the Casimir operator (negative of the group Laplacian) of $\SU(2)^N$ and 
$$
\mf W := \sum_{p \in \Lambda^2} ( W (p) + W(p)^*)
\,,
$$
where $W(p)$ is the quantum counterpart (multiplication operator) of $\tr a(p)$, called the Wilson loop operator. For details, see \cite{qcd3}, \cite{GR2}, \cite{KS}. Recall that the representative functions of spin $j$ on $SU(2)$ are eigenfunctions of the Casimir operator of $\SU(2)$ corresponding to the eigenvalue
$$
\epsilon_j = 4 j(j+1)
\,.
$$
It follows that the invariant representative functions  are eigenfunctions of ${\mathfrak C}$ corresponding to the eigenvalues
\beq
\label{Eigenvalue-Casimir}
\epsilon_{\ul j} = \epsilon_{j^1} + \cdots + \epsilon_{j^N}
\,. 
\eeq
The summand $\mf W$ is a $G$-invariant multiplication operator, whose matrix elements can be calculated as 
explained above. In a multi-index notation $I, J, \ldots $, with real quasicharacters denoted by $\chi_I$, one obtains the following eigenfunction equation for $H$:
\beq
\label{EP-H}
\sum_{J} 
 \left\{
\left({g^2}{2\delta} \epsilon_J - {\cal E} \right) \delta_J^K
 - 
\frac{1}{g^2 \delta} \sum_{I \in \mc I} W^I
\left(C_{IJ}^K + C_{IK}^J\right)\right\} \psi^J
 = 
0
 \,,
\eeq
for all multi-indices $K$. Here, we have written $\epsilon_J$ for the eigenvalue of the Casimir operator $\mf C$ corresponding to the eigenfunction $\chi_J$, given by \eqref{Eigenvalue-Casimir}. The numbers $C_{IJ}^K$ are the structure constants in the multiplication 
law of the quasicharacters and $W^I$ are the expansion coefficients for $\mf W$ in the basis $\{\chi_I\}$. Thus, we are left with a homogeneous system of linear equations for the eigenfunction coefficients $\psi^J$. The eigenvalues $\cal E$ are determined by the requirement that the determinant of this system must vanish. Note that the sum over $I$ in \eqref{EP-H} is finite, because there are only finitely many nonvanishing $W^I$. Moreover, it turns out that also the sum over $J$ is finite for every fixed $K$. Thus, we have reduced the eigenvalue problem for the Hamiltonian to a problem in linear algebra. Combining this with well-known asymptotic properties of $3nj$ symbols, see \cite{BL2} (Topic 9) and further references therein, we obtain an algebraic setting which allows for a computer algebra supported study of the spectral properties of $H$.


\subsubsection{A stratum relation for $G=\SU(2)$}
\label{A-GET-AA-SU2}

The case $N=1$ has been treated in detail in \cite{HRS}. In the case $N > 1$, there are $2^N+1$ secondary strata; $2^N$ of them have orbit type $G$ and one has orbit type $\mr U(1) \times \mr U(1)$ (the subgroup of diagonal matrices). The strata of orbit type $G$ consist of isolated points and hence the corresponding subspaces of localized wave functions are one-dimensional. They can be constructed without using relations, see \cite[Rrem.\ 5.4]{FuRS}. For the stratum $\mc S$ of orbit type $T$, a system of relations satisfying the zero locus and the radical ideal condition is given by 
\al{\label{G-SU2-Rel1}
&&
r_{ij}(\ul a) & := \tr([a_i,a_j]^2)
\,, &&
1 \leq i < j \leq N
\,,
&&
\\ \label{G-SU2-Rel2}
&&
r_{ijk}(\ul a) & := \tr([a_i,a_j]a_k)
\,, &&
1 \leq i < j < k \leq N
\,.
&&
}
Let us consider the case $N=2$. Here, the only reduction tree is the cherry, and a labelling is given by two leaf labels $j^1$ and $j^2$ and a root label $j \in \langle j^1,j^2 \rangle$. For simplicity, the corresponding modified quasicharacter will be denoted by 
$$
\hat\chi^\CC_{j^1,j^2,j}
\,.
$$
According to \eqref{DecLaw-R-SU2}, the multiplication law \eqref{MultLaw-SU2} reads
\beq\label{G-SU2-N2-Mult}
\hat\chi^\CC_{j_1^1,j_1^2,j_1} \cdot \hat\chi^\CC_{j_2^1,j_2^2,j_2}
=
\bpma
j_1^1 & j_1^2 & j_1
\\
j_2^1 & j_2^2 & j_2
\\
j_3^1 & j_3^2 & j_3
\epma^2
\hat\chi^\CC_{j_3^1,j_3^2,j_3}
\,.
\eeq
By \eqref{G-SU2-Norm}, the norms are
\beq\label{G-SU2-N2-Norm}
\|\hat\chi^\CC_{j^1,j^2,j}\|
=
(\hbar\pi)^{3/2} \sqrt{\frac{d_j}{d_{j^1} d_{j^2}}}
\
\mr e^{\hbar\beta^2(d_{j^1}^2 + d_{j^2}^2)/2}\,.
\eeq
The system \eqref{G-SU2-Rel1}, \eqref{G-SU2-Rel2} reduces to the single relation 
$$
r(a_1,a_2) = \tr([a_1,a_2]^2)
\,.
$$
In \cite{FJRS} it has been shown that
$$
r^\CC
= 
\hat\chi^\CC_{1,0,1} 
+ \hat\chi^\CC_{0,1,1} 
+ 3 \hat\chi^\CC_{1,1,0} 
- 2 \hat\chi^\CC_{1,1,1} 
- 3 \hat\chi^\CC_{0,0,0} 
\,.
$$
By \eqref{G-SU2-N2-Mult}, then
\ala{
r^\CC
\cdot
\hat\chi^\CC_{j^1,j^2,j} 
& =
\sum_{j'^1,j'^2,j'}
\left(
\bpma
1 & 0 & 1
\\
j^1 & j^2 & j
\\
j'^1 & j'^2 & j'
\epma^2
+ 
\bpma
0 & 1 & 1
\\
j^1 & j^2 & j
\\
j'^1 & j'^2 & j'
\epma^2
+ 3 
\bpma
1 & 1 & 0
\\
j^1 & j^2 & j
\\
j'^1 & j'^2 & j'
\epma^2
\right.
\\
& \hspace{3cm} \left.
- 2 
\bpma
1 & 1 & 1
\\
j^1 & j^2 & j
\\
j'^1 & j'^2 & j'
\epma^2
- 3 \ \delta_{j^1,j'^1} \ \delta_{j^2,j'^2} \ \delta_{j,j'} 
\right)
\hat\chi^\CC_{j'^1,j'^2,j'}
\,.
}
From this, we can read off the matrix elements via \eqref{G-ME} and \eqref{G-MEK}. Expressing the $9\lambda$ symbols in terms of Wigner's $9j$ symbols according to \eqref{G-9j} and reducing the $9j$ symbols containing a zero entry to Wigner's $6j$ symbols according to the rule
$$
\bBma
0 & k & k 
\\ 
j^1 & j^2 & j
\\
j'^1 & j'^2 & j'
\eBma
=
\frac{\delta_{j^1 j'^1}}{\sqrt{d_{j^1} d_k}}
(-1)^{j+k+j'^1+j'^2}
\bBma
k & j' & j 
\\ 
j^1 & j^2 & j'^2
\eBma
,
$$
we obtain
\ala{
(\hat r^\CC)_{j^1,j^2,j}^{j'^1,j'^2,j'}
= &
\sqrt{\frac{d_j d_{j'^1} d_{j'^2}}{d_{j^1} d_{j^2} d_j'}}
\mr e^{\hbar \beta^2(d_{j^1}^2 + d_{j^2}^2 - d_{j'^1}^2 - d_{j'^2}^2)/2}
\Bigg(
\delta_{j^2 j'^2} \, d_j \, d_{j'^1} 
\bBma
1 & j'^1 & j^1 
\\ 
j^2 & j & j'
\eBma^2
\\
& +
\delta_{j^1 j'^1} \, d_j \, d_{j'^2} 
\bBma
1 & j' & j 
\\ 
j^1 & j^2 & j'^2
\eBma^2
+
3 \, \delta_{j j'} \, d_{j^2} \, d_{j'^1} 
\bBma
1 & j'^2 & j^2 
\\ 
j & j^1 & j'^1
\eBma^2
\\
&
-
6 \, d_j \, d_{j'^1} \, d_{j'^2}
\bBma
1 & 1 & 1
\\
j^1 & j^2 & j
\\
j'^1 & j'^2 & j'
\eBma^2
- 3 \ \delta_{j^1,j'^1} \ \delta_{j^2,j'^2} \ \delta_{j,j'} 
\Bigg)
\,.
}
These coefficients vanish unless $(1,j^1,j'^1)$, $(2,j^2,j'^2)$ and $(1,j,j')$ form triads.


\subsubsection{A stratum relation for $G=\SU(3)$}
\label{A-GET-AA-SU3}


In the case $N>1$, there exists $3^N+3$ secondary strata; $3^N$ for orbit type $G$ and one for each of the orbit types $\mr U(2)$ (embedded e.g.\ as the lower right $2 \times 2$-block), $\mr U(1) \times \mr U(1)$ (the subgroup of diagonal matrices) and $\mr U(1)$ (embedded e.g.\ as the lower right corner). In case $N=1$, the last stratum is not present and the stratum of orbit type $\mr U(1) \times \mr U(1)$ is principal. As for $\SU(2)$, let us consider the case $N=2$ and let us focus on the stratum $\mc S$ of orbit type $\mr U(1) \times \mr U(1)$. Using McCoy's Theorem \cite{Radjavi}, one can show that $\mc S$ is the zero locus of the $15$ invariant functions
\ala{
r_1(a_1,a_2) & = \tr([a_1,a_2] a_1 a_2)
\,,
\\
r_2(a_1,a_2) & = \tr([a_1,a_2] a_1^2 a_2)
\,, & 
r_3(a_1,a_2) & = \tr([a_1,a_2] a_2^2 a_1)
\,,
\\
r_4(a_1,a_2) & = \tr\big(([a_1,a_2] a_1)^2\big)
\,, & 
r_5(a_1,a_2) & = \tr\big(([a_1,a_2] a_2)^2\big)
\,,
\\ 
r_6(a_1,a_2) & = \tr([a_1,a_2]^3)
\,,
\\
r_7(a_1,a_2) & = \tr\big(([a_1,a_2] a_1^2)^2\big)
\,, & 
r_8(a_1,a_2) & = \tr\big(([a_1,a_2] a_2^2)^2\big)
\,,
\\ 
r_9(a_1,a_2) & = \tr\big(([a_1,a_2] a_1 a_2)^2\big)
\,,
\\ 
r_{10}(a_1,a_2) & = \tr\big(([a_1,a_2] a_1^2 a_2)^2\big)
\,, &
r_{11}(a_1,a_2) & = \tr\big(([a_1,a_2] a_1 a_2 a_1)^2\big)
\,,
\\ 
r_{12}(a_1,a_2) & = \tr\big(([a_1,a_2] a_2 a_1^2)^2\big)
\,, & 
r_{13}(a_1,a_2) & = \tr\big(([a_1,a_2] a_2^2 a_1)^2\big)
\,,
\\ 
r_{14}(a_1,a_2) & = \tr\big(([a_1,a_2] a_2 a_1 a_2)^2\big)
\,, & 
r_{15}(a_1,a_2) & = \tr\big(([a_1,a_2] a_1 a_2^2)^2\big)
\,.
}
Let us discuss the matrix elements of the multiplication operator $\hat r_1^\CC$. For that purpose, we need some representation theoretic basics about $\SL(3,\CC)$. The highest weights correspond to pairs $(n,m)$ of nonnegative integers, each representing twice the spin of an appropriate $\SU(2)$ subgroup. For example, $(1,0)$ corresponds to the identical representation, $(0,1)$ to the contragredient of the identical representation (the representation induced on the dual space) and $(1,1)$ to the adjoint representation. We write $\lambda_{n,m}$ for the highest weight corresponding to $(n,m)$, $D^{n,m}$ for the corresponding irrep and $d_{n,m}$ for its dimension. Recall that
$$
d_{n,m} = \frac 1 2 \, (n+1)(m+1)(n+m+2)
\,.
$$
As for $\SU(2)$, due to $N=2$, the only reduction tree is the cherry, and a labelling is given by two leaf labels, $\lambda_{n_1,m_1}$ and $\lambda_{n_2,m_2}$, and a root label, $(\lambda_{n,m},k) \in \llangle \lambda_{n_1,m_1},\lambda_{n_2,m_2} \rrangle$. Thus, the elements of 
$
\mc L^\vee\big((\lambda_{n_1,m_1},\lambda_{n_2,m_2}),\lambda_{n,m}\big)
$
are enumerated by the multiplicity counter $k$. Therefore, the quasicharacters are labelled by the triple $(\lambda_{n_1,m_1},\lambda_{n_2,m_2},\lambda_{n,m})$ and two multiplicity counters $k',k$. For simplicity, we will denote
$$
\big(\hat\chi^\CC_{n_1m_1,n_2m_2,nm}\big)^{k'}_k
:=
\dcq
{\boldsymbol\vee}
{((\lambda_{n_1,m_1},\lambda_{n_2,m_2}),(\lambda_{n,m},k))}
{((\lambda_{n_1,m_1},\lambda_{n_2,m_2}),(\lambda_{n,m},k'))}
\,.
$$
In addition, we will omit the brackets and the multiplicity counters $k$, $k'$ in case the multiplicity is $1$. Let us compute the norms. By \eqref{G-norm}, 
$$
\left\|\big(\hat\chi^\CC_{n_1m_1,n_2m_2,nm}\big)^{k'}_k\right\|^2
=
\frac{d_{n,m}}{d_{n_1,m_1} d_{n_2,m_2}}
\,
(\hbar \pi)^8 
\, 
\mr e^{\hbar (|\lambda_{n_1,m_1} + \rho|^2 + |\lambda_{n_2,m_2} + \rho|^2)}
\,.
$$
To compute the exponent, let $\alpha_1$ and $\alpha_2$ be simple roots of $\sl(3,\CC)$. Then,
$$
\rho = \alpha_1 + \alpha_2
\,,\qquad
\lambda_{n,m} = \frac{2n+m}{3} \, \alpha_1 + \frac{n+2m}{3} \, \alpha_2
$$
and hence
$$
|\lambda_{n,m} + \rho|^2
=
\frac{(n+1)^2 + (m+1)^2 + (n+1)(m+1)}{3} \, |\alpha_1|^2
\,,
$$
where we have used that $|\alpha_1| = |\alpha_2|$ and $\frac{2 \langle \alpha_1 , \alpha_2 \rangle}{|\alpha_1|^2} = -1$ (Cartan matrix). Since $\sl(3,\CC)$ is simple, the invariant scalar product in $\su(3)$ can be written in the form
$$
\langle X,Y \rangle = - \frac{1}{2\beta^2} \tr(XY)
$$
with a scaling parameter $\beta > 0$. Then,
$$
|\alpha_1|^2 = 4 \beta^2
\,,
$$
so that, altogether, we obtain 
\beq\label{G-SU3-N2-Norm}
\left\|\big(\hat\chi^\CC_{n_1m_1,n_2m_2,nm}\big)^{k'}_k\right\|
=
\sqrt{\frac{d_{n,m}}{d_{n_1,m_1} d_{n_2,m_2}}}
\,
(\hbar \pi)^4 
\, 
\mr e^{\hbar \beta^2 (\zeta_{n_1,m_1} + \zeta_{n_2,m_2})}
\eeq
with
$$
\zeta_{n,m} 
:= 
\frac 2 3 \, \left((n+1)^2 + (m+1)^2 + (n+1)(m+1)\right)
\,.
$$
According to \cite{Coleman}, the Clebsch-Gordan series for twofold tensor products reads
\al{\nonumber
D^{n,m} \otimes D^{n',m'}
=
\bigoplus_{i=0}^{\min(n,m')}
\,
\bigoplus_{j=0}^{\min(n',m)}
\Bigg( &
D^{n+n'-i-j,m+m'-i-j}
\\ \nonumber
& 
\oplus\!\!\!
\bigoplus_{k=1}^{\min(n-i,n'-j)}
D^{n+n'-i-j-2k,m+m'-i-j+k}
\\ \label{G-SU3-CGS}
&
\oplus\!\!\!
\bigoplus_{k=1}^{\min(m-j,m'-i)}
D^{n+n'-i-j+k,m+m'-i-j-2k}
\Bigg)
.
}
For the ordinary Clebsch-Gordan coefficients, we introduce the shorthand notation
$$
\ocg{n_1m_1}{n_2m_2}{nm}{k}{\mu_1}{\mu_2}{\mu}
:=
\ocg{\lambda_{n_1,m_1}}{\lambda_{n_2,m_2}}{\lambda_{n,m}}{k}{\mu_1}{\mu_2}{\mu}
\,,
$$
with the convention that $k$ can be omitted whenever the multiplicity is $1$. For the computation of these coefficients, see \cite{AKHD}. 

Now, we can express $r_1$ in terms of the modified quasicharacters. Putting
$$
t_1(a_1,a_2) := \tr\big((a_1a_2)^2\big)
\,,\qquad
t_2(a_1,a_2) := \tr(a_1^2 a_2^2)
\,,
$$
we can write
$$
r_1 := t_1 - t_2
\,.
$$
For $t_1$, we find 
\ala{
t_1(a_1,a_2)
& =
\sum_{\mu_i \in \weight(\lambda_{1,0})}
D^{1,0}_{\mu_1\mu_2}(a_1) 
D^{1,0}_{\mu_2\mu_3}(a_2) 
D^{1,0}_{\mu_3\mu_4}(a_1) 
D^{1,0}_{\mu_4\mu_1}(a_2)
\\
& =
\sum_{\mu_i \in \weight(\lambda_{1,0})}
\bigbra{\bra{\lambda_{1,0},\mu_1} \otimes \bra{\lambda_{1,0},\mu_3}}
D^{1,0}(a_1) \otimes D^{1,0}(a_1)
\bigket{\ket{\lambda_{1,0},\mu_2} \otimes \ket{\lambda_{1,0},\mu_4}}
\\
& \hspace{2cm}
\cdot
\bigbra{\bra{\lambda_{1,0},\mu_2} \otimes \bra{\lambda_{1,0},\mu_4}}
D^{1,0}(a_2) \otimes D^{1,0}(a_2)
\bigket{\ket{\lambda_{1,0},\mu_3} \otimes \ket{\lambda_{1,0},\mu_1}}
\\
& =
\sum_{\mu_i \in \weight(\lambda_{1,0})}
~
\sum_{\lambda_i = \lambda_{2,0},\lambda_{0,1}}
~
\sum_{\nu_i,\nu_i' \in \weight(\lambda_i)}
\ocgs{10}{10}{\lambda_1}{\mu_1}{\mu_3}{\nu_1}
\,
\ocgs{10}{10}{\lambda_1}{\mu_2}{\mu_4}{\nu_1'}
\,
\ocgs{10}{10}{\lambda_2}{\mu_2}{\mu_4}{\nu_2}
\,
\ocgs{10}{10}{\lambda_2}{\mu_3}{\mu_1}{\nu_2'}
\\
& \hspace{9.25cm}
\cdot 
D^{\lambda_1}_{\nu_1\nu_1'}(a_1) 
\,
D^{\lambda_2}_{\nu_2\nu_2'}(a_2)
\,,
}
where we have used that the ordinary Clebsch-Gordan coefficients can be chosen to be real. We have 
$$
\ocgs{10}{10}{01}{\mu_2}{\mu_1}{\mu'}
=
-\ocgs{10}{10}{01}{\mu_1}{\mu_2}{\mu'}
\,,\qquad
\ocgs{10}{10}{20}{\mu_2}{\mu_1}{\mu}
=
\ocgs{10}{10}{20}{\mu_1}{\mu_2}{\mu}
\,.
$$
Thus, by orthogonality,
\ala{
\sum_{\mu_i \in \weight(\lambda_{1,0})}
\ocgs{10}{10}{20}{\mu_1}{\mu_3}{\mu}
\ocgs{10}{10}{20}{\mu_2}{\mu_4}{\nu}
\ocgs{10}{10}{20}{\mu_2}{\mu_4}{\mu'}
\ocgs{10}{10}{20}{1}{\mu_3}{\mu_1}{\nu'}
& =
\delta_{\mu,\nu'} \, \delta_{\nu,\mu'}
\,,
\\
\sum_{\mu_i \in \weight(\lambda_{1,0})}
\ocgs{10}{10}{01}{\mu_3}{\mu_1}{\mu}
\ocgs{10}{10}{01}{\mu_2}{\mu_4}{\nu}
\ocgs{10}{10}{01}{\mu_2}{\mu_4}{\mu'}
\ocgs{10}{10}{01}{\mu_3}{\mu_1}{\nu'}
& =
- \delta_{\mu,\nu'} \, \delta_{\nu,\mu'}
\,,
}
whereas the mixed terms vanish. Hence,
\beq\label{G-SU3-N2-t1}
t_1(a_1,a_2)
=
\sum_{\mu,\nu \in \weight(\lambda_{2,0})} D^{2,0}_{\mu\nu}(a_1) D^{2,0}_{\nu\mu}(a_2)
-
\sum_{\mu,\nu \in \weight(\lambda_{0,1})} D^{0,1}_{\mu\nu}(a_1) D^{0,1}_{\nu\mu}(a_2)
\,.
\eeq
To express $t_1$ in terms of modified complex quasicharacters, we compute the expansion coefficients with respect to the basis of complex quasicharacters. According to Corollary \rref{F-ReprF}, it suffices to compute the scalar products of the restriction of $t_1$ to $G^2$ with the normalized real quasicharacters. According to point \rref{i-S-CCG-qc} of Proposition \rref{S-CCG},
\ala{
\big(\chi_{n_1m_1,n_2m_2,nm}\big)^{k'}_k(a_1,a_2)
& =
\sqrt{\frac{d_{n_1,m_1} \, d_{n_2,m_2}}{d_{n,m}}}
\sum_{\mu_i,\mu_i' \in \weight(\lambda_{n_i,m_i})}
\sum_{\mu \in \weight(\lambda_{n,m})}
\ocg{n_1m_1}{n_2m_2}{nm}{k}{\mu_1}{\mu_2}{\mu}
\\
& \hspace{3cm}
\cdot
\ocg{n_1m_1}{n_2m_2}{nm}{k'}{\mu_1'}{\mu_2'}{\mu}
D^{n_1,m_1}_{\mu_1'\mu_1}(a_1) D^{n_2,m_2}_{\mu_2'\mu_2}(a_2)
\,.
}
Using 
$$
\bigbraket{D^{n_1,m_1}_{\mu_1\nu_1}}{D^{n_2,m_2}_{\mu_2\nu_2}}_{L^2(G^2)}
=
\frac{1}{d_{n_1,m_1}} 
\, 
\delta_{n_1,n_2} \, \delta_{m_1,m_2} 
\, 
\delta_{\mu_1,\mu_2} \, \delta_{\nu_1,\nu_2}
\,,
$$
we obtain
\ala{
\bigbraket{\big(\chi_{n_1m_1,n_2m_2,nm}\big)^{k'}_k}{t_1{}_{\res G^2}}
& =
\frac
{\delta_{n_1,2} \delta_{m_1,0} \delta_{n_2,2} \delta_{m_1,0}}
{d_{2,0} \sqrt{d_{n,m}}}
\,
\sum_{\mu,\mu' \in \weight(\lambda_{2,0}) \atop \nu \in \weight(\lambda_{n,m})}
\ocg{20}{20}{nm}{k}{\mu'}{\mu}{\nu}
\ocg{20}{20}{nm}{k'}{\mu}{\mu'}{\nu}
\\
& \phantom{=}
-
\frac
{\delta_{n_1,0} \delta_{m_1,1} \delta_{n_2,0} \delta_{m_1,1}}
{d_{0,1} \sqrt{d_{n,m}}}
\,
\sum_{\mu,\mu' \in \weight(\lambda_{0,1}) \atop \nu \in \weight(\lambda_{n,m})}
\ocg{01}{01}{n,m}{k}{\mu'}{\mu}{\nu}
\ocg{01}{01}{n,m}{k'}{\mu}{\mu'}{\nu}
\,.
}
By \eqref{G-SU3-CGS}, we have
\beq\label{G-SU3-N2-Zlg}
D^{2,0} \otimes D^{2,0} = D^{4,0} \oplus D^{2,1} \oplus D^{0,2}
\,,\qquad
D^{0,1} \otimes D^{0,1} = D^{0,2} \oplus D^{1,0}
\,.
\eeq
Hence, $k,k'=1$ and the relevant values of $(n,m)$ are $(4,0)$, $(2,1)$ and $(0,2)$ for the first sum and $(0,2)$ and $(1,0)$ for the second sum. Using 
\ala{
\ocgs{20}{20}{40}{\mu'}{\mu}{\nu}
& =
\ocgs{20}{20}{40}{\mu}{\mu'}{\nu}
\,, &
\ocgs{20}{20}{21}{\mu'}{\mu}{\nu}
& =
-\ocgs{20}{20}{21}{\mu}{\mu'}{\nu}
\,, &
\ocgs{20}{20}{02}{\mu'}{\mu}{\nu}
& =
\ocgs{20}{20}{02}{\mu}{\mu'}{\nu}
\\
\ocgs{01}{01}{02}{\mu'}{\mu}{\nu}
& =
\ocgs{01}{01}{02}{\mu}{\mu'}{\nu}
\,, &
\ocgs{01}{01}{10}{\mu'}{\mu}{\nu}
& =
\ocgs{01}{01}{10}{\mu}{\mu'}{\nu}
}
and taking the sums over weights, we finally arrive at 
$$
t_1
=
\hat\chi^\CC_{20,20,40}
- \hat\chi^\CC_{20,20,21}
+ \hat\chi^\CC_{20,20,02}
- \hat\chi^\CC_{01,01,02}
+ \hat\chi^\CC_{01,01,10}
\,.
$$
A similar calculation leads to 
\newpage
\ala{
t_2(a_1,a_2)
& =
\sum_{\mu_i \in \weight(\lambda_{1,0})}
\sum_{\lambda_i = \lambda_{2,0},\lambda_{0,1}}
\sum_{\nu_i,\nu_i' \in \weight(\lambda_i)}
\ocgs{10}{10}{\lambda_1}{\mu_1}{\mu_2}{\nu_1}
\ocgs{10}{10}{\lambda_1}{\mu_2}{\mu_3}{\nu_1'}
\ocgs{10}{10}{\lambda_2}{\mu_3}{\mu_4}{\nu_2}
\ocgs{10}{10}{\lambda_2}{\mu_4}{\mu_1}{\nu_2'}
\\
& \hspace{9cm}
D^{\lambda_1}_{\nu_1\nu_1'}(a_1)
D^{\lambda_2}_{\nu_2\nu_2'}(a_2)
}
and
\ala{
\bigbraket{\big(\chi_{n_1m_1,n_2m_2,nm}\big)^{k'}_k}{t_2{}_{\res G^2}}
& =
\sum_{\lambda_i = \lambda_{2,0},\lambda_{0,1}}
\frac
{\delta_{\lambda_1,\lambda_{n_1,m_1}} \, \delta_{\lambda_2,\lambda_{n_2,m_2}}}
{\sqrt{d_{\lambda_1} \, d_{\lambda_2} \, d_{n,m}}}
\,
\sum_{\mu_i \in \weight(\lambda_{1,0})}
\,
\sum_{\nu \in \weight(\lambda_{n,m})}
\,
\sum_{\nu_i,\nu_i' \in \weight(\lambda_i)}
\\
 & \hspace{1.5cm}
\ocgs{10}{10}{\lambda_1}{\mu_1}{\mu_2}{\nu_1}
\ocgs{10}{10}{\lambda_1}{\mu_2}{\mu_3}{\nu_1'}
\ocgs{10}{10}{\lambda_2}{\mu_3}{\mu_4}{\nu_2}
\ocgs{10}{10}{\lambda_2}{\mu_4}{\mu_1}{\nu_2'}
\ocgs{\lambda_1}{\lambda_2}{nm}{\nu_1}{\nu_2}{\nu}
\ocgs{\lambda_1}{\lambda_2}{nm}{\nu_1'}{\nu_2'}{\nu}
\,.
}
Here, in addition to \eqref{G-SU3-N2-Zlg}, the Clebsch-Gordan series 
$
D^{2,0} \otimes D^{0,1} = D^{2,1} \oplus D^{1,0}
$
occurs, showing that, still, $k,k'=1$ throughout. Evalution of the sums using the online calculator \cite{AKHD} yields 
$$
t_2
=
\hat\chi^\CC_{20,20,40}
- \frac 1 2 \hat\chi^\CC_{20,20,02}
- \frac 1 2 \hat\chi^\CC_{20,01,21} 
- \frac 1 2 \hat\chi^\CC_{01,20,21}
+ \frac 1 2 \hat\chi^\CC_{20,01,10}
+ \frac 1 2 \hat\chi^\CC_{01,20,10}
+ \frac 1 2 \hat\chi^\CC_{01,01,02}
$$
so that, altogether,
\al{\nonumber
p
& =
- \hat\chi^\CC_{20,20,21}
+ \frac 3 2 \, \hat\chi^\CC_{20,20,02}
+ \frac 1 2 \left(\hat\chi^\CC_{20,01,21} + \hat\chi^\CC_{01,20,21}\right)
\\ \label{G-SU3-N2-p}
& \hspace{3cm}
- \frac 1 2 \left(\hat\chi^\CC_{20,01,10} + \hat\chi^\CC_{01,20,10}\right)
- \frac 3 2 \, \hat\chi^\CC_{01,01,21}
+ \frac 3 2 \, \hat\chi^\CC_{01,01,10}
\,.
}
To illustrate the computation of the matrix elements of $\hat r_1^\CC$, we restrict attention to the case where $(r_1,s_1) = (r,s) = (0,1)$ and $(r_2,s_2) = (0,0)$, thus deriving a specific row of the matrix. First, we compute the coefficients $M$ in \eqref{G-MEK}. As an example, consider the contribution of the term $\hat\chi^\CC_{20,20,21}$ to \eqref{G-SU3-N2-p}. According to Corollary \rref{TreeProductRule},
$$
\hat\chi^\CC_{20,20,21} \cdot \hat\chi^\CC_{01,00,01}
=
\sum
R(\boldsymbol\vee)^{(20,20,21),(01,00,01),k}_{(n_1m_1,n_2m_2,nm,l'),\ul k}
\big(\hat\chi^\CC_{n_1m_1,n_2m_2,nm}\big)^{l'}_l
R(\boldsymbol\vee)^{(n_1m_1,n_2m_2,nm,l),\ul k}_{(20,20,21),(01,00,01),k}
\,,
$$
where the sum is taken over $n_1,m_1,n_2,m_2,n,m,\ul k,k$ satisfying 
\ala{
(\lambda_{n_1,m_1},k^1) & \in \llangle \lambda_{2,0},\lambda_{0,1} \rrangle
\,, &
(\lambda_{n_2,m_2},k^2) \in \llangle \lambda_{2,0},\lambda_{0,0} \rrangle
\,,
\\
(\lambda_{n,m},k) & \in \llangle \lambda_{2,1},\lambda_{0,1} \rrangle
\,, &
(\lambda_{n,m},l) , (\lambda_{n,m},l') \in \llangle \lambda_{n_1,m_1},\lambda_{n_2,m_2} \rrangle
\,.
}
Clearly, the second condition yields $(n_2,m_2)=(2,0)$ with $k^2=1$. Since 
$$
D^{2,0} \otimes D^{0,1} = D^{2,1} \oplus D^{1,0}
\,,
$$
the first condition implies that $(n_1,m_1) = (2,1)$ or $(1,0)$ with $k^1=1$. Since 
$$
D^{2,1} \otimes D^{0,1} = D^{2,2} \oplus D^{3,0} \oplus D^{1,1}
\,,
$$
the third condition implies $(n,m) = (2,2)$, $(3,0)$ or $(1,1)$ with $k=1$. Then, due to 
$$
D^{2,1} \otimes D^{2,0}
= 
D^{4,1} \oplus D^{2,2} \oplus D^{3,0} \oplus D^{0,3} \oplus D^{1,1}
\,,\qquad
D^{1,0} \otimes D^{2,0} = D^{3,0} \oplus D^{1,1}
\,,
$$
the last condition implies that $l=l'=1$ and that $(n_1m_1,n_2m_2,nm)$ can take the values 
$$
(21,20,22)
\,,\qquad
(21,20,30)
\,,\qquad
(21,20,11)
\,,\qquad
(10,20,30)
\,,\qquad
(10,20,11)
\,.
$$
Since in addition $R(\boldsymbol\vee)$ is real and thus coincides with its transpose, we obtain
\beq\label{G-SU3-N2-202021}
\hat\chi^\CC_{20,20,21} \cdot \hat\chi^\CC_{01,00,01}
=
\sum_{(n_1m_1,nm)} 
W_{n_1m_1,nm}^2 \,\, \hat\chi^\CC_{n_1m_1,20,nm}
\,,
\eeq
where the sum runs over the tuples
$$
(n_1m_1,nm)
=
(21,22) ,~ (21,30) ,~ (21,11) ,~ (10,30) ,~ (10,11)
\,.
$$
Here, according to Theorem \rref{T-R}, we have expressed $R(\boldsymbol\vee)$ in terms of the $9\lambda$ symbols 
\beq\label{G-SU3-W}
W_{n_1m_1,nm}
=
\bpma
\lambda_{2,0} & \lambda_{2,0} & \lambda_{2,1} & 1
\\
\lambda_{0,1} & \lambda_{0,0} & \lambda_{0,1} & 1
\\
\lambda_{n_1,m_1} & \lambda_{2,0} & \lambda_{n,m} & 1
\\
1 & 1 & 1
\epma
\,.
\eeq
According to Remark \rref{Bem-CCG}, the $9\lambda$ symbols can be expressed, in turn, in terms of ordinary Clebsch-Gordan coefficients. Since the latter are real and since
$$
\ocg{\lambda}{\lambda_{0,0}}{\lambda}{1}{\mu_1}{\mu_2}{\mu_3}
=
\delta_{\mu_1,\mu_3}
\,,
$$
we obtain
$$
W_{n_1m_1,nm}
=
\sum_{\mu_1^1,\mu_1^2,\mu_1^\rt,\mu_3^1,\mu_2^1}
\ocgs{20}{20}{21}{\mu_1^1}{\mu_1^2}{\mu_1^\rt}
\,
\ocgs{21}{01}{nm}{\mu_1^\rt}{\mu_2^1}{\mu_3^\rt}
\,
\ocgs{20}{01}{n_1m_1}{\mu_1^1}{\mu_2^1}{\mu_3^1}
\,
\ocgs{n_1m_1}{20}{nm}{\mu_3^1}{\mu_1^2}{\mu_3^\rt}
\,,
$$
for any given $\mu_3^\rt \in \weight(\lambda_{n,m})$. The ranges of the summation variables are 
$$
\mu_1^1,\mu_1^2 \in \weight(\lambda_{2,0})
\,,\qquad
\mu_1^\rt \in \weight(\lambda_{2,1})
\,,\qquad
\mu_3^1 \in \weight(\lambda_{n_1,m_1})
\,,\qquad
\mu_2^1 \in \weight(\lambda_{0,1})
\,.
$$
Using the tables of isoscalar factors in \cite{Kaeding} and the tabulated values of $\SU(2)$ Clebsch-Gordan coefficients to compute the Clebsch-Gordan coefficients in \eqref{G-SU3-W}, we find 
$$
W_{21,22} = 1
\,,\quad
W_{21,30} = - \frac{\sqrt{3}}{2}
\,,\quad
W_{21,11} = - \frac{\sqrt{6}}{4}
\,,\quad
W_{10,30} = \frac 1 2 
\,,\quad
W_{10,11} = \frac{\sqrt{10}}{4}
\,.
$$
This can also be checked using the online calculator \cite{AKHD}. Thus, 
$$
\hat\chi^\CC_{20,20,21} \cdot \hat\chi^\CC_{01,00,01}
=
\hat\chi^\CC_{21,20,22}
- \frac{\sqrt{3}}{2} \, \hat\chi^\CC_{21,20,30}
- \frac{\sqrt{6}}{4} \, \hat\chi^\CC_{21,20,11}
+ \frac 1 2 \, \hat\chi^\CC_{10,20,30}
+ \frac{\sqrt{10}}{4} \, \hat\chi^\CC_{10,20,11}
\,.
$$
The corresponding contribution to the factor $M$ in \eqref{G-MEK} can be obtained from the right hand side by replacing the modified quasicharacters by appropriate Kronecker deltas. By analogy, we determine the contributions of the other terms in \eqref{G-SU3-N2-p}. Putting all of this together, and computing the norm ratios using \eqref{G-SU3-N2-Norm}, we finally arrive at the following formula for the elements of the row of $\hat r_1^\CC$ under consideration:
\ala{
(\hat r_1^\CC & )^{n_1m_1,n_2m_2,nm,k'k}_{01,00,01}
\\
= &
- \frac{3}{\sqrt{30}} \, z^{11} \, \delta_{(),(21,20,22)}
- \frac{1}{4} \, z^{11} \, \delta_{(),(21,20,30)}
+ \frac{1}{2} \, z^{11} \, \delta_{(),(21,20,03)}
+ \frac{3}{8\sqrt{5}} \, z^{11} \, \delta_{(),(21,20,11)}
\\
& 
+ \frac{3}{2\sqrt{15}} \, z^8 \, \delta_{(),(21,01,22)}
+ \frac{1}{\sqrt{6}} \, z^8 \, \delta_{(),(21,01,30)}
- \frac{7}{12\sqrt{10}} \, z^8 \, \delta_{(),(21,01,11)}
+ \frac{\sqrt{3}}{4} \, z^8 \, \delta_{(),(02,20,22)}
\\
& 
- \frac{1}{12\sqrt{2}} \, z^8 \, \delta_{(),(02,20,11)}
- \frac{1}{12} \, z^8 \, \delta_{(),(02,20,00)}
+ \frac{\sqrt{5}}{12} \, z^5 \, \delta_{(),(10,20,30)}
+ \frac{1}{24} \, z^5 \, \delta_{(),(10,20,11)}
\\
& 
- \frac{\sqrt{5}}{2} \, z^5 \, \delta_{(),(02,01,03)}
+ \frac{1}{4} \, z^5 \, \delta_{(),(02,01,11)}
- \frac{7}{12\sqrt{2}} \, z^2 \, \delta_{(),(10,01,11)}
+ \frac{1}{6} \, z^2 \, \delta_{(),(10,01,00)}
\,.
}
Here, 
$$
z = \mr e^{4 \hbar \beta^2 / 3}
\,.
$$
and the double bracket $()$ in the Kronecker delta symbols is a shorthand notation for $(n_1m_1,n_2m_2,nm)$.


\section{Outlook}
\label{outlook}


For future work, the following tasks will be interesting.
\ben
\item
There is a deep relation between rooted binary trees and trivalent graphs \cite{BL1,Yutsis}. It would be interesting to study whether our methods may be reformulated 
in terms of trivalent graph theory.

\item
For a systematic investigation of $G= \SU(3)$, it seems reasonable to start with continuing the analysis of the strata subspaces and their orthoprojectors for $N=2$, the simplest situation where all the strata are present.

\item
In future work, the spectral problem of the quantum Hamiltonian for both $G = \SU(2)$ and  $G= \SU(3)$ should be 
studied along the lines explained above.  
\een

\section*{Acknowledgements}

M.S.\ acknowledges funding by Deutsche Forschungsgemeinschaft under the grant SCHM 1652/2-1.

\section*{Data availability statement}

The data that support the findings of this study are available from the corresponding author upon reasonable request.


\begin{appendix}
\section{A direct proof of Corollary \ref{F-TPR}}
\label{sec:Appendix}
\renewcommand{\theequation}{A--\arabic{equation}}


Here, we prove this corollary by means of the composite Clebsch-Gordan coefficients. 

Let $T$ be a coupling tree. For simplicity, we assume that the leaves of $T$ are numbered $1,\dots,N$. For $i=1,2,3$, let $\ul\lambda_i$ be a leaf labelling of $T$, let $\lambda_i \in \langle \ul\lambda_i \rangle$ and let $\alpha_i \in \mc L^T(\ul\lambda_i,\lambda_i)$. Let $\ul k$ be an assignment of a positive integer to every leaf of $T$ and let $k$ be a positive integer. Assume that $(\alpha_3,\ul k,k) \in \llangle \alpha_1,\alpha_2 \rrangle$. 

Under the identification $H_{\ul\lambda_1 \!\cdot\! \ul\lambda_2} = H_{\ul\lambda_1} \otimes H_{\ul\lambda_2}$, the vectors $\ket{T \bcdot T ; \join{\alpha_1}{\alpha_2}{(\lambda,k)},\mu}$, where $(\lambda,k) \in \llangle \lambda_1,\lambda_2 \rrangle$ and $\mu \in \weight(\lambda)$, can be expanded with respect to the tensor product vectors 
$\ket{T;\alpha_1,\mu_1} \otimes \ket{T;\alpha_2,\mu_2}$, where $\mu_i \in \weight(\lambda_i)$, and vice versa:
\al{\nonumber
\bigket{T \bcdot T ; \join{\alpha_1}{\alpha_2}{(\lambda,k)},\mu}
& =  
\sum_{\mu_1\in\weight(\lambda_1)}
\,
\sum_{\mu_2\in\weight(\lambda_2)}
\ocg{\lambda_1}{\lambda_2}{\lambda}{k}{\mu_1}{\mu_2}{\mu}
\ket{T;\alpha_1,\mu_1} \otimes \ket{T;\alpha_2,\mu_2}
\,,
\\ \label{TreeJoinStates}
\ket{T;\alpha_1,\mu_1} \otimes \ket{T;\alpha_2,\mu_2}
& = 
\sum_{(\lambda,k)}
\,
\sum_{\check\mu=1}^{m_\lambda(\hat\mu)}
\big(\ocg{\lambda_1}{\lambda_2}{\lambda}{k}{\mu_1}{\mu_2}{\mu}\big)^\ast
\bigket{T \bcdot T;\join{\alpha_1}{\alpha_2}{(\lambda,k)},\mu}
\,,
}
where $\hat\mu:=\hat\mu_1+\hat\mu_2$ and $\mu:=(\hat\mu,\check\mu)$ and where the sum is over $(\lambda,k) \in \llangle \lambda_1,\lambda_2 \rrangle$ such that $\hat\weight(\lambda)$ contains $\hat\mu$. Similarly, under the identification 
$$
H_{\ul\lambda_1 \!\ast\! \ul\lambda_2} 
= 
\bigotimes_{n=1}^N
\big(H_{\lambda_1^n} \otimes H_{\lambda_2^n}\big)
\,,
$$
the vectors $\bigket{T^\vee;\ld{\alpha_1}{\alpha_2}{\alpha_3,\ul k},\mu}$, where $\mu \in \weight(\lambda_3)$, can be expanded with respect to the vectors 
$$
\bigket{\boldsymbol\vee;\big(\lambda_1^1,\lambda_2^1,(\lambda_3^1,k^1)\big) , \mu^1}
\otimes \cdots \otimes
\bigket{\boldsymbol\vee;\big(\lambda_1^N,\lambda_2^N,(\lambda_3^N,k^N)\big) , \mu^N}
\,,
$$
where $\mu^n \in \weight(\lambda_3^n)$ for all $n$. Here, $\big(\lambda_1^n,\lambda_2^n,(\lambda_3^n,k^n)\big)$ stands for the labelling of $\boldsymbol\vee$ assigning $\lambda_1^n$, $\lambda_2^n$ to the leaves and $(\lambda_3^n,k^n)$ to the root. To find the expansion coefficients, we first expand with respect to the tensor product basis,
\beq\label{G-Zlg-Dupl}
\ket{T^\vee;\ld{\alpha_1}{\alpha_2}{\alpha_3},\mu}
= 
\sum_{\ul\mu_1\in\weight(\lambda_1)}
\,
\sum_{\ul\mu_2\in\weight(\lambda_2)}
\ccg{T^\vee}{\ld{\alpha_1}{\alpha_2}{\alpha_3}}{\ul\mu_1 \!\ast\! \ul\mu_2}{\mu}
\,\,
\bigket{\ul\lambda_1 \!\ast\! \ul\lambda_2 \ \ul\mu_1 \!\ast\! \ul\mu_2}
\,.
\eeq
In view of \eqref{eq:CompositeCG}, we may factorize the composite Clebsch-Gordan coefficient as 
$$
\ccg{T^\vee}{\ld{\alpha_1}{\alpha_2}{\alpha_3}}{\ul\mu_1 \!\ast\! \ul\mu_2}{\mu}
=
\sum_{\check\mu_3^1=1}^{m_{\lambda_3^1}(\hat\mu_3^1)}
\cdots
\sum_{\check\mu_3^N=1}^{m_{\lambda_3^N}(\hat\mu_3^N)}
\ccg{T}{\alpha_3}{\ul\mu_3}{\mu}
\,
\prod_{n=1}^N
\ocg{\lambda^n_1}{\lambda^n_2}{\lambda^n_3}{k^n}{\mu_1^n}{\mu_2^n}{\mu_3^n}
\,,
$$
where $\hat\mu_3^n:=\hat\mu_1^n+\hat\mu_2^n$ and $\mu_3^n:=(\hat\mu_3^n,\check\mu_3^n)$. Decomposing the sum accordingly, from \eqref{G-Zlg-Dupl} we obtain
\ala{
&
\ket{T^\vee;\ld{\alpha_1}{\alpha_2}{\alpha_3},\mu}
\\
& \hspace{1cm}
= 
\sum_{\ul\mu_3\in\weight(\ul\lambda_3)} 
\ccg{T}{\alpha_3}{\ul\mu_3}{\mu}
\left(
\bigotimes_{n=1}^N
\left(
\sum_{\mu_1^n\in\weight(\lambda_1^n)}
\,
\sum_{\mu_2^n\in\weight(\lambda_2^n)}
\,
\ocg{\lambda^n_1}{\lambda^n_2}{\lambda^n_3}{k^n}{\mu_1^n}{\mu_2^n}{\mu_3^n}
~
\ket{\lambda_1^n \ \mu_1^n} \otimes \ket{\lambda_2^n \ \mu_2^n}
\right)
\right)
.
}
Since
$$
\sum_{\mu_1^n\in\weight(\lambda_1^n)}
\,
\sum_{\mu_2^n\in\weight(\lambda_2^n)}
\,
\ocg{\lambda^n_1}{\lambda^n_2}{\lambda^n_3}{k^n}{\mu_1^n}{\mu_2^n}{\mu_3^n}
~
\ket{\lambda_1^n \ \mu_1^n} \otimes \ket{\lambda_2^n \ \mu_2^n}
=
\bigket{\boldsymbol\vee;\big(\lambda_1^n,\lambda_2^n,(\lambda_3^n,k^n)\big) , \mu_3^n}
\,,
$$
we may rewrite
\al{\nonumber
\ket{T^\vee;\ld{\alpha_1}{\alpha_2}{\alpha_3},\mu}
& = 
\sum_{\ul\mu_3 \in \weight(\ul\lambda_3)} 
\ccg{T}{\alpha_3}{\ul\mu_3}{\mu}
\,\,
\bigket{\boldsymbol\vee;\big(\lambda_1^1,\lambda_2^1,(\lambda_3^1,k^1)\big) , \mu_3^1}
\otimes \cdots 
\\ \label{LeafDuplicateStates}
& \hspace{3.5cm}
\cdots \otimes
\bigket{\boldsymbol\vee;\big(\lambda_1^N,\lambda_2^N,(\lambda_3^N,k^N)\big) , \mu_3^N}
\,.
}
Now, we use \eqref{Form-chi}, \eqref{TreeJoinStates} and \eqref{G-D-U} to calculate
\ala{
&
\dcq{T}{\alpha_1^{\phantom{\prime}}}{\alpha_1'}(\ul a)
\cdot
\dcq{T}{\alpha_2^{\phantom{\prime}}}{\alpha_2'}(\ul a)
\\
& = 
\sum_{\mu_1\in\weight(\lambda_1)}
\,
\sum_{\mu_2\in\weight(\lambda_2)}
\bigbra{T;\alpha_1',\mu_1} D^{\ul\lambda_1}(\ul a) \bigket{T;\alpha_1,\mu_1}
\,
\bigbra{T;\alpha_2',\mu_2} D^{\ul\lambda_2}(\ul a) \bigket{T;\alpha_2,\mu_2}
\\
& =
\sum_{\mu_1\in\weight(\lambda_1)}
\,
\sum_{\mu_2\in\weight(\lambda_2)}
\bigbra{\,
\bra{T;\alpha_1',\mu_1} \otimes \bra{T;\alpha_2',\mu_2}
\,}
D^{\ul\lambda_1\!\cdot\!\ul\lambda_2}(\ul a)
\bigket{\,
\ket{T;\alpha_1,\mu_1} \otimes \ket{T;\alpha_2,\mu_2}
\,}
\\
& =
\sum_{\mu_1\in\weight(\lambda_1)}
\,
\sum_{\mu_2\in\weight(\lambda_2)}
\,
\sum_{(\lambda,k) , (\lambda',k')}
\,
\sum_{\check\mu=1}^{m_\lambda(\hat\mu_1+\hat\mu_2)}
\,
\sum_{\check\mu'=1}^{m_{\lambda'}(\hat\mu_1+\hat\mu_2)}
\ocg{\lambda_1}{\lambda_2}{\lambda'}{k'}{\mu_1}{\mu_2}{(\hat\mu_1+\hat\mu_2,\check\mu')}
\,
\big(
\ocg{\lambda_1}{\lambda_2}{\lambda}{k}{\mu_1}{\mu_2}{(\hat\mu_1+\hat\mu_2,\check\mu)}
\big)^\ast
\\
& \hspace{0.75cm}
\cdot 
\Bigbra{
T \bcdot T;\join{\alpha_1'}{\alpha_2'}{(\lambda',k')},(\hat\mu_1+\hat\mu_2,\check\mu')
}
D^{\ul\lambda_1 \!\cdot\! \ul\lambda_2} (\ul a)
\Bigket{
T \bcdot T;\join{\alpha_1}{\alpha_2}{(\lambda,k)},(\hat\mu_1+\hat\mu_2,\check\mu)
}
\\
& =
\sum_{(\lambda,k),(\lambda',k')\in\llangle\lambda_1,\lambda_2\rrangle}
\,
\sum_{\mu\in\weight(\lambda)}
\,
\sum_{\mu'\in\weight(\lambda')}
\,
\sum_{\mu_1\in\weight(\lambda_1)}
\,
\sum_{\mu_2\in\weight(\lambda_2)}
\,
\ocg{\lambda_1}{\lambda_2}{\lambda'}{k'}{\mu_1}{\mu_2}{\mu'}
\,
\big(\ocg{\lambda_1}{\lambda_2}{\lambda}{k}{\mu_1}{\mu_2}{\mu}\big)^\ast
\\
& \hspace{0.75cm}
\cdot
\Bigbra{
T \bcdot T;\join{\alpha_1'}{\alpha_2'}{(\lambda',k')},\mu'
}
\Pi D^{\ul\lambda_1 \!\ast\! \ul\lambda_2} (\ul a) \Pi^{-1}
\Bigket{
T \bcdot T;\join{\alpha_1}{\alpha_2}{(\lambda,k)},\mu
}
\\
& =
\sum_{\lambda,k,k'}
\,
\sum_{\mu\in\weight(\lambda)}
\Bigbra{T \bcdot T;\join{\alpha_1'}{\alpha_2'}{(\lambda,k)},\mu}
\Pi D^{\ul\lambda_1 \!\ast\! \ul\lambda_2} (\ul a) \Pi^{-1}
\Bigket{T \bcdot T;\join{\alpha_1}{\alpha_2}{(\lambda,k)},\mu}
\,,
}
where the sum is over $\lambda\in\langle\lambda_1,\lambda_2\rangle$ and $k=1,\dots,m_{(\lambda_1,\lambda_2)}(\lambda)$. In the last step, we have used 
$$
\sum_{\mu_1\in\weight(\lambda_1)}
\,
\sum_{\mu_2\in\weight(\lambda_2)}
\,
\ocg{\lambda_1}{\lambda_2}{\lambda'}{k'}{\mu_1}{\mu_2}{\mu'}
\,
\big(\ocg{\lambda_1}{\lambda_2}{\lambda}{k}{\mu_1}{\mu_2}{\mu}\big)^\ast
=
\delta_{\lambda\lambda'} \, \delta_{kk'} \, \delta_{\mu\mu'}
\,.
$$
Inserting the unit operator 
$$
\unit
=
\sum_{\alpha_3,\ul k} \sum_{\mu_3}
\bigket{T^\vee;\ld{\alpha_1}{\alpha_2}{\alpha_3,\ul k},\mu_3}
\bigbra{T^\vee;\ld{\alpha_1}{\alpha_2}{\alpha_3,\ul k},\mu_3}
$$
twice and using the relations
\begin{align*}
\bigbra{T \!\cdot\! T;\join{\alpha_1'}{\alpha_2'}{(\lambda,k)},\mu}
\Pi
\bigket{T^\vee;\ld{\alpha_1'}{\alpha_2'}{\alpha_3',\ul k'},\mu_3'}
= &
\delta_{\lambda,\lambda_3'} \, \delta_{\mu,\mu_3'}
\,
\rcneu{T}{\alpha_1'}{\alpha_2'}{\alpha_3'}{\ul k'}{k}
\,,
\\
\bigbra{T^\vee;\ld{\alpha_1}{\alpha_2}{\alpha_3,\ul k},\mu_3}
\Pi^{-1}
\bigket{T \bcdot T;\join{\alpha_1}{\alpha_2}{(\lambda,k)},\mu}
=&
\delta_{\lambda,\lambda_3} \, \delta_{\mu,\mu_3}
\,
\rctneu{T}{\alpha_1}{\alpha_2}{\alpha_3}{\ul k}{k}
\,,
\end{align*}
we obtain
\begin{align*}
\dcq{T}{\alpha_1^{\phantom{\prime}}}{\alpha_1'}(\ul a)
\,
\dcq{T}{\alpha_2^{\phantom{\prime}}}{\alpha_2'}(\ul a)
& =
\sum_{\alpha_3,\alpha_3'}
\sum_{\ul k,\ul k'}
\sum_{k}
\rcneu{T}{\alpha_1'}{\alpha_2'}{\alpha_3'}{\ul k'}{k}
\,
\rctneu{T}{\alpha_1}{\alpha_2}{\alpha_3}{\ul k}{k}
\\
& \hspace{0.25cm} 
\cdot
\left(
\sum_\mu
\bigbra{T^\vee;\ld{\alpha_1'}{\alpha_2'}{\alpha_3',\ul k'},\mu}
D^{\ul j_1\!\ast\!\ul j_2}(\ul a)
\bigket{T^\vee;\ld{\alpha_1}{\alpha_2}{\alpha_3,\ul k},\mu}
\right)
,
\end{align*}
where $\alpha_3$ and $\alpha_3'$ assign the same highest weight $\lambda_3$ to the root and $\mu \in \weight(\lambda_3)$. We plug in \eqref{LeafDuplicateStates} and use unitarity of the isomorphisms \eqref{G-TePr-Zlg}, as well as \eqref{G-CCG-Entw} and \eqref{Form-chi}, to rewrite the sum over $\mu$ as
\ala{
&
\sum_\mu
\bigbra{T^\vee;\ld{\alpha_1'}{\alpha_2'}{\alpha_3',\ul k'},\mu}
D^{\ul j_1\!\ast\!\ul j_2}(\ul a)
\bigket{T^\vee;\ld{\alpha_1}{\alpha_2}{\alpha_3,\ul k},\mu}
\\
& \hspace{0.5cm}
=
\sum_{\mu} 
\,
\sum_{\ul\mu\in\weight(\ul\lambda_3)}
\,
\sum_{\ul\mu'\in\weight(\ul\lambda_3')}
\big(\ccg{T}{\alpha_3'}{\ul\mu'}{\mu}\big)^\ast
\,
\ccg{T}{\alpha_3}{\ul\mu}{\mu}
\\
& \hspace{1cm}
\cdot
\left(
\prod_{n=1}^N 
\bigbra{\boldsymbol\vee;\big(\lambda_1^n,\lambda_2^n,(\lambda_3'^n,k'^n)\big) , \mu'^n}
D^{\lambda_1^n} \otimes D^{\lambda_2^n}(a_n)
\bigket{\boldsymbol\vee;\big(\lambda_1^n,\lambda_2^n,(\lambda_3^n,k^n)\big) , \mu^n}
\right)
\\
& \hspace{0.5cm}
=
\delta_{\ul\lambda_3 \ul\lambda_3'}
\,
\delta_{\ul k \ul k'}
\,
\sum_{\mu} 
\,
\sum_{\ul\mu\in\weight(\ul\lambda_3)}
\,
\sum_{\ul\mu'\in\weight(\ul\lambda_3')}
\big(\ccg{T}{\alpha_3'}{\ul\mu'}{\mu}\big)^\ast
\,
\ccg{T}{\alpha_3}{\ul\mu}{\mu}
\,
\bigbra{\ul\lambda_3 \ \ul\mu'} D^{\ul\lambda_3}(\ul a) \bigket{\ul\lambda_3 \ \ul\mu}
\\
& \hspace{0.5cm}
=
\delta_{\ul\lambda_3 \ul\lambda_3'}
\,
\delta_{\ul k \ul k'}
\,
\sum_{\mu} 
\left\langle T;\alpha_3',\mu \right|
D^{\ul\lambda_3}(\ul a)
\left| T;\alpha_3,\mu \right\rangle
\\
& \hspace{0.5cm}
=
\delta_{\ul\lambda_3 \ul\lambda_3'}
\,
\delta_{\ul k \ul k'}
\,
\dcq{T}{\alpha_3^{\phantom{\prime}}}{\alpha_3'}(\ul a)
.
}
Plugging this in, we finally obtain the assertion.

\end{appendix}

\end{document}